\documentclass[AMA,STIX1COL]{WileyNJD-v2}

\usepackage[colorinlistoftodos]{todonotes}

\usepackage{siunitx}
\usepackage{chngcntr}
\usepackage{centernot}
\usepackage{xcolor}
\usepackage{amssymb}
\usepackage{pifont}
\usepackage{multirow}
\usepackage{lscape}
\usepackage{longtable}
\usepackage{amsmath}

\DeclareMathOperator*{\argmax}{argmax} 
\DeclareMathOperator*{\argmin}{argmin} 

\articletype{MAIN PAPER}%
\received{<day> <Month>, <year>}
\revised{<day> <Month>, <year>}
\accepted{<day> <Month>, <year>}

\begin{document}

\title{Modern approaches for evaluating treatment effect  heterogeneity from clinical trials and observational data}

\author[1]{Ilya Lipkovich}
\author[2]{David Svensson}
\author[3]{Bohdana Ratitch}
\author[4]{Alex Dmitrienko}

\authormark{LIPKOVICH \textsc{et al}}

\address[1]{ \orgname{Eli Lilly and Company}, \orgaddress{\city{Indianapolis}, \state{Indiana}, \country{USA}}}
\address[2]{ \orgname{AstraZeneca}, \orgaddress{\city{Gothenburg},  \country{Sweden}}}
\address[3]{ \orgname{Bayer Inc.}, \orgaddress{\city{Mississauga}, \state{ON}, \country{Canada}}}
\address[4]{ \orgname{Mediana}, \orgaddress{\city{San Juan}, \state{Puerto Rico}, \country{USA}}}

\corres{Ilya Lipkovich,  Eli Lilly and Company, Indianapolis, IN 46285, USA\\
\email{ilya.lipkovich@lilly.com}}

\abstract[Abstract]{
In this paper we review recent advances in statistical methods for the evaluation of the heterogeneity of treatment effects (HTE), including subgroup identification and estimation of individualized treatment regimens, from randomized clinical trials and observational studies. We identify several types of approaches using the features introduced in Lipkovich, Dmitrienko and D'Agostino\cite{Lipkovich2017} that distinguish the recommended principled methods from basic methods for HTE evaluation that typically rely on rules of thumb and general guidelines (the methods are often referred to as common practices). We discuss the advantages and disadvantages of various principled methods as well as common measures for evaluating their performance. We use simulated data and a case study based on a historical clinical trial to illustrate several new approaches to HTE evaluation.
}
\keywords{Subgroup identification, Personalized medicine, Individualized treatment regimen}
\maketitle

\section{Introduction}\label{introduction}

Exploration of subgroups  of the overall clinical trial population and HTE evaluations in general have attracted a lot of attention in the clinical trial literature over the past 10-15 years. Most often, the following two settings are considered in clinical trials: 
\begin{itemize}
    \item Estimation of treatment effects in a relatively small number of pre-defined subgroups (typically based on key demographic or clinical covariates) as per regulatory guidance.\cite{FDASIA2018,EMA2019subgconf,FDASIA2018,FDA2022}
    \item Data-driven assessments of treatment effect heterogeneity. 
\end{itemize}    
This paper focuses on the latter setting. We view data-driven subgroup identification and analysis as statistical problems for evaluating the heterogeneity of treatment effect across subsets of the overall population and argue for principled data-driven subgroup identification approaches.\cite{Ruberg2015, Lipkovich2017}
 
HTE evaluations are primarily presented under the headings of ``precision/personalized/stratified medicine'' generating a certain ``subgroup analysis mythology''.  It can be viewed as a tension among ``common practices'', ``good practices'' and rigorous methods for subgroup evaluation. The ``common practices'' are often criticized as a clinical trial sponsor's attempt to make questionable claims of efficacy in selected patient subgroups, often as a ``salvaging strategy'' in the absence of a convincing treatment effect in the overall population (see systematic reviews\cite{Sune2012,Wallach2017} and references therein). These poor practices are often contrasted with ``good practices'' of subgroup analysis that have been developed under various guidelines as a reaction to these practices.\cite{Brookes2001, ROTHWELL2005, Sun2010} Even though these guidelines are more principled, they are still quite vague, lack theoretical justification and do not address all relevant issues. Here we review the numerous and substantial challenges of subgroup assessments in clinical trials/observational studies and state-of-the-art methods for HTE evaluation. 

Examples of ``common practices'' in subgroup analysis include:
\begin{itemize}
    \item The contention that multiplicity does not need to be controlled in the subgroup analysis since these analyses are for internal decision making.
    \item Uncertainty of multi-stage subgroup assessment strategies is often not fully accounted for in the last stage; for example, ``preliminary data looks'' involving decisions to remove insignificant interaction terms are ignored, rendering the final inferences in the selected subgroups invalid (see Zhao et al \cite{Zhao2018} for discussion).
    \item Subgroup assessments are not based on a pre-defined algorithm and often involve human ad hoc decisions that are rarely captured and reported.
    \item Subgroup assessments often rely on simple covariate-by-treatment interaction tests in parametric models as  a ``gatekeeper'' for identifying predictve covariates. These tests are often implemented using a one-covariate at-a-time strategy that has been subject to criticism.\cite{Ruberg2010, Lipkovich2017,Kent2018, VanderWeele2019,Dmitrienko2020}
    \item The treatment effect is not permitted to be tested in any subgroup unless the effect in the overall population is significant (this condition is often labeled ``consistency'') which is at odds with the needs of personalized/precision medicine.
\end{itemize}

Modern data-driven approaches to evaluating individual causal effects and subgroup analyses emerged as the cross-pollination of three fields: causal inference, machine learning (ML) and multiple hypothesis testing. While these areas are typically represented by distinct groups of researchers, the needs of personalized medicine have fostered collaboration and interdisciplinary efforts in this area over the last decade. 

\textit{The connection between personalized/precision/stratified medicine and causal inference} is not obvious. Some researchers may see the need for statistical methods in personalised medicine largely in the context of disease progression models under no treatment (or under standard of care) given individual prognostic factors. Patients whose prognosis is poor could be assigned to a more aggressive treatment/higher dose, especially, once early signs of critical changes in the key outcomes have been observed.\cite{Frohlich2019} The distinction between \textit{prognostic} factors (i.e., predictors of clinical outcomes under no treatment) and \textit{predictive} factors (i.e., predictors of treatment effect or treatment effect modifiers) is not well established in the literature on precision medicine.\cite{Kent2018, Rekkas2020} Consequently, it is not well understood that treatment decisions should be driven by predictive rather than purely prognostic factors. In this paper, we emphasize that individualized treatment decisions should be based on predicted individual treatment effects (ITEs), that is, hypothetical  differences in outcomes for a patient  who would undergo multiple treatments, as if simultaneously present in ``parallel worlds''. This reflects a fundamental challenge of causal inference: only one of multiple treatment outcomes is measured on each patient, corresponding to the treatment actually taken, and thus needs to rely on advanced methods, e.g., a missing data framework. This is different from a conventional setting of predictive learning, often used for prognostic modeling, where the outcomes of interest are fully observed in a training sample. 

\textit{Connections of personalized medicine with statistical/machine learning} stem from the fact that the covariate space for estimating individual or patient-level treatment effects may be high-dimensional, making variable/model selection and regularization critical. A na\"ive application of machine learning methods for predictive learning to compute ITEs, say, by tuning predictive models separately for each trial arm, may lead to a large regularization bias. Estimating ITEs often involves a multi-stage strategy. If each stage is optimized separately, the final estimation target may be far from  optimal. One challenge of estimating ITEs is that one may need to regularize prognostic and predictive effects differently. Furthermore, evaluating individual treatment effects and identifying patient subsets from non-randomized trials by ML is complicated by the need to model the propensity (treatment choice) process resulting in additional nuisance parameters. Unsurprisingly, high expectations about bringing ML to solve key problems in personalized medicine were overoptimistic causing negative reactions in medical and scientific communities (see Fr\"ohlich et al\cite{Frohlich2019} and references therein).    

\textit{The connection between personalized medicine and multiple hypothesis testing} also have to be understood in the context of data-driven methods that are typically required for estimating causal parameters, often involving high-dimensional data. Statistical inferences following an application of ML (known as post-selection inferences) are challenging due to multiplicity and model selection inherent in tuning complex black box models to high-dimensional data. Fundamentally, the challenge of post-selection inferences is that the set of hypotheses of interest is not pre-specified but data-driven. Therefore, most of the methods for addressing multiplicity developed in the statistical literature (such as traditional multiplicity adjustment methods in clinical trials\cite{Dmitrienko2013}) will not be applicable. Often, the distribution of the test statistics under the null hypothesis of no effect following model selection is analytically intractable and researchers have to resort to resampling methods\cite{Meinshausen2010,Dezeure2015} (despite recent theoretical breakthroughs in selective inference for LASSO and related methods,\cite{Lee2016, Tibshirani2016, Taylor2018} and regression trees\cite{Neufeld2022}). 
These challenges are exacerbated in the context of personalized medicine where, unlike in supervised learning, the target quantities for evaluating HTEs (such as ITEs) cannot be evaluated even in a training sample. Furthermore, while the notions of statistical significance and inference often play a secondary role in predictive modeling, statistical applications in personalized medicine require that analysis objectives should be stated within the estimand and hypothesis testing framework. Lipkovich et al\cite{Lipkovich2018} presented an overview of multiplicity issues in exploratory subgroup analysis. We review recent advances in the development of inferential methods for HTE evaluation in Section~\ref{sec.post.inf}.

While several overviews of existing methodologies and tutorials have appeared in recent years,\cite{BICA, Sies2019, UPLIFT2016, UnifiedSurvey,JacobsCATE,Hoogland2021,Segal2023,Salditt2023, Sechidis2024} they are typically tailored to one specific area or application. The categorization of methods in many reviews appear to be based on arbitrary features, making it harder to see key innovations which are often masked under variations of non-essential elements. Some reviews may simply list different methods for outcomes modeling (which is merely a component of the overall strategy) as distinct methods for HTE assessment. 
Some of these are more high-level and theoretical, while others are focusing on modeling options and software in a practical setting while leaving out important classes of recent approaches. Some papers discuss statistical pros and cons for various approaches but do not consider machine learning, and some papers do not cover inferential aspects (such as bias adjustment). Some reviews discuss strategies for evaluating HTE in randomized trials\cite{Sechidis2024} whereas others focus on the real-world setting\cite{Segal2023}.

This tutorial provides an overview of recent methods for assessing HTE in the context of medical applications (personalized medicine) and covers both randomized controlled trials (RCT) and observational studies. Although we do not aim to evaluate the performance of all individual methods, we included a simulation study under different scenarios to illustrate their performance with several evaluation metrics and a case study to illustrate key challenges of evaluating HTE. We are building on a previous review paper\cite{Lipkovich2017} and provide a unifying account of recent innovations in the development of methods for HTE evaluation, such as subgroup identification and estimation of individualized treatment regimens (ITR), scattered across the diverse literature. In our presentation we try to make connections with different fields and segments of literature on causal inference, machine learning, and post-selection inference.    

The rest of the tutorial is organized as follows. Section \ref{problem.statement} places HTE evaluation within the causal inference framework; Section~\ref{key.features} discusses the key elements of data-driven methods for HTE evaluation; Section~\ref{tech.details} provides technical details on the recent methodological advances in various domains of HTE evaluation using machine learning methods, and Section~\ref{case.study} illustrates selected methods using a case study with synthetic data representing a randomized clinical trial (the R code is provided in supplementary materials along with the synthetic data example); Section~\ref{sec.post.inf} provides an overview of existing approaches to formal statistical inference for HTE evaluation, which is a challenging and often ignored topic; Section~\ref{simdata.description} contains a simulation study with several scenarios of increasing complexity that pinpoints the many challenges of HTE evaluation; the paper  concludes with a summary and discussion in Section~\ref{summary.discussion}.

\section{Causal framework for HTE evaluation}\label{problem.statement}


We begin by introducing a general framework for assessing HTE. It is worth pointing out that several application fields make use of the HTE assessment methods: biostatistics (estimating individualized treatment regimens, the focus of our review), public policy makers (e.g., evaluating who would benefit most in terms of future earnings from a certain type of a training program), and market researchers (e.g., identifying what type of consumers should be targeted with a specific form of advertisement). Moreover, the input data for HTE evaluation/policy research comes either from randomized clinical trials or observational studies which require different approaches and involve different research communities. As a result, contributions come from a diverse set of journals concerned with pharmaceutical statistics, econometrics, machine learning, etc. Different researchers use different terminology that often refers to similar mathematical concepts. As an example, the gain resulting from  treatment assignment based on a data-driven treatment regimen can be conceptualized as a ``value function''\cite{Zhang2021} in medical research, ``welfare function''\cite{Kitagawa2018} in public policy, or ``uplift''\cite{Radcliffe2007, Zaniewicz2013, Soltys2014} in market research. Sometimes an existing method is rediscovered by different communities and appears under different names. 

We use a formal approach based on the causal language to define the heterogeneous treatment and subgroup effects, which provides a unifying framework for RCT and observational studies. First, we express the individual treatment effect for a binary or continuous outcome $Y$ in terms of potential outcomes (PO)\cite{Neyman1923, Rubin1974} as 
\[
\Delta_i=Y_i(1)-Y_i(0), 
\]
where $Y_i(t)$, $t\in \{0,1\}$, is a PO that would have been observed, had the $i$th subject been treated (possibly contrary to the fact) with the treatment $T=t$ (in a RCT, $t=0$ and $t=1$ correspond to the control and experimental treatments, respectively). We assume \textit{consistency} of the potential and observed outcomes implied by a more general \textit{Stable Unit Treatment Value Assumption}\cite{Rubin1978} (SUTVA):
\[
Y_i=Y_i(T_i)=Y_i(1)T_i+Y_i(0)(1-T_i),
\]
where $T_i$ is the treatment actually received and $Y_i$ is the observed outcome for the $i$th subject. Typically, we are interested in modeling the heterogeneity of ITE as a function of observed patient characteristics leading to the conditional average treatment effect (CATE), defined as
\[
\Delta(x_i)=E(Y_i(1)-Y_i(0)|X=x_i),
\]
where $x_i=(x_{1i},\cdots,x_{pi})$ is a vector of $p$ biomarkers, denoted by $X_1,\cdots,X_p$, for the $i$th subject. Dropping the patient index, let $m(t,x)=E(Y(t)|X=x), t \in \{0,1\}$, and then define $\Delta(x)=m(1,x)-m(0,x)$. Note that under strong treatment ignorability, ensured by randomization in RCTs and assumed in observational studies conditional on the covariates (Rosenbaum and Rubin \cite{rosenbaum1983propscore}), we can replace the POs with the conditional expectations of the observable random variables, i.e., $m(t,x)=E(Y|T=t, X=x)$. Note that we can represent the response surface without loss of generality as 
\begin{equation}\label{eq.meanresp}
m(t,x)=h(x)+\frac{1}{2}\Delta(x)(2t-1), t \in \{0,1\},
\end{equation}
where $h(x)$ is the main covariate effect, i.e.,  
\begin{equation}\label{eq.maineff}
h(x)=\frac{1}{2}\left( m(1,x)+m(0,x) \right).
\end{equation}
In observational data with non-randomized treatments, identifying causal effects, such as the average treatment effect (ATE) and CATE, requires additional assumptions. First, we need to assume treatment ingnorability conditional on the observed covariates (a.k.a. unconfoundedness or ``no unmeasured confounders''), i.e., $T \perp \{Y(1),Y(0) \} \mid X$. Second, we often need to estimate the propensity function $\pi(x)=\text{Pr}(T=1\mid X=x)$ from the observed data. To be able to make valid inferences, we have to assume \textit{positivity}, i.e., the propensities are bounded away from 0 and 1, $ 0 < \pi(x) < 1$. Occasionally we will use a general treatment assignment function $\pi(t,x)=\text{Pr}(T=t\mid X=x)$ allowing us to write some expressions more compactly, with propensity $\pi(x) \equiv \pi(1,x)$. 

It is sometimes important to distinguish a CATE \textit{estimator} from ITE \textit{estimates}. The former is more general and represents an algorithm for finding an estimate of the expected ITE, i.e., $\widehat{\Delta}(x)$, given a subject's covariate vector $x$, even if we do not have access to outcome data for that  subject. The latter simply means that we have somehow computed an estimate $\widehat{\Delta}_i$ for the $i$th subject with the covariate profile $x_i$. Suppose that the $i$th subject belongs to the experimental treatment arm, then we could compute $\widehat{\Delta}_i$ by matching him/her with a subject (or multiple subjects) from the control arm having similar covariate profile(s) using the nearest neighbor or other methods (see, for example, a matching tree approach of Zhang et al\cite{ZHANG2021MT}). We can estimate (impute) the missing potential outcome $Y_i(0)$ by ``borrowing'' outcomes from the matched subjects, $\widehat{Y}_i(0)$, and then compute $\widehat{\Delta}_i=Y_i-\widehat{Y}_i(0)$. Such imputation methods do not yet give us a general algorithm for predicting CATE for an arbitrary patient with known covariates $X$, rather they allow us to predict ITEs by imputing missing (counterfactual) PO's for a set of subjects with an observed outcome for one of the candidate treatments.   

Let us assume we obtained a good estimate of CATE, $\widehat{\Delta}(x)$. The next step in subgroup identification may proceed in two directions:  
\begin{itemize}
\item Define a subgroup as a set of all subjects with a positive treatment effect, i.e., $\widehat{S}(x)=\{x: \widehat{\Delta}(x) > \delta\}$ with a pre-defined $\delta$, e.g., $\delta=0$. 
\item Define subgroups using interpretable rules, e.g., subgroups may be obtained by fitting a regression tree to the estimated $\widehat{\Delta}(x)$ as the outcome variable. In this case, subgroups would be defined as terminal nodes of the resulting tree having a large average treatment effect. 
\end{itemize}
The first approach is closely related to developing individualized treatment assignment rules (regimens) that, given a subject's covariate profile $X=x$, choose an optimal (in some sense) treatment, $D(x) \in {0,1}$. The second approach is more heuristic and does not ensure that each subject in the selected subgroup(s) would have a positive individualized treatment contrast. However, it may be attractive in producing easy-to-interpret subgroups based on one or more biomarkers.

While this tutorial focuses on quantifying individual treatment effects via the mean function of the conditional distribution of potential outcomes, $E(Y(t)|X)$, some researchers argue that modeling the $p$th quantiles $F^{-1}_{Y(t)|X}(p)$ of conditional distributions $Y(t)|X$, may provide a more natural framework for studying treatment heterogeneity.\cite{Giessing2021} This work will be not be presented here.   

\section{Key elements of data-driven methods for HTE evaluation}\label{key.features}
As advocated for in Lipkovich at al, \cite{Lipkovich2017} data-driven methods for treatment effect heterogeneity evaluation should support several key features: 
\begin{itemize}
 \item The Type I error rate/false discovery rate for the entire subgroup search strategy needs to be evaluated.
 \item Uncertainty of the entire subgroup search strategy needs to be accounted for.
 \item Complexity control to prevent data overfitting and covariate selection bias (e.g., in tree-based methods when choosing among covariates with different numbers of candidate splits) needs to be implemented.
 \item Reproducibility of the identified subgroups needs to be assessed.
 \item ‘Honest’ estimates of treatment effects in the identified subgroups needs to be obtained.
\end{itemize}

Accordingly, when reading papers on HTE evaluation and subgroup identification, it is important to understand how the authors address these features. Additionally, one needs to pay attention to the following aspects of the underlying methodology.    

\textit{
Does the method apply to RCTs only, observational studies only or both?} For non-randomized data, there is a complex interplay between confounders (predicting both treatment assignment and outcome) and modifiers of treatment effect (predictors of CATE), making  model selection more challenging. More formally, if the confounders $Z$ are distinct from the modifiers $X$ in the sense that some variables in $Z$ are not in $X$, the assumption of strong treatment ignorability does not hold with conditioning on $X$ alone, therefore $m(t,x) \ne E(Y|T=t,X=x)$ resulting in misspecified estimates of ITE, $m(t,x)$, using na\"ive machine learning methods that may easily overlook confounders.\cite{Vegetabile2021} Valid estimation of ITE may require marginalizing over confounders,\cite{Sugasawa2021} e.g., estimating $m(t,x) = E\{E(Y|T=t,X=x, Z=z)\}$, where the outer expectation is computed with respect to the distribution of the covariates in $Z$ that are not part of the set $X$ ($Z\backslash X$) in the overall population. 

\textit{What is the size of the covariate space ($P$) that a proposed method can handle?} Some methods assume a very low-dimensional setting, and some target the identification of a single continuous biomarker (STEPP\cite{BonettiSTEPP2000}, Han et al \cite{Han2022}). At the other extreme, the number of candidate predictors $P$ may be larger than the sample size $n$ and grow as a function of $n$ \cite{chernozhukov2018}. 

\textit{What is the complexity of the ``model space'' and how is it controlled to prevent over-fitting?} For HTE methods that involve multiple modeling steps, this is not obvious. For example, consider a strategy that first estimates $\Delta(x)=E(Y|T=1,X=x)-E(Y|T=0,X=x)$ by positing penalized regressions (e.g., LASSO) within each treatment arm and then fitting a shallow regression tree with up to 2 levels to estimate $\Delta(x)$ as the outcome variable. This results in subgroups defined by simple biomarker signatures with up to 2 variables, e.g., $\widehat{S}(x)=\{x:X_1 \le c_1, X_3 > c_2\}$, yet the models used as building blocks in the first step are much more complex.  Complexity control applied at each step separately may not guarantee optimality of the entire modeling strategy. In addition to balancing between the bias and variance of the estimated quantities, obtaining interpretable solutions is desirable in the context of HTE evaluation and is sometimes an important consideration in the complexity control strategy. Instead of forcing interpretable low-complexity models, as mentioned above, some approaches may be amenable to utilizing explainability tools for ``black-box'' models, such as variable importance scores or low-dimensional graphs depicting marginal effects.

\textit{What elements of the strategy are pre-defined versus data driven?} In a data-driven HTE evaluation process, subgroups and predictive biomarkers are meant to be discovered rather than pre-specified and constitute possible outputs from the analysis procedure. Nevertheless, all aspects of data preparation and analysis methodology are expected to be pre-specified, including details on the models used at various stages of the analysis strategy, related hyper-parameter settings and tuning parameters, feature selection, etc. It is recommended to follow best practices for planning and reporting HTE evaluation{Walsh2021,Stevens2020}.

\textit{What output is produced by the analysis method?} 
The key types of findings from HTE evaluation methods can include (i) a list of biomarkers that the method identified as predictive or a list of candidate biomarkers with associated variable importance scores or rankings, (ii) individual   benefit scores (i.e., estimated ITEs) as a function of covariates, (iii) optimal treatment regimens as a function of covariates, (iv) subgroup signatures or classifications as a function of covariates.

\textit{What formal inferences are performed (if at all)?} For example, inferences on HTE can address (i) the global hypothesis of treatment effect heterogeneity, i.e., $H_0: \Delta_i = \Delta$; often it is stated as lack of HTE explainable by covariates, i.e., that CATE is constant across the covariate space: $H_0: \Delta(X) = \Delta$; (ii) individual treatment effects, i.e., providing confidence intervals for $\Delta(X)$; (iii) treatment effects in the identified subgroup(s), ($E(Y(1)-Y(0)|\widehat{S}(X)=1)$); (iv) the value of the estimated treatment regimen, i.e., the outcome expected if everyone would receive the treatment consistent with the estimated regimen, $E(Y(\widehat{D}(X))$).   

Resampling-based methods are commonly employed to account for the uncertainty of model selection and address multiplicity issues arising in subgroup identification and related data-driven methods for HTE evaluation. In general, it is important to distinguish between the roles of resampling in different (although often interrelated) contexts such as (i) tuning hyper-parameters of ML methods via cross-validation, (ii) controlling the Type I error rate or false discovery rate of ML strategies and computing p-values for the hypotheses of interest, (iii) computing bias-corrected point estimates and associated confidence intervals for various quantities of interest, and (iv) evaluating the performance of ML-based procedures based on a single data set, i.e., in the absence of any knowledge about the true parameters and independent test data.

\section{Building blocks of methods for HTE evaluation}\label{tech.details}

As proposed in Lipkovich et al \cite{Lipkovich2017}, it is convenient to conceptualize methods for subgroup identification using the following four categories:

\begin{itemize}
\item Global outcome modeling, $m(t,x)=E(Y|T=t, X=x)$.   
\item Direct modeling of individual treatment effects conditional on observed covariates (CATE), $\Delta(x)$.
\item Direct modeling of ITR, $D(x)=I\{\Delta(x) > 0\}$.
\item Direct subgroup search, with a subgroup $S(x)$ having a ``simple structure'' in the covariate space, e.g., a rectangular set.
\end{itemize}

\begin{figure*}[!ht] 
\centering
\includegraphics[scale=0.9] {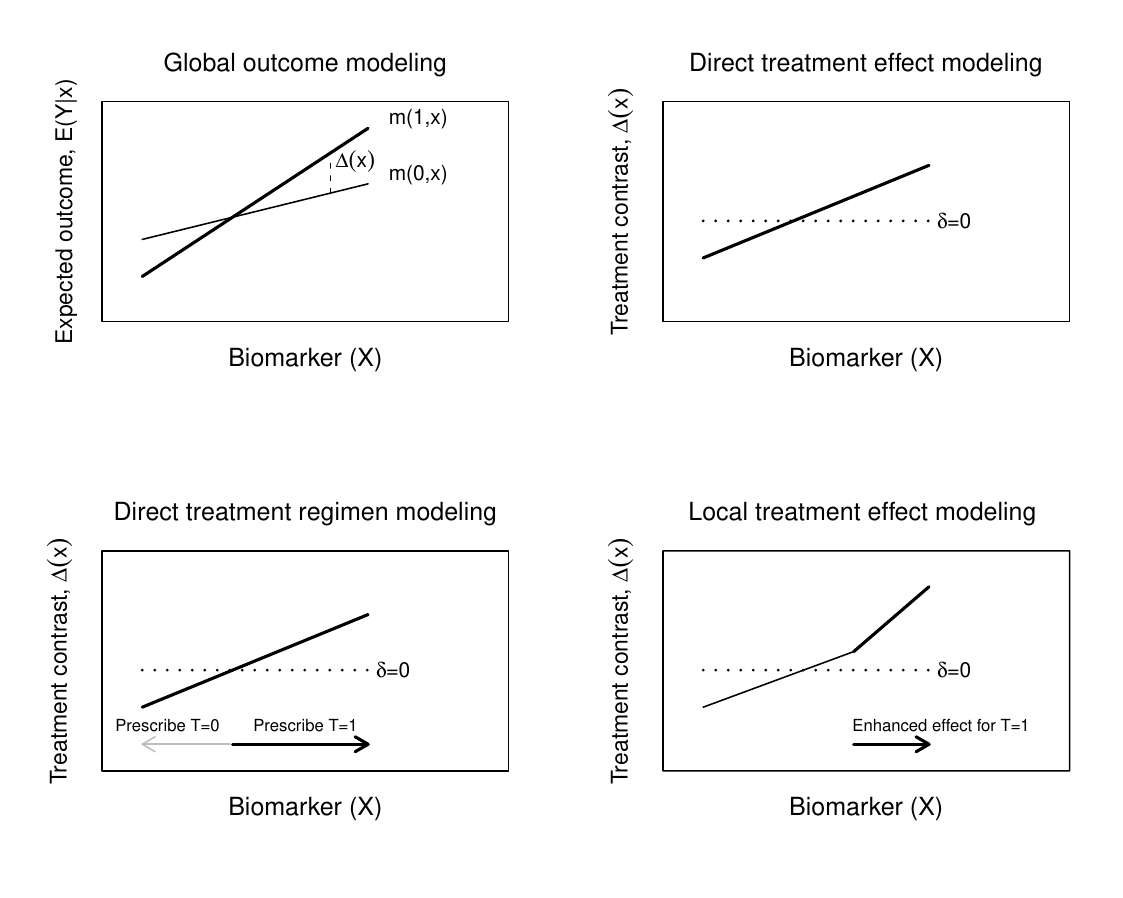}
\caption{Typology of HTE evaluation methods}
  \label{fig.typology}
\end{figure*}

This typology is presented graphically in Figure~\ref{fig.typology} and further detailed in the subsections below with key contributions made since the previous review\cite{Lipkovich2017}. It should be noted that, in recent years, hybrid approaches have been developed to combine methods from different classes, e.g., to combine global outcome modeling methods with global treatment effect modeling, or methods for direct modeling of  treatment effect can be combined with a subsequent search of ITR using the CATE estimated at the first stage. 

To help the reader better understand the methods and their applications, Table~\ref{tab.appendix.keyfeatures} in the Appendix summarizes the key features of selected methods.

\subsection{Estimating CATE via global outcome modeling}\label{global.outcome.mod}
This class of methods combines a number of multi-stage strategies, recently dubbed \textit{metalearners}{Kuenzel2019}, to emphasize that ``meta-algorithms decompose estimating the CATE into several regression problems that can be solved with any regression or supervised learning method''. This name, now widely accepted,  is not very descriptive, tempting researchers to declare any strategy for estimating CATE via some off-the-shelf software a ``novel metalearner''.

A significant method that serves as a precursor of several classes of metalearers was the Virtual Twins method by Foster, Taylor and Ruberg.\cite{Foster2011} The idea was to estimate first the response surface as a function of the candidate predictors and treatment indicator, $m(t,x)$, using random forests (RF)\cite{Ishwaran2008} or another black-box method, such as gradient boosting (XgBoost),\cite{Chen2016} neural networks, or Bayesian additive regression trees (BART)\cite{Sparapani2021}. At the second stage, ITEs are predicted for each subject as 
\begin{equation}\label{eq.indirect.ITE}
\widehat{\Delta}(x)=\widehat{m}(1,x)-\widehat{m}(0,x)
\end{equation}
and used as the response variable for a tree-based regression against the same candidate predictors. The terminal nodes of the tree represent interpretable subgroups described in terms of covariates  (``biomarker signatures'') and those with an enhanced treatment effect can be used for the development of personalized/stratified medicine in subsequent drug development programs. Alternatively, subgroups can be constructed by selecting subjects with estimated $\widehat{\Delta}(X)$ exceeding a threshold; in this case researchers often undertake additional steps to obtain more interpretable subgroups.

Depending on whether the response $m(t,x)$ is learned at the first stage as a single model on the combined data from the two arms or separately by treatment arms, the procedure is referred to as an S-learner ("S" stands for \textit{single} regression) or a T-learner (``T'' stands for \textit{two} regressions).\cite{Kuenzel2019} To emphasise that the response surface is obtained separately in each trial arm, we will be sometimes refer to the estimate for the arm $t \in \{0,1\}$ as $m_t(x)$, where $m_t(x) \equiv m(t,x)$. 

Both S- and T-learning methods may fail to capture the predictive effects accurately. For example, a S-learner based on a RF model may fail to capture important treatment-by-covariate interaction effects if the fit is dominated by splits on prognostic factors. To ensure the predictive effects are not missed, researchers\cite{Foster2011,Hermansson2021} proposed including in the set of candidate predictors for RF treatment-by-covariate interactions $X \times T$, which in some scenarios proved helpful. Some researchers argued for separately fitting outcome models for the treated and control groups while taking into account the lack of balance between the groups (in the setting with non-randomized treatment assignments). For example, Shalit et al\cite{Shalit2017} proposed a neural network version of T-learning where the outcome functions $m_0(X)$ and $m_1(X)$ are learned simultaneously by a network with separate but jointly trained compartments (``heads'') for the two functions, while the covariates are transformed from their original space into the so-called \textit{representation} space by a one-to-one $p$-dimensional function $\Phi(X)$ learned from the data; therefore the solution is obtained via $m_t(x)=m(t,\Phi(x)), t \in \{0,1\}$. The loss function incorporates a penalty term for the covariate misbalance in the representation space $\Phi(X)$ between the treated and control groups, in hopes that identifying and incorporating in the solution such a balanced space would indirectly improve estimation of the target causal parameter, $\Delta(x)$. On the other hand, fitting models separately in each arm in T-learning may induce spurious treatment-by-covariate interactions in the estimated CATE. The presence of regularization bias in CATE based on modeling outcomes in each trial arm via ML methods was discussed in Chernozhukov at al. \cite{Chernozhukov2020} It was argued\cite{Kuenzel2019} that such bias is particularly likely when the sample sizes in the two arms are drastically different. For example, when analyzing observational data, there may be many more subjects in the control than in the treatment arm. As a result, a non-parametric ML estimator for the response in the control arm, $\widehat{m}(0,x)$, may be of much higher complexity than that for the treatment group, $\widehat{m}(1,x)$, and the final estimator of CATE obtained as their difference may have spurious ``wiggles'' contributed by the estimator from the control arm. 

Motivated by such a scenario,  K\"unzel et al \cite{Kuenzel2019}  proposed a method called X-learner, which is a ``hybrid'' estimator of CATE formed as a weighted average of two estimators $\widehat{\Delta}_0(x)$ and $\widehat{\Delta}_1(x)$ constructed using a multi-step procedure. For $\widehat{\Delta}_1(x)$, first, ``coarse'' estimates of ITE are computed for subjects in the treatment arm by contrasting their observed potential outcomes $Y_i(1)=Y_i, i\in \{i: T_i=1\}$ with the counterfactual outcomes $\widehat{Y}_i(0)$ predicted from the response model $m_0(x)$ estimated from the control arm. That is, $\widehat{\Delta}_i=Y_i-\widehat{m}_0(X_i)$, $i \in \{i: T_i=1 \}$. At the second step, $\widehat{\Delta}_i, i \in \{i: T_i=1 \}$ are modeled as a response variable using the treatment arm alone, resulting in the estimator $\widehat{\Delta}_1(x)$ that can be evaluated on any subject as a function of the candidate covariates $X$. Similarly, the second estimator $\widehat{\Delta}_0(x)$ is constructed utilizing the observed outcomes from subjects in the control arm and their counterfactual responses predicted from a model for $m_1(x)$ estimated in the treatment arm, $\widehat{\Delta}_i=\widehat{m}_1(X_i)-Y_i$, $i \in \{i: T_i=0\}$. Then $\widehat{\Delta}_0(x)$ is obtained by regressing $\widehat{\Delta}_i$ on the covariates using observations from the control arm. The final estimator for the X-learner is obtained as
\begin{equation}
\widehat{\Delta}(x)=w(x)\widehat{\Delta}_0(x)+(1-w(x))\widehat{\Delta}_1(x), 
\end{equation}
where the weight function is often taken as the estimated propensity score, $w(x)=\widehat{\pi}(x)= \widehat{\text{Pr}}(T=1|X=x)$ or a constant probability of treatment assignment (for RCT). The reason why the X-learner was expected to improve on a T-learner is that in the described situation where the control arm is much larger than the treated arm, each of the two estimators $\widehat{\Delta}_1(x)$ and $\widehat{\Delta}_0(x)$ are based on contrasting the observed and counterfactual outcomes predicted from the alternative arm rather than on contrasting the predicted potential outcomes from models fitted to different arms. With the X-learner, the differences in the complexity of models fitted to each respective arm would be ``smoothed out'' within each of the two estimators, $\widehat{\Delta}_0(x)$ and $\widehat{\Delta}_1(x)$.

This example underscores challenges of tuning model complexity (regularization) when fitting prognostic models while the interest lies in the assessment of predictive effects. X-learning tries to tackle this issue using a rather informal trick. Some researchers proposed more structured approaches by separate regularization of prognostic and predictive effects within the S-learning framework. Imai and Ratkovic \cite{Imai2013} developed a penalized parametric regression model with two different penalty parameters, one for prognostic effects and the other for predictive effects (treatment-by-covariate interactions). The idea is that the latter may need harsher penalties. Similar considerations led to a procedure based on Bayesian additive trees\cite{HahnBART} with different priors placed on predictive and prognostic effects.

It is worth noting that X-learning can be seen as a generalization of a 3-step imputation-based procedure for building a CATE estimator. At the first step, estimators $\widehat{m}_0(x)$ and $\widehat{m}_1(x)$ of the mean outcomes in each treatment arm are obtained. At the second step, ITEs are imputed as $\widehat{\Delta}_i=T_i(Y_i-\widehat{m}_0(X_i))+(1-T_i)(\widehat{m}_1(X_i)-Y_i)$. Finally, at the third step, the imputed values $\widehat{\Delta}_i$ are used as outcomes for a generic ML method to obtain $\widehat{\Delta}(x)$.\cite{HuangShu2022}


\subsection{Direct modeling of CATE}\label{direct.ITE.mod}
A natural response to many challenges of tuning complexity parameters for predictive and prognostic effects is to restate the problem so that the need to estimate prognostic effects is obviated. There are several clever approaches developed over the past 10 years that allow us to model CATE directly without fitting in the prognostic part of the outcome model. The obvious advantage of these methods is a protection  against the misspecification of prognostic effects. However, a disadvantage that can be overlooked is that CATE estimates by such methods are often prone to large variability, as we will see later. Therefore an early optimism about such methods was gradually replaced with an understanding that incorporating some estimates of prognostic effects is unavoidable and a more meaningful goal would be to develop methods for combining prognostic and predictive effects to minimize regularization bias and ensure robustness to model misspecification.  

\subsubsection{Tree-based direct estimators of CATE}
Perhaps the earliest proposal of a method for direct estimation of ITE was to modify the splitting criteria of tree-based methods so that the resulting child nodes maximize a measure of treatment homogeneity within each node. Notable examples were Interaction Trees (IT) by Su et al, \cite{Su2009} Lo at al,\cite{Loh2015} an adaptation of model-based recursive partitioning\cite{Zeileis2008} to subgroup identification by Seibold, Zeileis, and Hothorn \cite{Seibold2016,Seibold2018} and Thomas, Bornkamp, and Seibold.\cite{Thomas2018} As an example, IT used as a splitting criterion the test statistic for evaluating interactions formed as treatment-by-candidate splits indicator variables. In  case of a binary outcome, the test statistic is a standardized difference of the treatment differences in proportions between the left and right nodes for a given split scaled by the pooled estimate of the standard error: 
\[
G=\frac{\left( (p_{L1}-p_{L0})-(p_{R1}-p_{R0})\right)^2}{\bar{p}(1-\bar{p}) \left(\frac{1}{n_{L1}}+\frac{1}{n_{L0}} + \frac{1}{n_{R0}} + \frac{1}{n_{R1}}\right)},
\]
where $p_{L1}$, $p_{L0}$ are the proportions of patients with the outcome $Y=1$ for the treatment and control arms in the left node, and $p_{R1}$, $p_{R0}$ are the same quantities for the right node. Also,  $n_{L1}, n_{L0}, n_{R1}, n_{R0}$ are the associated sample sizes and $\bar{p}$ is the pooled estimate of the proportion in the four groups. In general, the splitting criterion is based on the likelihood ratio statistics for the treatment-by-split interaction from a local model fitted to the the parent node that includes the treatment indicator, split indicator and treatment-by-split interaction. 

Another line of research is represented by procedures for causal trees that simultaneously addressed treatment effect heterogeneity and lack of randomization. Su et al \cite{Su2012} proposed the idea of constructing causal trees using the concept of ``facilitating scores'', see also Causal Interaction Trees (CIT) by Yang, Dahabreh, and Steingrimsson. \cite{Yang2021} Athey and Imbens \cite{AtheyImbens2016} developed causal trees using different ideas further extended to causal random forests\cite{WagerAthey2018} and generalized random forests (GRF).\cite{AtheyGRF2019} The splitting criteria for causal trees were motivated by minimizing the squared-error loss in regression trees. This is equivalent to choosing a split that maximizes the variance of the estimated treatment effects $\widehat{\Delta}(x_i)$ for samples in the training set. In addition to using a special type of splitting criteria encouraging treatment effect heterogeneity in child nodes, causal trees/forests introduced several innovations. First, they ``rethought'' random forests as a type of local smoother producing non-parametric estimates of CATE from the terminal nodes across multiple trees. Secondly, they introduced the concept of ``honesty'' for tree construction by dividing the full data set into two halves, and using one half (training set) for splitting and the second half (test set) for estimating CATE. Thirdly, they developed pointwise confidence intervals for $\Delta(x)$  by leveraging the seminal work on inference for bagging\cite{Efron2014} and random forests.\cite{Wager2014} The ideas behind Causal Forest evolved over time adopting various innovations, for example, a doubly robust estimator of CATE motivated by the R-learner method (discussed further in this section) is implemented in the R package for generalized random forests \texttt{grf}.\cite{AtheyGRF2019} 

\subsubsection{Parametric and semi-parametric direct estimators of CATE}
The second class of methods stems from the ideas of the modified covariate methods of Tian et al,\cite{Tian2014} extended to the modified loss method in Chen et al \cite{Chen2017} (see also a review and several proposals in Powers et al \cite{Powers2018}). To motivate these methods, we first review an old idea of using a \textit{modified outcome} approach that emerged under different names in numerous publications. This is based on an inverse probability of treatment weighted (IPW) transformation that can be applied to a continuous or binary outcome $Y$: 
\begin{equation}
Y_i^*=  Y_i \frac{T_i}{\pi(X_i)}- Y_i\frac{1-T_i}{1-\pi(X_i)} = Y_i  \frac{T_i-\pi(X_i)}{\pi(X_i)(1-\pi(X_i))} 
\end{equation}
The key insight is that $E(Y^*|X=x)= \Delta(x)$. 
It follows that an estimator of CATE can be obtained by any method of predictive modeling of $Y^*$, including parametric or non-parametric (e.g., tree-based) approaches.\cite{AtheyImbens2016} In practice, for studies with non-randomized treatments, the propensity function $\pi(x)$ needs to be estimated from the data. 

It is instructive, as will be seen from the following, to express the modified outcome via the treatment indicator $A=2T-1 \in \{-1,1\}$.
\[
Y_i^*=Y_i \frac{A_i}{A_i \pi(X_i) + (1-A_i)/2}.
\]
An alternative way of obtaining a CATE estimator is by using a weighted squared loss with the outcome multiplied  by $2A$ and subject weights
\begin{equation}\label{eq.weights}
  W_i=\frac{1}{A_i \widehat{\pi}(X_i) + (1-A_i)/2}. 
\end{equation} 
It can be verified that the resulting population minimizer is $\Delta(x)$, i.e.,
\begin{equation}\label{eq.minsqloss}
 \argmin_{g(x)} E\left(W_i\left(2A_i Y_i-g(X_i)\right)^2 |X_i=x\right) = \Delta(x).
\end{equation}
It is easy to see that (\ref{eq.minsqloss}) can be written equivalently as 
\begin{equation}\label{eq.minsqloss1}
 \argmin_{g(x)} E\left(4W_i\left(Y_i-\frac{A_i}{2} g(X_i)\right)^2 |X_i=x\right) = \Delta(x).
\end{equation}

This observation immediately suggests estimating CATE by modifying the weighted loss function on the $g(x)$ part rather than by modifying the outcome, leading to the minimization of a weighted empirical loss: 
\[
\frac{1}{n}\sum_{i=1}^n W_i\left(Y_i-\frac{A_i}{2} g(X_i)\right)^2.
\] 


When $g(x)$ is a linear function, this representation is equivalent to simply multiplying each covariate by $A_i/2$. Hence the name, the modified covariate method (MCM). In general, modifying the weighted loss by working with the $g$-function rather than directly working with the outcome is beneficial because it can be done for a variety of loss functions besides the squared loss, thus allowing to extend the modified loss/covariate approach across different types of outcomes such as binary endpoints, count-type endpoints and time-to-event endpoints.\cite{Chen2017,Huling2021}

For example, for a binary outcome coded as $Y \in \{0,1\}$, CATE can be estimated by minimizing the modified weighted empirical loss 
\[
\frac{1}{n}\sum_{i=1}^n W_i L\left(Y_i,\frac{2T_i-1}{2} g(X_i)\right),
\]
where $W_i=\frac{T_i}{\pi(X_i)}+\frac{1-T_i}{1-\pi(X_i)}$ is the inverse probability of treatment weights and the loss, $L(u,v)$, can be an appropriate loss function for the binary outcome such as a logistic regression loss $L(u,v)=-uv+\text{log}(1+\exp(-v))$. In this case, the population minimizer $g(x)$ is related to $\Delta(x)=P(Y=1|X=x,T=1)-P(Y=1|X=x,T=0)$ as
\[
\Delta(x)=\frac{\text{exp}(g(x))-1}{\text{exp}(g(x))+1}.
\]

Methods for the direct estimation of CATE using the modified loss/covariates are often illustrated using linear or generalized linear models for $g(x)$. However, various forms of penalized regression, non-parametric modeling as well as more advanced machine-learning methods, such as tree-based methods, including random forest and gradient boosting,\cite{Sugasawa2019} can be used as well. The described modified loss method uses inverse probability weighting and will be also referred to in this tutorial as W-learning (weighted learning). 

Another method for directly estimating CATE via modified loss, called \textit{A-learning} in Chen et al, \cite{Chen2017} is based on minimizing the modified loss $E(Y-(T-\pi(X)) g(X))^2$. It is easy to see that its population minimizer for the squared loss is $g(x)=\Delta(x)$. This property holds because the centered interaction $(T-\pi(X)) g(X)$ is orthogonal to the main effect, $h(X)$ in (\ref{eq.maineff}).

It is important to note that direct estimation methods are prone to high variability resulting in poor efficiency, which can be improved by augmenting the estimating equations with an additional zero-expectation term. For example, the estimating equations for (\ref{eq.minsqloss1}) under a squared loss (after taking the derivative with respect to $g(x)$ and using $A^2=1)$ takes the following form:
\begin{equation}\label{eq.minsqloss2}
 E \left( \frac{1}{\pi(X) - (1-A)/2} \left(Y- \frac{A}{2} g(X)\right) |X=x \right) = 0.
\end{equation}
It can be shown that the solution of the equation (\ref{eq.minsqloss2}) does not change if we add a term $\frac{b(X)}{\pi(X) - (1-A)/2}$ for an arbitrary function of covariates $b(X)$. 


Now the goal is to choose the function $b(X)$ that minimizes the variance of the estimating equations.\cite{Chen2017} In practice, such function can be taken as the estimated main effect $\widehat{h}(X)=\frac{1}{2}(\widehat{m}(0,X)+\widehat{m}(1,X))$. This requires fitting a response model for $E(Y|T=t,X=x)$ that the direct estimation methods try to avoid in the first place. Therefore, while na\"ive meta-learners of CATE may carry large regularization bias and should be replaced by more ``targeted'' estimators, they should still incorporate information on the main (prognostic) effects in some form. The outlined method of augmented modified loss via W-learning corresponds to the method of modified outcome when  continuous responses are transformed into doubly robust augmented inverse probability weighted scores \eqref{eq.dr.scores} (considered in the next section).\cite{Kennedy2020}

Another proposal for outcome transformation (corresponding to A-learning with augmentation) is based on the so-called Robinson's transformation\cite{Robinson1988} of an outcome variable that involves simultaneously centering the response and treatment indicator around their estimated expected values. Specifically, consider 
\begin{equation}\label{eq.robinson}
Y^*_i=\frac{Y_i-m(X_i)}{T_i-\pi(X_i)}, 
\end{equation} 
where $m(x)=E(Y|X=x)$ is the overall response function, capturing the main effect of covariates on the outcomes in the pooled data. It is easy to show that $E(Y^*|X=x)=\Delta(x)$, therefore a simple approach similar to the modified outcome is to estimate CATE by regressing $Y^*$ on the covariates (for continuous or binary response). We note that such ``residualization'' of the marginal outcome and treatment effects was advocated recently in the literature on estimating the overall treatment effect from observational data under the name of double/debiased machine learning\cite{chernozhukov2018} as well as the literature on HTE, see, for example, the work of Athey et al \cite{AtheyGRF2019} on generalized random forests. 

Centering both the response and treatment ensures Neyman orthogonality, making the estimation of the target parameters of interest ``locally insensitive to the value of the nuisance parameters (i.e., parameters of regression models for the outcomes and treatment assignment), which allows one to plug-in noisy estimates of these parameters''.\cite{chernozhukov2018} Specifically, as shown in Chernozhukov et al, \cite{chernozhukov2018} double/debiased estimates of the average treatment effect are ``doubly robust'' requiring that only one of the two nuisance functions should be consistent for the target estimate to be consistent. If both models are consistent, the procedure ensures much improved rates of convergence for the target estimator even if both estimators of the nuisance parameters converge at fairly slow rates. This helps mitigate regularization biases induced by tuning hyper-parameters involved in learning $m(x)$ and $\pi(x)$. An important role in removing biases due to overfitting is played by cross-validated predictions of the nuisance function. Specifically, the plugged-in values $\widehat{m}^{-i}(X_i)$ and $\widehat{\pi}^{-i}(X_i)$ for the outcome and propensity functions, respectively, are based on models fitted on subsamples excluding the $i$th observation. This was termed \textit{cross-fitting} in the econometric literature.\cite{chernozhukov2018} 

Going back to estimating CATE, the transformation (\ref{eq.robinson}) leads to the following data representation 
\begin{equation}\label{eq.robinson1}
Y_i-m(X_i)=(T_i-\pi(X_i)) \Delta(X_i)+\epsilon_i, 
\end{equation} 
where the plug-in estimates of nuisance parameters $m(x)$ and $\pi(x)$ are obtained from some off-the-shelf machine learning methods (with a cross-fitting step) following the estimation of $\Delta(x)$. These ideas were first conceptualized in the proposal by Zhao et al \cite{Zhao2018, Zhao2019} and further generalized in \textit{R-learning} of Nie and Wager.\cite{Nie2020} The R-learning proceeds in two steps. At the first step, cross-fitted predicted values $\widehat{m}^{-i}(X_i)$ and $\widehat{\pi}^{-i}(X_i)$ are obtained for each subject as follows. The data set is divided into $K$ equally sized folds and predictions for each observation are obtained by fitting outcome and treatment models using only the data from the $K-1$ folds that do not include the $i$th subject. Data fitting can be performed using any machine learning method. Cross-validation is typically required to tune one or more hyper-parameters, e.g., the learning rate (shrinkage), the number of trees,  tree depth, etc for boosting, within each fold of the cross-fitting. At the second stage, individualized treatment effects are estimated by penalized empirical loss minimization. Using a squared loss, 
\begin{equation}\label{eq.rlearn}
\begin{aligned}
\widehat{\Delta}(\cdot)&=\argmin_\Delta \left(\widehat{L}_n\{\Delta(\cdot)\} + \Lambda_n\{\Delta(\cdot)\} \right) \\
\widehat{L}_n\{\Delta(\cdot)\} &=\frac{1}{n}\sum_{i=1}^n \left( Y_i-\widehat{m}^{-i}(X_i) - (T_i-\widehat{\pi}^{-i}(X_i)) \Delta(X_i) \right)^2. 
\end{aligned}
\end{equation} 
Here $\Lambda_n\{\Delta(\cdot)\}$ is a regularizer on the complexity of the $\Delta(\cdot)$ specific to a given machine learning method. It was shown\cite{Nie2020} that the cross-fitting procedure controls the convergence rates of the target CATE estimator independently of learning the nuisance parameters, as if those were provided by an approximate ``oracle''. The two-stage procedure enables the separation of the two challenges inherent in CATE estimation by machine learning. It removes confounding and spurious effects by controlling the correlations between prognostic and propensity effects, and facilitates an accurate representation of $\Delta(X)$ by choosing an appropriate penalty and optimization algorithm for (\ref{eq.rlearn}). Note that minimizing penalized loss (\ref{eq.rlearn}) is equivalent to minimizing penalized weighted loss based on the outcome  transformation (\ref{eq.robinson}) with the individual weights set as $W_i=(T_i-\pi(X_i))^2$.

There is a stream of literature on ``direct'' estimation of the ratio-based  CATE defined as $\Delta(x)=\frac{E(Y(1)|X=x)}{E(Y(0)|X=x)}$ for count data\cite{Yadlowsky2021} and zero-inflated outcomes.\cite{Yu2022} These methods use doubly-robust estimation strategies and, like the R-learning method, gain efficiency via positing relatively simple parametric models for the target parameters capturing HTE while allowing the nuisance functions $m_0(x), m_1(x)$, and $\pi(x)$ to be modeled non-parametrically via ML. Direct modeling of the ratio estimand is also possible via the Modified Loss approaches.\cite{Huling2021}

A different approach to directly estimating the predictive (or treatment benefit) score recently proposed in Huang et al \cite{Huang2022} relies on maximizing a measure of the predictive performance such as a difference in the areas under ROC curves between treated and control subjects for a binary outcome variable. Somewhat similar ideas were developed earlier in the machine learning literature under the name of ``differential prediction'', where the goal was to directly optimize the uplift measure, the difference in the areas under lift curves via support vector machines (SVM)-based method.\cite{Kuusisto2014}

\subsection{Direct modeling of ITR}\label{direct.ITR.mod}

Individualized Treatment Regimens (ITR, the last term of the acronym is often replaced with ``regimes'', ``rules'', or ``recomendations'') became increasingly popular recently as the most accurate quantification of the task of personalized medicine concerned with determining the right treatment at the right time for the right patient. Perhaps ITRs are better known under the heading of Dynamic Treatment Regimes (DTR),\cite{tsiatis2020book} where ``dynamic'' refers to two types of adaptations that may be implemented by the treatment assignment rules: (1) incorporating individual patient characteristics (measured prior to the treatment initiation at baseline), and (2) accounting for evolving patient outcomes from previously assigned treatments. Here we limit our review to the first type of adaptations, i.e., estimating optimal ITRs for a \textit{single} decision point, when the treatments are assigned only once at baseline and remain unchanged throughout the entire study period. The second type of adaptions is relevant only for treatment regimens with \textit{multiple} ($ > 1$) decision points when the optimal regimen is a sequence of treatment assignments as functions of past outcomes and treatments. Some authors reserve ``dynamic'' only for this type of adaptations,\cite{Murphy2001,Chakraborty2014} whereas others\cite{tsiatis2020book} apply the term to both single and multiple-stage strategies. While logically single-stage ITRs precede multiple-stage ITRs as their special case, historically the literature on DTRs was motivated by longitudinal observational trials with time-varying treatments\cite{Robins1986,murphy2003} and Sequential Multiple Assignment Randomized Trials (SMART).\cite{Murphy2005} It was not until the seminal work by Qian and Murphy \cite{qian2011} when the attention of the statistical community was drawn to the special and, by all means, non-trivial case of estimating optimal ITR from a single-stage study.

Qian and Murthy \cite{qian2011} introduced an important concept of the value $V(D)$ associated with an individualized treatment assignment rule $D(X)$. The value is defined as the expected potential outcome 
\begin{equation}\label{eq.value}
V(D)=E\{ Y(D(X)) \}, 
\end{equation}
assuming every patient is treated according to the rule $D(X)$ that maps the multivariate vector of covariates $X$ onto the treatment set, here represented by two options $t \in \{0,1\}$. The expectation in (\ref{eq.value}) is taken with respect to the joint distribution of the outcomes $Y$, covariates $X$, and treatment $T$, given $T=D(X)$. The potential outcome associated with the treatment regimen $D(X)$ is well-defined for an arbitrary covariate vector $X=x$ as
\begin{equation}\label{eq.PO.ITR}
Y(D(x))=Y(1)D(x)+Y(0)(1-D(x)).
\end{equation}

It is easy to see that the optimal treatment regimen, maximizing the value function, is the one that assigns the treatment $T=1$ to the patients with individualized treatment effects $\Delta(x) > 0$, and assigns $T=0$ to the patients with $\Delta(x) < 0$. Qian and Murthy \cite{qian2011} proposed a two-stage procedure, which at the first stage estimated (by penalized regression) the outcome functions $m_0(x)$ and $m_1(x)$ and, at the second stage, estimated ITR for a patients with $X=x$ as $\widehat{D}(x)=I(\widehat{m}_1(x) >\widehat{m}_0(x))$. This essentially makes the estimation of ITR a by-product of any method of estimating CATE by a global outcome modeling method in Section~\ref{global.outcome.mod}. 

Similarly, any method of direct estimation of $\Delta(x)$ from Section~\ref{direct.ITE.mod} can be adopted as a method of ITR by simply taking the sign of the estimated $\widehat{\Delta}(x)$. 
These ideas were developed into the D-learning of Qi and Liu \cite{Qi2018} that is closely related to the modified outcome/covariate methods.\cite{Tian2014, Chen2017}

Another subclass of methods for \textit{directly} estimating ITR (i.e., without the need of estimating the main effects) is by identifying/searching for a regimen $\widehat{D}(X)$ that directly maximizes the value function or its approximation. The key approaches are the outcome-weighted learning (OWL) by Zhao et al \cite{zhao2012} and by Zhang et al.\cite{Zhang2012} These approaches cast estimating the optimal ITR as a weighted classification problem. 

It is easy to see that the value of $D(X)$ can be expressed via observables under the assumption of \textit{ignorable} treatment assignment as 
\begin{equation}\label{eq.value.ignore}
    V(D)=E\left(\frac{I(T=D(X))}{A \pi(X) + (1-A)/2} Y \right),
\end{equation}
where $A=2T-1$. In other words, the value of the regimen $D(X)$ is expressed via the observed outcomes $Y$ for the sub-population of patients whose assigned treatment was consistent with that regimen, inversely weighted by the probability of being assigned to their treatments. Zhao et al \cite{zhao2012} observed that maximizing the empirical value function is equivalent to minimizing the weighted classification loss, with the weights proportional to the outcomes ($Y$):
\begin{equation}
\label{empiricalloss}
\widehat{D}_{opt}(X)= \underset{D}{\text{argmin}} \;
 \frac{1}{n}\sum_{i=1}^{n} I(T_i\ne D(X_i))W_i, 
\end{equation}
where the weights $W_i$ are given by
\begin{equation}\label{eq.weights.owl}
  W_i=\frac{Y_i}{A_i \widehat{\pi}(X_i) + (1-A_i)/2}. 
\end{equation} 
Therefore, $(\ref{empiricalloss})$ can be seen as the minimization of an outcome-weighted classification loss. Assuming that larger values of the outcome are beneficial, patients who did well on their assigned treatment would incur higher misclassification costs, making it likely for the optimal treatment assignment to be the same as the one actually received. Conversely, those who did poorly on their assigned treatment would incur low misclassification costs, which would encourage a decision to ``switch'' them to the alternative treatment.

In practice, the 0-1 loss in $(\ref{empiricalloss})$ is difficult to deal with due to its discontinuity and non-convexity. Therefore it was proposed\cite{zhao2012} to replace it with a smooth surrogate loss function $\varphi$ for the binary outcome $A=2T-1 \in \{-1,1\}$ with respect to a predictor function $g(X|\beta)$ that is indexed by the parameters $\beta$: 
\begin{equation}
\label{eq.eloss.mod.smooth}
 \widehat{\beta}= \underset{\beta}{\text{argmin}}
 \frac{1}{n}\sum_{i=1}^{n}\varphi(A_i, g(X_i\mid \beta))W_i.
\end{equation}
This representation casts the estimation of ITR as a supervised learning problem that can be tackled using off-the-shelf software. The optimal regimen is expressed via estimated coefficients $\beta$ as $\widehat{D}_{opt}(X)=I(g(X|\widehat{\beta}) > 0)$. A weighted loss in (\ref{eq.eloss.mod.smooth}) can be further regularized by adding a penalty term as a function of $\beta$, for example, the lasso or elastic net. The surrogate loss function in \eqref{eq.eloss.mod.smooth} can be more compactly written in terms of a single argument, the margin $U=A \times g(X|\beta)$, with a larger margin indicating good classification. The loss $\varphi(u)$ can take various forms, including the hinge loss, $\varphi(u)=(1-u)_{+}$, from the support vector machines used in the original OWL proposal,\cite{zhao2012} logistic loss, $\varphi(u)=\text{log}(1+\text{exp}(-u))$, proposed in Xu et al \cite{Xu2015} and other types of loss functions.\cite{Huang2014} Furthermore, decision rules can be modeled using a linear form, tree-structured form, \cite{laber2015} gradient boosting, \cite{Sugasawa2019} and others. 

An important limitation of the OWL framework is that, when the 0-1 loss is replaced with the surrogate loss, the resulting rule ends up depending on such trivial changes as simply adding a constant to the outcome. To address this problem, the residual weighted learning (RWL)\cite{Zhou2017} and Augmented Outcome-Weighted Learning (AOL)\cite{Liu2018} methods were developed. The key idea was to subtract the mean effect $m(x)$ (estimated by any appropriate method, e.g., a non-parametric method) from the outcome, which renders the estimated ITR more robust. Replacing the outcomes $Y$ with the residuals $\widetilde{Y}=Y-\widehat{m}(X)$, however, results in negative weights for some patients. Zhou et al \cite{Zhou2017} handled that by replacing the surrogate loss by a nonconvex loss function, namely, the smoothed ramp loss. Liu et al \cite{Liu2018} showed that standard tools for OWL can be applied to residual-based outcomes by redefining weights as 
\[
\widetilde{W}_i=\frac{|\widetilde{Y}_i|}{A_i \widehat{\pi}(X_i) + (1-A_i)/2}, 
\]
and further setting the treatment indicator in (\ref{eq.eloss.mod.smooth}) to $\widetilde{A}_i=A_i \text{sign}(\widetilde{Y}_i)$.

Another proposal that casts ITR as a weighted classification problem (by Zhang et al \cite{Zhang2012} building on their framework for estimating optimal ITR by directly maximizing its value\cite{Zhang2012A}) is based on doubly robust scores $\widehat{\Delta}(X_i)$ obtained by an augmented inverse propensity weighting (AIPW) \cite{AtheyWager2019, Kennedy2020} method that combines the outcome regression with propensity: 
\begin{equation}\label{eq.dr.scores}
\begin{aligned}
    \widehat{\Delta}_{AIPW}(X_i)=&\widehat{m}_1(X_i)-\widehat{m}_0(X_i) + \frac{T_i(Y_i-\widehat{m}_1(X_i))}{\widehat{\pi}(X_i)}-\frac{(1-T_i)(Y_i-\widehat{m}_0(X_i))}{1-\widehat{\pi}(X_i)}\\
        =&\widehat{m}_1(X_i)-\widehat{m}_0(X_i) + \frac{T_i-\widehat{\pi}(X_i)}{\widehat{\pi}(X_i)(1-\widehat{\pi}(X_i))} \left(Y_i-\widehat{m}(T_i,X_i) \right),
\end{aligned}    
\end{equation}
where  $\widehat{m}_1(x) \equiv \widehat{m}(1,x)$ and $\widehat{m}_0(x) \equiv \widehat{m}(0,x)$ are based on the outcome regression (\ref{eq.indirect.ITE}).

It is easy to see from (\ref{eq.PO.ITR}) that the value of a regimen can be written as $E\{Y(D(X))\}=E\{D(X)(Y(1)-Y(0))+Y(0)\}=E\{D(X)\Delta(X)\}+E\{Y(0)\}$, where the expectation is taken with respect to the distribution of $(Y,X)$. Because the last term does not depend on the regimen $D(X)$, an optimal ITR can be estimated as 
\begin{equation}
\label{eq.dr.obj}
\widehat{D}_{opt}(X)= \underset{D}{\text{argmax}} \;
 \frac{1}{n}\sum_{i=1}^{n} D(X_i)\widehat{\Delta}_{AIPW}(X_i). 
\end{equation}
Using the representation $\Delta(x)=|\Delta(x)| \text{sign}(\Delta(x))=|\Delta(x)| (2I(\Delta(x)>0)-1)$ and the fact that $D(x)^2=D(x)$, after some elementary algebra, we can rewrite $D(x)\Delta(x)=-|\Delta(x)|(I(\Delta(x)>0)-D(x))^2+ |\Delta(x)|I(\Delta(x)>0)$\cite{Zhang2012}. This suggests that the optimization of (\ref{eq.dr.obj}) with respect to $D(x)$ can be implemented by minimizing the weighted classification loss with the binary outcome set as $Z_i=I(\widehat{\Delta}_{AIPW}(X_i)>0)$ and the individual weights set to the absolute values, $W_i=|\widehat{\Delta}_{AIPW}(X_i)|$. Intuitively, patients with small absolute values of ITE receive small weights because  either of the two treatments would result in similar benefits for those and misclassification is unimportant.  

It can be further shown\cite{Zhang2012,Wu2022} that the optimal regimen $D_{opt}(X)=I(g(X\mid \beta)>0)$ for this type of a weighted classification loss approach can be also obtained via minimizing a smooth surrogate loss function $\varphi(\cdot)$, where the parameters $\beta$ are estimated as  
\begin{equation}
\label{eq.dr.smmooth}
\nonumber \widehat{\beta}= \underset{\beta}{\text{argmin}}
 \frac{1}{n}\sum_{i=1}^{n} \varphi \left(\text{sign}(\widehat{\Delta}_{AIPW}(X_i)) \times g(X_i\mid \beta)\right) |\widehat{\Delta}_{AIPW}(X_i)|. 
\end{equation}

In the spirit of direct optimization of the value function, van der Laan \cite{vanderLaan2013,Montoya2023} proposed estimating an optimal ITR semi-parametrically using a flexible Super-learner framework\cite{vanderLaan2007}. First, a library of candidate estimators of $\Delta(x)$ is created by regressing the doubly-robust scores (\ref{eq.dr.scores}) on all or a subset of the designated covariates (candidate treatment effect modifiers) under various modeling strategies that can include parametric regression models and highly data-driven ML algorithms. This produces a series of estimators $\widehat{\Delta}^{(j)}_{AIPW}(x), j=1\cdots J$ that are then combined in a single weighted estimator $\widehat{\Delta}^{(\alpha)}_{AIPW}(x)=\sum_{j=1}^J \alpha_j \widehat{\Delta}^{(j)}_{AIPW}(x)$, where $\alpha_{j} >0$ denote weights. The final ITR is $\widehat{D}^{(\alpha)}(x)=I(\widehat{\Delta}^{(\alpha)}_{AIPW}(x)>0)$. The weights are estimated using a cross-validated loss function that directly optimizes the value function of the final ITR, which is very similar to the loss function proposed from weighted classification perspectives.\cite{Zhang2012} 

To summarize, various methods of direct ITR estimation by searching for a regimen maximizing the value function within a certain class can be constructed by combining different representations of the outcome, loss functions, complexity penalties, and decision rule types. 

Several notable extensions of methods for ITR estimatiion have been proposed in recent years. One important extension was to allow an arbitrary number of treatment arms. Qi and Liu \cite{Qi2018} extended their D-learning method to $K>2$ treatments by estimating all pairwise $K(K-1)/2$ CATE functions $\Delta_{i,j}(x)$ comparing the treatment arms $i,j =1,..,K$. Zhang and Liu \cite{Zhang2014mult} proposed an elegant method, called multi-armed angle-based direct ITR learning, that allows one to estimate the individual treatment preferences simultaneously rather than in a pairwise fashion (see also Zhang et al \cite{Zhangl2020mow} and Qi et al \cite{Qi2020}). It represents multiple treatments as vertices in a $K-1$ dimensional Euclidean space based on a $K$-vertex simplex structure with the origin as the center. For each patient, $K-1$ decision functions are estimated as a single vector in the same space and the optimal treatment is then selected as the vertex having the smallest angle with that vector. The family of D-learning methods was recently generalized within the stabilized direct learning (SDL) framework.\cite{Shah2023} 

Estimating ITR within a restricted class of functions indexed by the parameters $\beta$ may lead to the estimated optimal rule based on $g(X|\widehat{\beta})$ not attaining the true optimal rule $D_{opt}(X)$. However, the hope is that the value loss due to restricting the search space (part of the misspecification error) would be relatively small\cite{Zhang2012A}. Moreover, an important line of research deals with incorporating in an optimal treatment rule various constraints that may be related to the cost of treatments, side effects or other considerations such as simplicity, feasibility, and interpretability. For example, there are methods for estimating ITR that control the rate of adverse events at a pre-specified level\cite{Wang2018, Doubleday2022} or take into consideration cost-effectiveness trade-offs\cite{Xu2022}. Another type of constraint can be ensuring ``fairness'' of ITR with respect to certain disadvantaged subpopulations\cite{Ethan2023}. One type of constraint imposed on ITR can be that they have a ``simple'' structure that could be easily interpreted by the stake-holders, e.g., a linear function,\cite{Wu2022} a tree, \cite{laber2015,AtheyWager2021} a set of decision rules,\cite{Zhang2015} or another type of restricted class of regimens.\cite{Zhang2012} For example, Laber and Zhao \cite{laber2015} proposed a tree-based procedure for estimating an optimal ITR that uses a value-based splitting criterion: each parent node is split into two daughter nodes by maximizing the purity of the resulting split. Specifically, for two candidate treatments, the purity measure is an estimator of the value (standardized by the size of the parent node) of the rule that assigns all patients in the left node to one of the treatment options and all patients in the right node to the other. While Laber and Zhao used an IPW estimator of the value based on \eqref{eq.value}, Tao, Wang, and Almirall \cite{Tao2018} proposed an improved tree-based reinforcement learning  procedure (T-RL) that applies in more general settings with multiple treatments and multiple decision points to estimate optimal dynamic ITR with the purity measure based on a doubly robust AIPW estimator of the value associated with an ITR, $D(X)$\cite{Zhang2012A}
\begin{equation}\label{eq.value.aipw}
        \widehat{V}_{AIPW}(D)=\frac{1}{n}\sum_{i=1}^{n} \left(\frac{C_i Y_i}{\widehat{\pi}(D_i,X_i)} - \frac{C_i-\widehat{\pi}(D_i,X_i)}{\widehat{\pi}(D_i,X_i)} \widehat{m}(D_i,X_i)\right),
\end{equation}
where $C_i$ is the indicator of whether the treatment $T_i$ actually received coincides with the treatment $D_i$ dictated by the ITR, $C_i=I(T_i=D_i)$. Note that the value estimator \eqref{eq.value.aipw} incorporates information from all $n$ patients including those whose actual treatment was not consistent with the regimen, thus having improved precision compared with the IPW estimator. The T-RL method is implemented as an R package available at GitHub \texttt{Team-Wang-Lab/T-RL}. 

Athey and Wager \cite{AtheyWager2021} developed a procedure for searching through \textit{interpretable} treatment policies within a restricted class of fixed-depth decision trees, implemented in an R package \texttt{policytree}.\cite{Sverdrup2020} They use a representation of the objective function similar to \eqref{eq.dr.obj} based on cross-fitted estimates of doubly robust scores \eqref{eq.dr.scores}, where the optimal regime $D(X)$ is sought in the constrained space. This work was recently extended in an approach called CAPITAL\cite{CaiCAPITAL2022} and is also discussed in the next section. 

We conclude the subsection on ITR by acknowledging the many conceptual challenges associated with optimizing treatment selection for a given patient, such as defining a class of feasible treatment regimens and choosing clinically meaningful optimization criteria (see a recent discussion article\cite{Tran2022} and references therein). 

\subsection{Direct subgroup identification}\label{direct.SID}
The term ``subgroup identification'' has been used historically as a broad term and it may cover multiple domains, including procedures for identifying subgroups of patients with better prognosis if treated or untreated, irrespective of predictive effects. However, these procedures would be outside the scope of this paper, as we are focused on methods for identifying subgroups of patients with differential treatment effects. Within this domain, subgroup identification may still apply to fairly diverse settings. For example, many would consider methods for ITR estimation in this category, since they produce the rules for selecting a subgroup of patients who should be treated using each individual treatment, e.g., partitioning the overall population into $K$ subpopulations for $K$ available treatment arms. Others would apply the term ``subgroup'' only when the resulting subpopulations are presented as interpretable rules, e.g., produced by decision trees where each rule is a union of the regions defined based on pre-specified biomarkers. Here we define subgroup identification as a class of statistical procedures that aim at selection of subgroups of patients with enhanced treatment effects (or other desirable features) only for the selected treatment(s) of interest, typically for the new experimental treatment. We emphasize that the subgroup membership indicator $\widehat{S}(X)$ returned by a subgroup identification  procedure is a function of baseline covariates (a ``biomarker signature'')  and  therefore should be applicable to any external data set as long as it contains the biomarkers $X$.  

Instead of estimating the response function $\Delta(x)$ over the entire covariate space and then identifying subpopulations where $\widehat{\Delta}(x)$ is large, we can search directly for such ``interesting regions''. Early implementations of this approach used the ideas of the ``bump hunting'' method\cite{Friedman1999} (a.k.a. PRIM, Patient Rule Induction Method) developed for a target function $f(x)$ that represented an outcome variable and adopted it for the target function being a treatment contrast, $f(x) \equiv \Delta(x)$.\cite{Kehl2006, Chen2015} 

The motivation behind the original PRIM algorithm was to by-pass the problem of estimating the entire response surface that may result in smoothing out regions with irregular behavior that are actually of primary interest. The goal of the PRIM procedure is to find sub-regions of the covariate space with relatively high values of the target variable $Y$ without attempting to explicitly model these values. The sub-regions are described by simple rules formed as small and possibly disjoint rectangles in the covariate space and their union is considered an approximation of the sought subgroup $S$.

The sub-regions are produced using a greedy iterative search method with two steps at the core of each iteration: \textit{peeling} and \textit{pasting}. Starting from the full covariate space $B$ in the first iteration, a peeling step successively removes (peels off) small strips, each defined by a covariate $j$ as either $\{x \in  B: x_j < x_{j(\alpha)}\}$ or $\{x \in  B: x_j > x_{j(1-\alpha)}\}$, where $x_{j(\alpha)}$ denotes the $\alpha$-quantile. That is, the strips are selected from the boundaries of the considered covariate space and the best strip to peel off is chosen as the one that yields the largest target value (e.g., treatment effect) in the remaining region. The peeling sequence stops if the remaining region becomes smaller than a prespecified limiting value. This value as well as the $\alpha$-quantile are hyper-parameters that can be tuned using cross-validation. Therefore the subgroup search is controlled by hyper-parameters that help prevent overfitting  \textemdash in contrast with many brute-force approaches that ignore the stochastic nature of the outcome process and are likely to result in ``identifying'' irregularities that may never generalize to future data. Secondly, a pasting step is the reverse of the peeling procedure performed to readjust the outcomes of peeling. The region identified by peeling is then enlarged by repeatedly adding strips on the boundary until these additions fail to increase the target value. Once a pair of peeling and pasting steps has been completed, the entire process is repeated starting from the complement of covariate space not yet covered by the previously selected sub-regions. Some enhancements of this greedy procedure have been proposed, e.g., replacing the pasting step with a jittering step helping to avoid local optima.\cite{Polonik2010} 


The Subgroup Identification Differential Effect Search (SIDES)\cite{Lipkovich2011} and SIDEScreen\cite{Lipkovich2014,Lipkovich2017a} methods approached the general subgroup search problem utilizing ideas from tree-based regression. Unlike conventional algorithms of recursive partitioning\cite{Su2009,Dusseldorp2014,Loh2015} that split every node into child nodes and finally partition the entire covariate space into non-overlapping regions, SIDES pursues only one of the two child nodes at each split, abandoning the child with a weaker treatment effect. This algorithm results in constructing a single branch (terminal node) rather that a tree. The branch presents each subgroup as a ``biomarker signature'' involving up to $d$ biomarkers, where $d$ is the \textit{depth} parameter pre-specified by the user, typically $d \le 3$. The data within each intermediate node are split simultaneously on $1< m \le p$ best candidate covariates (the \textit{width} parameter), and as a result SIDES generates a collection or ensemble of branches representing the most promising subgroups. 

The adaptive SIDEScreen method\cite{Lipkovich2014,Lipkovich2017a} further capitalizes on the ensemble by computing variable importance scores (summarizing the predictive strength of each biomarker over all splits) and selecting only those biomarkers whose importance exceeds a threshold calibrated from the null distribution generated by permutations. At the second stage of SIDEScreen, the base SIDES method is applied to the selected biomarkers. The adjusted $p$-values are computed from an appropriate permutation distribution taking into account the uncertainty associated with the two-stage process. Specifically, at the first stage, the base SIDES algorithm\cite{Lipkovich2011} is applied to the data to generate a large number of candidate subgroups and compute the Variable Importance score  (VI) for each candidate covariate (biomarker) by aggregating the contributions of a given biomarker across the generated subgroups. At the second stage, the observed VI scores for the $p$ candidate covariates, denoted by $VI(X_i)$, $i=1,\dots,p$, are compared with the cutoffs computed from the reference distribution of the maximal VI score $\text{VI}_{max}$ under the null hypothesis of no predictive biomarker. Specifically, the cutoff is computed as $v=E_0(\text{VI}_{max})+k S_0(\text{VI}_{max})$, where $E_0(\text{VI}_{max})$ is the expected value of the maximal score under the null hypothesis and $S_0(\text{VI}_{max})$ is the null standard deviation. The biomarker $X_i$ is selected if $\text{VI}(X_i)> v$. The constant $k$ is calibrated to control the rate of falsely identifying at least one biomaker under the null at a desired level. If the VI score exceeds the cutoff for at least one candidate biomarker, the base SIDES algorithm is applied again to the biomarkers that passed the VI screening and the best subgroup is reported. The p-value for this subgroup is obtained using the null distribution for the maximal selected subgroup computed by replicating the two-stage procedure (involving the biomarker selection step) on each null set, thus accounting for the uncertainty associated with both stages of the adaptive SIDEScreen. The adjusted p-value for the treatment effect in the final selected subgroup is computed as the proportion of the null sets where this p-value is as or more significant than the treatment effect p-value for the best subgroup identified using the null data. The null sets for both stages (i.e., for the maximal VI score at the first stage and for the maximally selected subgroup at the second stage) are generated by randomly permuting the treatment indicator variable. Alternative permutation schemes can be considered depending on how the null distribution is conceptualized.\cite{Foster2016} 

Sequential-BATTing (Bootstrapping and Aggregating of Thresholds from Trees) by Huang et al\cite{Huang2017} are implemented in the R package \texttt{SubgrpID}. The idea is to search for the final subgroup as a product of individual biomarker thresholding rules
\[
S(X)=\prod_{j=1}^p I(s_jX_j \ge s_ja_j),
\]
where $s_j \in \{-1,1\}$ and $a_j$ is a cutoff value associated with the $j$th biomarker selected by using a bootstrap-based BATTing procedure, e.g., as the median value from a collection of estimates computed over a large number of bootstrap samples. Using the bootstrap as in the original bagging procedure\cite{Breiman1996} ensures robustness against random data perturbations for an inherently unstable procedure (here, threshold selection). Recently, the ideas of constraint policy search\cite{AtheyWager2021} were extended by Cai et al \cite{CaiCAPITAL2022} to define a method for interpretable tree-structured subgroup selection rules (SSR) called CAPITAL (optimal subgroup identification via constrained policy tree search). Their method searches to produce the largest sample sizes of patients benefiting from a given treatment while maintaining the average treatment effect in the selected subgroup(s) at a pre-defined $\delta$ while ensuring that for all subjects in the selected subgroup(s), their estimated ITEs exceed a certain threshold $\eta$ (in expectation). Specifically  $\eta$ is chosen to satisfy $E(\Delta(X) \mid \Delta(X)>\eta)=\delta$. 

TSDT (Treatment-Specific Subgroup Detection Tool) also uses bootstrap aggregation to construct subgroups, it applies recursive partitioning on each bootstrap sample, separately by each treatment group, to identify subgroups with superior treatment effect relative to the overall population. The final subgroups are formed by combining  similar subgroup signatures as those based on the same splitting variables.\cite{Shen2020, TSDTCRAN2022}

A number of subgroup search/identification ideas were proposed recently that use, as building blocks, methods described in the earlier sections. For example, the matching trees \cite{ZHANG2021MT} method applies regression trees to ITEs estimated as the outcome differences within matching pairs formed based on propensity scores combined with candidate treatment effect modifiers selected by subject matter experts. Then the tree is pruned and the final subgroups identified as tree nodes with non-overlapping confidence intervals for associated least squares means using the procedure of Lin et al. (2019) for constructing simultaneous confidence intervals in subgroups.\cite{Lin2019}

Another important stream of literature models subgroups as latent class variables with probabilities of subgroup membership estimated as functions of covariates jointly with other parameters of the data generation process. For example, Shen and He \cite{ShenHe2015} proposed a logistic-normal mixture model where the outcome is modeled via a linear mixed-effects regression with normal errors containing subgroup effects represented as $(T \times S)$ interactions, where $S$ is an unobserved (latent) subgroup with subgroup memberships modeled via a logistic regression as a function of candidate covariates, $P(S_i=1|X_i)$. Kim et al\cite{Kim2019} proposed a latent class model with multiple non-overlapping subgroups estimated via Bayesian mixtures. The subgroup membership model is estimated jointly with other parameters by factorizing the joint likelihood into a product of conditional likelihood given the subgroup and the probability of subgroup membership that is connected with the observed covariates via a multinomial probit function. This approach is applied to data from non-randomized studies by first selecting matching pairs using propensity-based matching and then applying the Bayesian latent class mixture to the resulting paired data. 

An approach that can also be considered within this class although not explicitly using a mixture representation is termed ``value guided subgroup identification''.\cite{Zhang2020, Zhang2021} It is closely related to direct ITR estimation presented in Section~\ref{direct.ITR.mod}: the resulting ITR here is  based on thresholding the estimated subgroup membership probability $\widehat{D}(X_i) =I \left(\widehat{P}(S_i=1|X_i)>0.5\right)$; the method estimates subgroup membership probabilities using gradient tree boosting directly maximizing their version of a value function $\tilde{V}(\widehat{D})$ reflecting the clinical interest associated with the subgroups to be identified. The proposed value function is different from the usual value of ITR in \eqref{eq.value} (outcomes expected if everyone's treatment is consistent with the estimated ITR) in that the authors subtract the expected outcomes from the above quantity if everyone's treatment is not consistent with the estimated ITR: 
\[
    \tilde{V}(\widehat{D})=E\left(\frac{I(T=\widehat{D}(X))}{A \pi(X) + (1-A)/2} Y \right)-E\left(\frac{I(T \neq \widehat{D}(X))}{A \pi(X) + (1-A)/2} Y \right).
\]


\section{Case study: Schizophrenia trial}\label{case.study}
We will illustrate applications of selected methods for HTE evaluation using a synthetic data set based on a Phase III clinical study for a treatment of schizophrenia. Patients diagnosed with schizophrenia or schizoaffective disorder were randomized to an antipsychotic medication ($n=149$) or placebo ($n=152$). 

In this case study, we focus on evaluating a continuous endpoint that represents change from baseline to Day 42 in the Positive and Negative Syndrome Score (PANSS) total score, with large negative values indicating a beneficial effect. The PANSS total score is a widely used outcome in schizophrenia trials.\cite{Kay1989} The overall treatment effect is not significant in this data set: the mean changes from baseline to Day 42 in PANSS total score for the experimental treatment and placebo are $-13.75$ and $-14.64$, respectively (placebo is numerically superior to the experimental treatment), and a two-sided $p$-value from the two-sample $t$-test is $p=0.615$.

This data set was synthesised to mimic a real study without artificially adding a subgroup effect that would be known to the authors. As such, it represents a common realistic setting, where the estimated treatment effect for the overall study population is close to zero, and there is an interest in exploring whether the experimental treatment might be beneficial to at least some subgroups of patients. 

Many methods discussed so far and their software implementations are well suited for data explorations but do not provide inferential tools for assessing whether a significant treatment effect heterogeneity is present. By interpreting the output of these methods in a qualitative manner, it is often tempting to conclude that predictive biomarkers and interesting subgroups exist even when they actually do not. Using this case study, we first present the results that would be commonly extracted from the discussed methods and invite the readers to reflect on what conclusions they would draw from such results. In Section~\ref{sec.post.inf}, we will return to the question of whether there is any treatment effect heterogeneity in this data set. In Section~\ref{simdata.description}, we present a simulation study that allows us to assess the ability of selected methods to detect treatment effect heterogeneity when it is known to exist. 

\subsection{Baseline variables}
The baseline variables to be used in this case study are listed in Table \ref{tab.bslCS3} and include demographic variables such as \textit{age}, \textit{gender}, \textit{race}, years from diagnosis (\textit{diagyears}), as well as several measures of disease status, such as the Clinical Global Impressions - Severity of illness scale (\textit{cgis}),\cite{Guy1976} PANSS positive (\textit{pansspos}), PANSS negative (\textit{panssneg}), and PANSS General Psychopathology (\textit{panssgen}) subscales. 
 

\subsection{Application of selected methods to the case study}\label{seq.apply.casestudy}

A  number of methods for global outcome modeling and direct modeling of CATE were applied to the case study with the outcome variable being a change from baseline in PANSS total score at Day 42 (\textit{panss42.change}): Causal Forest (\texttt{grf}); Bayesian Causal Forest (\texttt{bcf}); meta-learners T-, S-, X-, and R-learner (\texttt{rlearner}) using \texttt{xgboost} with custom code for tuning of hyper-parameters; modified loss methods (MLM), also known as modified covariate/outcome methods for continuous endpoints, (\texttt{personalized} 0.2.7) with the W-learning and A-learning methods using \texttt{xgboost} (loss=``sq\_loss\_xgboost'') as learners. For MLM with \texttt{xgboost}, the current package version does not yet support tuning hyper-parameters and we used a fixed-parameter setup with \texttt{tree.depth}=$5$, \texttt{shrinkage}=$0.01$, the subsample and the column sample fractions both set to $0.90$. The MLM methods were both applied with and without the augmentation option (via an elastic net penalty using the \texttt{glmnet} package), with the penalty tuned via $10$-fold cross-validation. Estimating the nuisance parameters for the augmented version of the W- and A-learning methods employed cross-fitting. However, for the R-learner, which also, in principle, should include cross-fitting, the current implementation of the \texttt{rlearner} package using \texttt{xgboost} base learners does not provide the cross-fitting functionality, and therefore may not harness the full potential of this meta-learning method. In our application of R-learning to the case study reported in the Supplementary  materials, we implemented cross-fitting for the outcome function. Note that, since the data in our case study are from an RCT, we did not need to estimate the propensity function from the data, and for methods that use the probability of treatment we provide the constant value for the probability of being assigned to the experimental treatment as $n_1/(n_0+n_1)=0.495$. In a more general setting of observational data, all the nuisance functions need to be estimated from the data. When implementing cross-fitting, this may pose additional challenges, e.g., should we estimate all nuisance functions from the same training set or use non-overlapping training sets for each nuisance function?\cite{Jacob2020} 

The details of R implementation can be found in the Supplementary materials. As a general note, the availability and maturity of the software tools for specific platforms and operating systems differ a lot and some packages may not be available for certain versions of R. The actual version of the package used for analysis is indicated in the R code.  

\begin{figure*}[!ht] 
\centering
\includegraphics[scale=0.6]{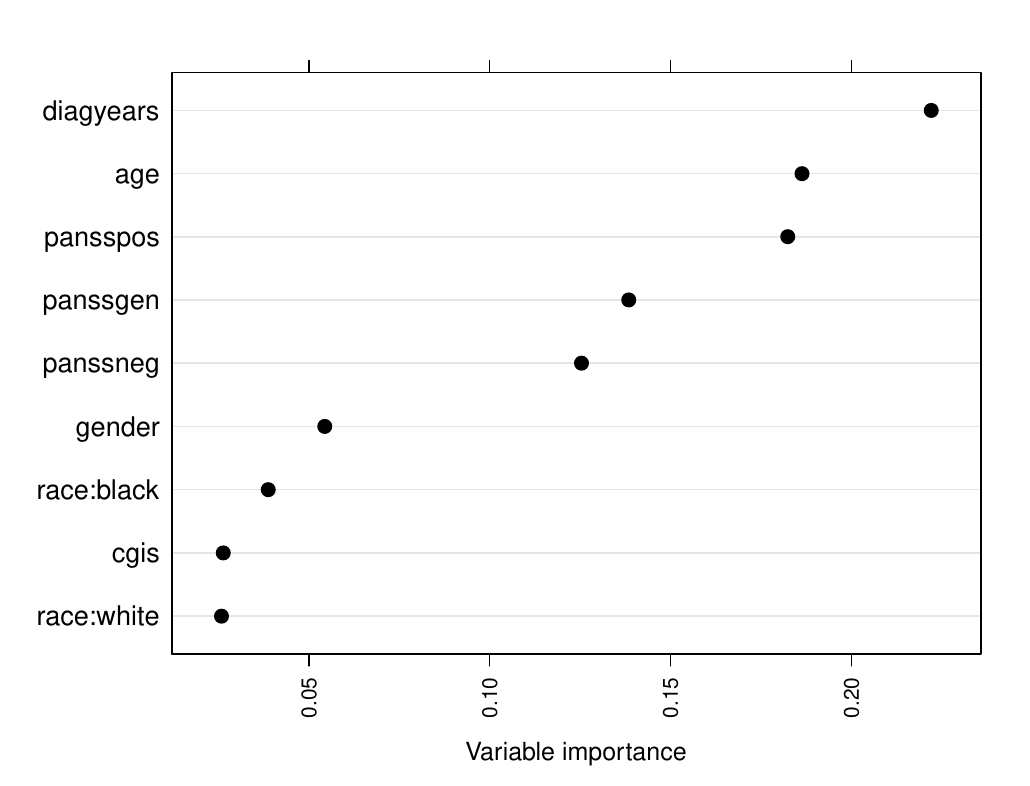} 
\caption{Ranking of the baseline covariates in the schizophrenia case study based on variable importance scores from Causal Forest with $10,000$ trees.}
  \label{fig.vipCFCS3}
\end{figure*}

All packages listed above allow the user to extract estimates of CATE for patients in the training data set. The only package that also provides variable importance (VI) scores for the candidate biomarkers is \texttt{grf} for Causal Forest. Variable importance scores are computed as weighted sums of how many times a given variable was split on at each tree level in the forest using the default options (max depth = 4 with weights exponentially decaying by the tree level as a power of -2). To facilitate the interpretation of the results across different methods, we present the patterns of estimated CATE versus the two biomarkers with the highest VI scores as determined by Causal Forest (Figure~\ref{fig.vipCFCS3}): the time since diagnosis (\textit{diagyears}) and patient's age (\textit{age}).

\begin{figure*}[!ht] 
\centering
\includegraphics[scale=0.6] {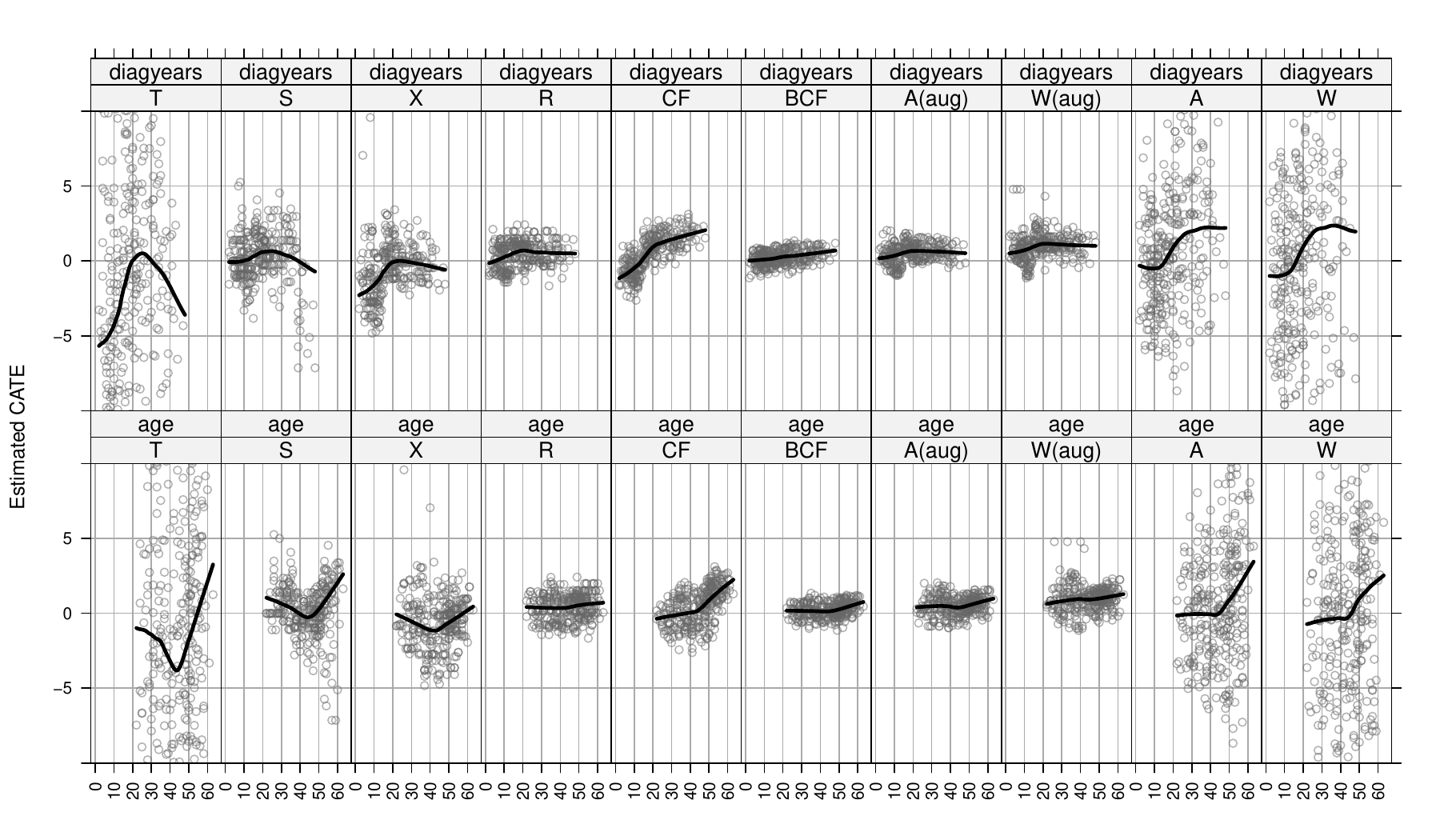} \caption{Estimated 
CATE for changes in PANSS total from baseline to Day 42 against the time since diagnosis (\texttt{diagyears}, in the first row) and patient's age (\texttt{age}, in the second row) for the selected approaches in the schizophrenia case study. Larger negative values of CATE indicates superiority of the experimental treatment over control.}\label{fig.CS3_COLORPLOT}
\end{figure*}

Figure~\ref{fig.CS3_COLORPLOT} displays the estimated CATE for three classes of methods: meta-learners (T, S, R, X), forests (CF and BCF), and modified loss (A-and W-learning with augmented and unaugmented versions) against the two variables, (\textit{diagyears} and \textit{age}). Note that the Y axis was truncated for some methods to ensure that outliers would not dominate the picture. The smoothing lines are loess fits based on all available data including points outside the displayed Y-range. There is a considerable difference in the variability of the estimated CATE across the three classes. Causal Forest and Bayesian Causal Forest exhibited considerably lower variability compared to the other methods. The W- and A-learning methods without augmentation produced very unstable results and many CATE estimates  were off-scale. The augmented versions of these two methods performed quite well and showed fairly low variability  comparable to CF. The plots suggests that younger patients (based on \textit{age}) and, more importantly, patients with a recent diagnosis (based on \textit{diagyears}) are likely to experience a beneficial treatment effect, i.e., a larger reduction in the PANSS total score compared to placebo (recall that larger positive values of the outcome variable, \textit{panss42.change} indicate worsening of symptom severity).

\begin{figure*}[!ht] 
\centering
\includegraphics[scale=0.5]{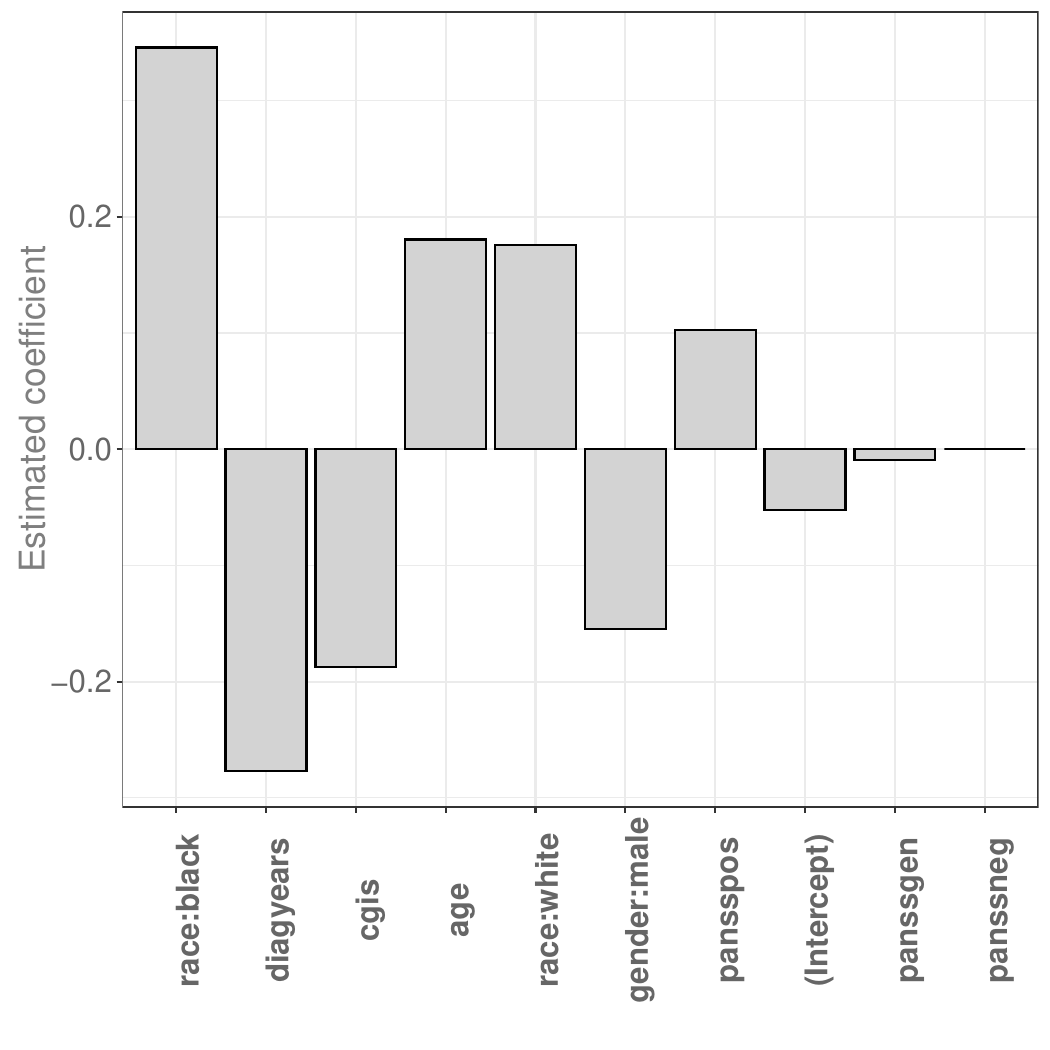} 
\caption{Estimated coefficients from OWL in the schizophrenia case study. See Figure \ref{fig.OWLprob} for the resulting probability of assignment to the active treatment.}
  \label{fig.owlRBFlogit}
\end{figure*}

\begin{figure*}[!ht] 
\centering
\includegraphics[scale=0.6, angle=90, trim=20 0 20 0,clip]{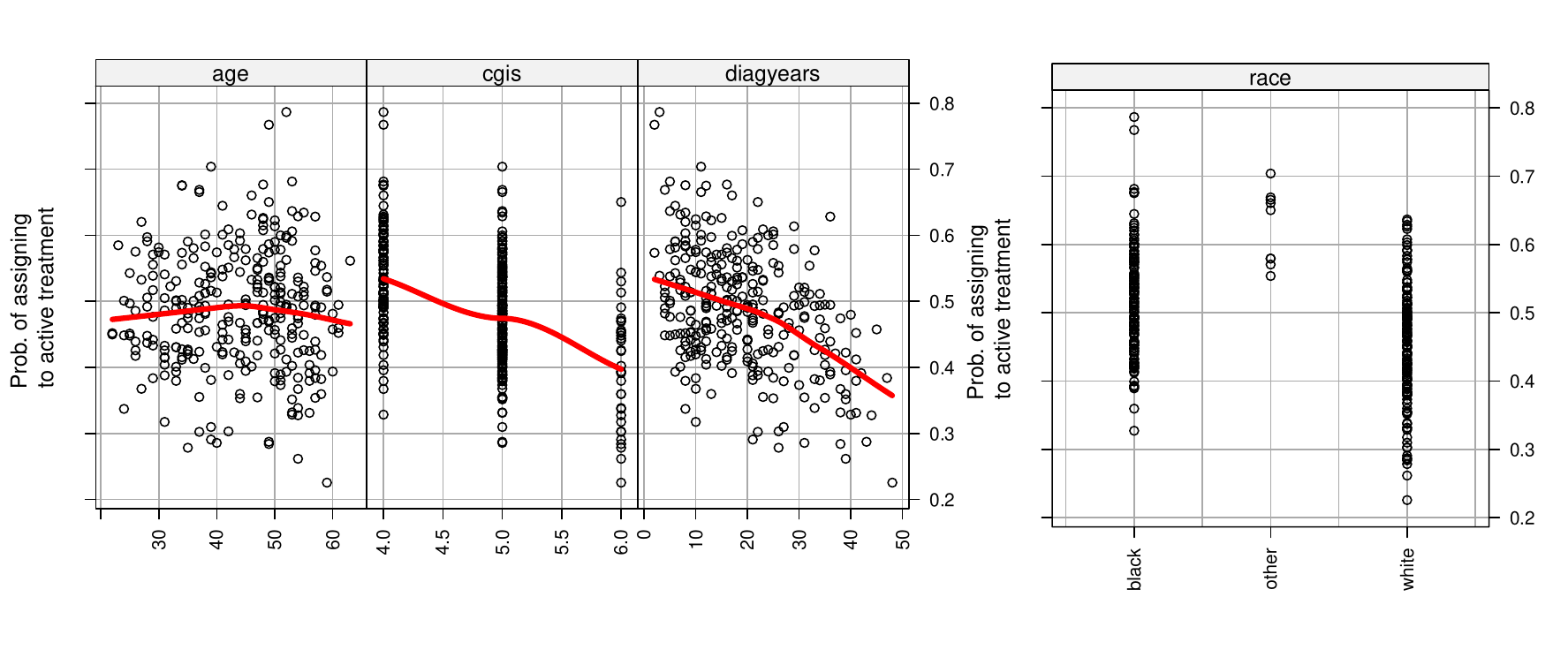} 
\caption{Estimated probability of assignment to the recommended active treatment as estimated by OWL  in the schizophrenia trial.}  
  \label{fig.OWLprob}
\end{figure*}

A method of direct ITR modeling, Outcome Weighted Learning, was applied using the \texttt{DTRlearn2} package. The baseline variables were standardized to ensure the mean of zero and variance of 1. The standardization procedure was also applied to the binary variables for \textit{gender} and \textit{race}. 
We used the following modeling options: Radial Basis Function (RBF) kernel as the learner, logistic loss function, and 10-fold cross-validation for tuning the hyper-parameters associated with the RBF learner. The estimated model coefficients are displayed in Figure~\ref{fig.owlRBFlogit} indicating that \textit{race}, \textit{cgis}, \textit{age} and \textit{diagyears} as the most important variables.
Figure~\ref{fig.OWLprob} presents the scatter plots for the estimated probability of assignment to the active treatment against these 4 covariates (e.g., $P > 0.5$ on the Y-axis indicates that the active treatment should be recommended).  

For subgroup identification, we applied three methods: the SIDES method including the Base SIDES\cite{Lipkovich2011} and the Adaptive SIDEScreen\cite{Lipkovich2014} (implemented in the \texttt{rsides} R package), the CAPITAL method\cite{CaiCAPITAL2022} (\texttt{policytree} R package), and the sequential-BATTing method (\texttt{SubgrpID} package). 

We used the base SIDES with with $depth=2$, $width=3$, $min\_subgroup\_size=30$, and required for a parent subgroup to be split in a way that the best child's p-value were no worse than that of the parent ($gamma=1$). To compute adjusted p-values we generated $M=10,000$ null sets by permuting treatment labels. This resulted in 12 subgroups with the following top 4 subgroups: (1) ${diagyears \le 13, (2) race= \text{Black\;or\;Other}}$, (3) ${diagyears \le 13, pansspos > 23}$, (4) ${gender=Female, diagyears \le 14}$, and ${diagyears \le 13, age>34}$. The  associated unadjusted one-sides p-values were 0.007, 0.008, 0.014, and 0.015, respectively. The permutation adjusted p-values were all greater than 0.5. 

\begin{figure}[htb!] 
\centering
\includegraphics[scale=0.6, trim=100 120 50 140,clip]{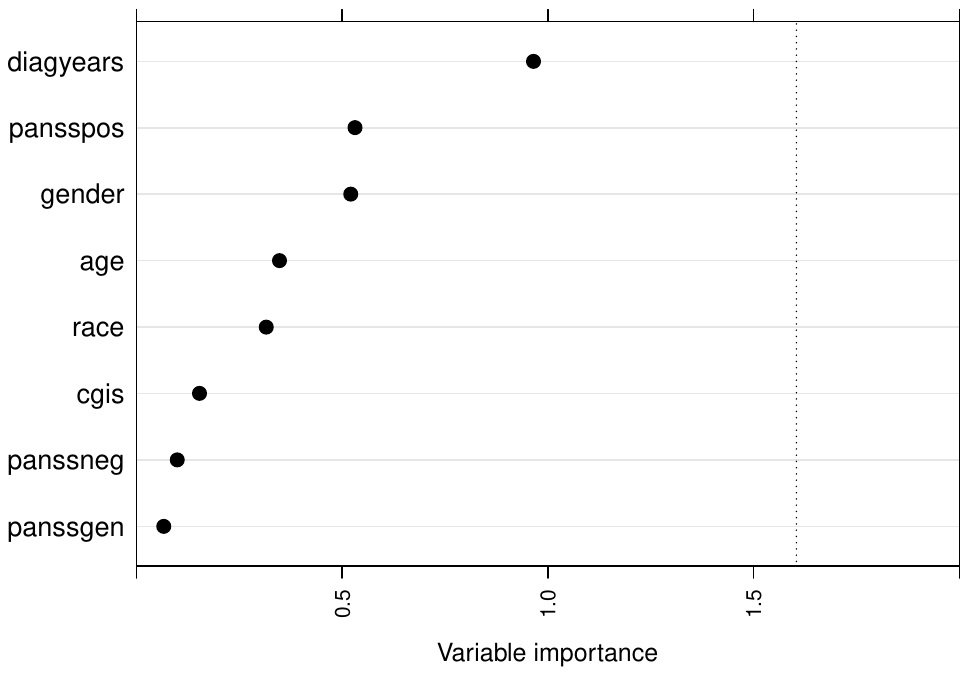} 
\caption{Variable importance plot for the application of the Adaptive SIDEScreen to the case study. The dotted vertical line indicates the threshold for maximal VI score = $E_0(\text{VI}_{max})+1S_0(\text{VI}_{max})$ computed from 1000 reference data sets.}
\label{fig.case.sides.vimp}
\end{figure}

We then used the  Adaptive SIDEScreen with the following options. For the first stage, the base SIDES was used with $depth=3$, $width=5$ to ensure a large pool of subgroups in order to compute the variable importance scores, shown in Figure~\ref{fig.case.sides.vimp}. The vertical dashed line indicates the cut-off $E_0(\text{VI}_{max})+1S_0(\text{VI}_{max})$ obtained from the null distribution of the maximal variable importance score based on 1000 reference data sets. As we can see, no variable reaches the cut-off indicating lack of predictive biomarkers in this data set.

For the CAPITAL method, the default options were used (as given in the example code in the software repository\cite{CAPITALBLOG} (thttps://github.com/HengruiCai/CAPITAL/blob/main/CAPITAL.R). Specifically, we used Random Forest (\texttt{randomForestSRC} R package) as the base learner for a T-learner to estimate CATE. Specifically, the predicted potential outcomes for actually observed treatments were obtained from the out-of-bag (OOB) samples (i.e., based on the trees fitted to bootstrap samples that did not contain these observations), whereas predictions for the counterfactual outcomes were the usual ``plugged-in'' (PIN) predictions from the RF model: 
\begin{equation}\label{eq.RF.pred}
\widehat{\Delta}_i=T_i(\widehat{\mu}^{oob}(X_i,1)-\widehat{\mu}^{pin}(X_i,0))+ (1-T_i)(\widehat{\mu}^{pin}(X_i,1)-\widehat{\mu}^{oob}(X_i,0)).
\end{equation}

\begin{figure*}[!ht] 
\centering
\includegraphics[scale=0.6, trim=0 200 0 0,clip]{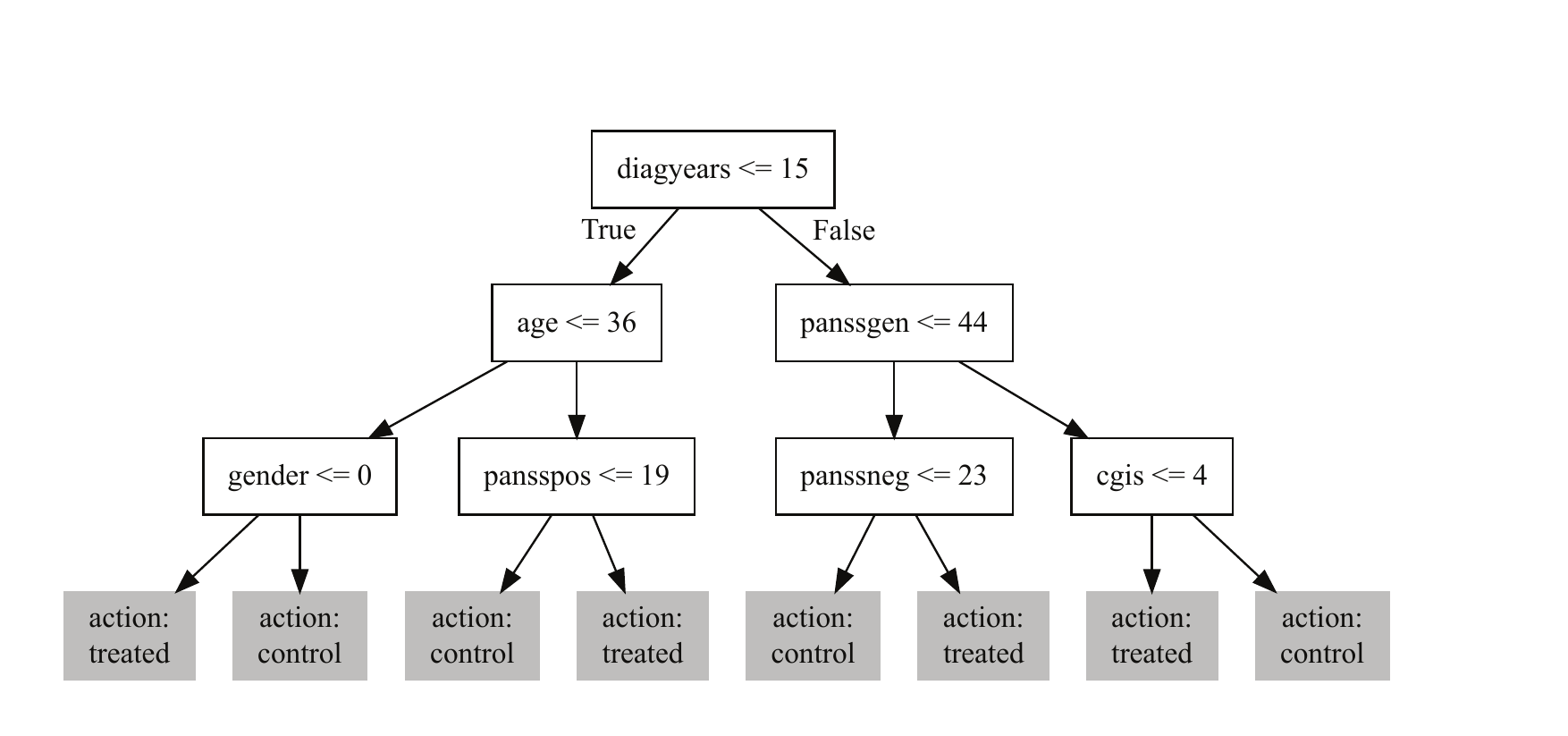} 
\caption{Subgroups identified using the CAPITAL method in the schizophrenia trial.}
  \label{fig.CS3_CAPITAL}
\end{figure*}

These estimates are further used as the basis for computing rewards associated with tree-splitting  (we used the default option of type 1 rewards). Alternatively, we could use as input for the procedure the doubly robust scores \eqref{eq.dr.scores} (e.g. computed by the Causal Forest). 
The threshold was set to $\delta=5$ (corresponding to the  minimal clinically relevant effect for \textit{panss42.change}) and tree $depth=3$.  
The resulting tree is displayed in Figure~\ref{fig.CS3_CAPITAL}, with the first split on \texttt{diagyears} and further splits on the variables \textit{age}, \textit{panssgen}, \textit{panssneg}, \textit{pansspos} and \textit{cgis}. The tree has eight terminal nodes , i.e., patient subgroups; some of these are rather small (see Table ~\ref{tab.CAPITAL}). For example, the subgroup $\{$\textit{diagyears} $\leq 15$, \textit{age}$>36$, \textit{pansspos}$ \leq 19\}$ only contains four observations.

The sequential-BATTing method was fitted using 5 CV-folds and 20 replications of cross-validation; it identified the subgroup $\widehat{S}=\{pansspos > 20\}$. This is a rather large subgroup with the prevalence of 91.4\% (the median $pansspos=26$). The na\"ive mean estimates of \textit{panss42.change} within the identified subgroup $\widehat{S}$ were $-14.9$ and $-14.4$ for the experimental treatment and  control arms, respectively, suggesting a small difference of $-0.5$ in favor of the experimental treatment. However, the cross-validation adjusted estimates (averaged over 20 CV replications) were $-13.5$ and $-16.0$, respectively, indicating a reversal of treatment effect in favor of the control arm and suggesting that the identified subgroup was spurious.



\section{Inference for HTE}\label{sec.post.inf}
In the previous section, we examined the results obtained from multiple methods for HTE evaluation applied to the case study and it may be tempting to conclude that there is evidence of treatment effect heterogeneity and that patients with certain characteristics may benefit from the experimental treatment more than others. In this section we will review several formal methods for testing hypotheses about the presence of HTE and illustrate them using existing packages on the case study. We will see that all the tests indicated lack of heterogeneity in our data, which underscores the danger of interpreting the CATE estimates in a qualitative manner without a principled assessment of the significance of any apparent heterogeneity patterns.  

In this section, we review several recent advances related to the challenging task of performing ``post-selection'' inferences for HTE, such as testing for the overall heterogeneity of TE, constructing confidence intervals for estimators of CATE, the value of ITRs, or treatment effects in selected subgroups. We summarized available software tools for performing inference on HTE in Table~\ref{tab.appendix.inference} of the Appendix.

\subsection{Testing for heterogeneity of treatment effect}\label{sec.post.inf.HTE}
For clinical trial sponsors and public decision makers, it is tempting to approach the potential heterogeneity of treatment effect using a two-stage ``gate-keeping'' strategy. First, carry out an overall test for the presence of HTE in the data set of interest and, if the null hypothesis of homogeneity is rejected at a certain  significance level, proceed to evaluating the treatment effects in candidate subgroups. This general approach is often implemented using a rather arbitrary combination of an interaction test nested within a family of parametric models (e.g., logistic regressions) with a proverbial alpha level of 0.1, accompanied by a series of tests within candidate subgroups (e.g., by assessing  different cut-off values of a continuous biomarker). Such strategies were generally considered not very powerful and robust, because of the low power of interaction tests and its reliance on parametric assumptions. With the advent of machine learning approaches to the assessment of treatment effect heterogeneity, several proposals were made for constructing similar global tests in a completely non-parametric fashion taking into account large sets of candidate covariates. 

Chernozhukov et al \cite{Chernozhukov2020} proposed a homogeneity test for comparing the average treatment effects across the $K$ subgroups $\{G_1, \dots, G_{K}\}$ based on sorted estimates of CATE, i.e., $\widehat{\Delta}(x)$. This method is known as the Sorted Group ATE (GATE). In this case, the homogeneity test is defined with respect to the following null hypothesis: $E(\Delta(X)|G_1)= \dots = E(\Delta(X)|G_K)$. The test is constructed under the monotonicity assumption, i.e., $E(\Delta(X)|G_1) \le \dots \le E(\Delta(X)|G_K)$. The key idea is to randomly split the data into a training set $(A)$ and a test set $(B)$, estimate $\Delta(x)$ by applying a generic ML algorithm on the set $A$, and use this as a ``proxy'' estimator to group patients in the set $B$ into subsets with similar CATE. The target GATE parameters are then estimated in the set $B$ using a simple linear ANCOVA-type model with $K$ terms for treatment-by-group interactions and prognostic score $\widehat{m}_0(x)=\widehat{E}(Y|T=0, X=x)$ (estimated by an ML method on the training set) added to the model to increase precision. The treatment variable in the interaction terms is included as a deviation from the probability of the treatment assignment, $T_i-\pi(X_i)$, which orthogonalizes the interaction relative to all other regressors that are functions of covariates. Inferences for the GATE effects (i.e., coefficients for the interaction terms in the ANCOVA model) can be performed in a straightforward manner, because all the ``heavy lifting'' (tuning and fitting ML models) is done on the data set $A$ \textit{independently} of the test set $B$. The key observation is that this simple linear projection from $A$ to $B$ dramatically reduces the variability in the GATEs have they been directly computed on the training set from the proxy estimator $\widehat{\Delta}(x)$. The authors argue that the proposed GATE estimator will work even if the underlying estimate of $\Delta(x)$ by a generic ML method is not consistent. The data splitting is repeated many times and the individual GATE estimates are combined by computing the medians of the point estimates and associated lower and upper confidence limits across the sub-samples. To account for the uncertainty due to random data splitting, the authors proposed adjusting the nominal confidence levels by simply doubling the alpha and demonstrated (by a theoretical argument) that it results in a valid coverage probability. These ideas are implemented in the R package \texttt{GenericML}.       

Recently Imai and Li \cite{ImaiLi2022} proposed a different procedure for testing the homogeneity of GATEs based on a generic ML algorithm, also taking advantage of data splitting (implemented in the R package \texttt{evalITR}). Similarly to the proposal in Chernozhukov et al, \cite{Chernozhukov2020} their method does not require a consistent estimator of CATE but it does not impose the monotonicity assumption and, unlike Chernozhukov et al, \cite{Chernozhukov2020} do not rely on a linear regression when making inferences for the GATEs, providing a fully non-parametric strategy. The hypothesis testing in Imai and Li \cite{ImaiLi2022} is based on a chi-squared-type test statistic for the homogeneity of treatment effects $\theta_k, k=1,..,K$ computed within each of the $K$ subgroups induced by a generic  ML estimator of CATE. Importantly, the variance of the test statistic accounts for uncertainty due to data partitioning given CATE (estimated from an independent data set) that serves merely as a fixed ``scoring rule''. The authors provide several alternative strategies of data splitting, including single \textit{sample  splitting} and (potentially more efficient) \textit{cross-fitting}. Under cross-fitting, the full data set is randomly divided into $L\ge 2$ subsets (folds) of equal size. Considering each fold sequentially as a test set and the remaining $L-1$ folds as a training set, a generic ML algorithm is applied to the training set to estimate CATE, $\widehat{\Delta}(x)$. Then a GATE-based homogeneity statistic is computed for the test set. As in Chernozhukov et al,\cite{Chernozhukov2020} because the CATE (scoring rule) is estimated from the training data independently of the test data, the asymptotic variance of the GATE statistic conditional on the scoring rule remains valid. However, combining the GATE statistics across the $L$ test folds induces additional sampling variability due to re-estimating CATE from the (overlapping) training sets that would be unaccounted for in a na\"ive estimator of variance formed by simply adding variances across all the test sets. In addition, this cross-fitting scheme changes the interpretation of GATEs. Now the scoring rule underlying GATEs is not fixed but rather is a function of a specific ML algorithm applied on all possible training samples of size $n(L-1)/L$ from the full data. Therefore, to perform correct inference for the GATE-based test under cross-fitting, an additional positive term in the variance is required to account for the uncertainty induced by sampling from the training data, which is offset by a negative term due to combining the test data across the folds. Fortunately, the latter will dominate the former provided the sample size is sufficiently large, resulting in net efficiency gains (comparing with the sample splitting for the same size of the test set). 

We applied the heterogeneity test of Imai and Li \cite{ImaiLi2022} to our case study using the function \texttt{hetcv.test} of the R package \texttt{evalITR}. The function assumes for each subject $i$ the cross-validated prediction of CATE, $\widehat{\Delta}_i^{(-i)}$, obtained by some ML method. We obtained predictions by randomly dividing the data into 10 folds (stratified by the treatment group) and running Causal Forest with 10,000 trees on each training set (formed by excluding each fold in turn as a testing set) and applying the estimated CF to compute CATE for each subject in the test set. We used the default number of groups $K=5$ (\textit{ngates} parameter) for heterogeneity testing. It is worth noting that the user also has to retain and provide as input the estimated CATEs for each subject from the respective training set (by data re-substitution), which is needed for computing the standard error for the test statistic. Therefore, the \texttt{hetcv.test} function serves as a wrapper over any arbitrary ML method as long as it can be applied in a cross-validated fashion. The resulting p-value for testing the null hypothesis of heterogeneity was $0.927$, indicating no evidence for heterogeneity of the treatment effect across the groups. 

Another approach for evaluating the global hypothesis of treatment effect heterogeneity suggested in Chernozhukov et al \cite{Chernozhukov2020} is based on the ``best linear predictor'' (BLP) of an ML proxy of the ITE. The idea is to fit on the test set $B$ a weighted linear model for $Y$ on the terms for $\widehat{m}_0(X_i)$, centered treatment effect $(T_i-\pi(X_i))$ and centered interaction  $(T_i-\pi(X_i))(\widehat{\Delta}(X_i)-\bar{\Delta})$, with $m_0(X_i)$ and  $\Delta(X_i)$ estimated using a generic ML method on the training set $A$; here $\bar{\Delta}$ is the average treatment effect. The subject weight here is $W(X_i)=\{\pi(X_i)(1-\pi(X_i))\}^{-1}$ to ensure the orthogonality of the centered terms for the treatment and interaction effects. The coefficient for the interaction term then captures the HTE, i.e., its equality of 1 indicates that the ML estimate of HTE is perfectly well-calibrated, and rejecting the hypothesis that the coefficient equals 0 provides evidence for the presence of true heterogeneity of treatment effect. 

The reviewed proposals assume randomized treatment, limiting the application to RCTs, although the setup in Chernozhukov et al \cite{Chernozhukov2020} is broader, assuming that the treatment assignment can be covariate-based as long as the propensity function $\pi(x)$ is known and bounded away from 0 and 1 (although intended for randomized experiments, their \texttt{GenericML} R function allows for the propensity score to be estimated from observed data using ML methods); see also a proposal for testing for the presence of systematic variation in TE in randomized trials in Ding et al. \cite{Ding2016,Ding2019}

Athey and Wager \cite{AtheyWager2019} adopted the idea of a BLP test from Chernozhukov et al \cite{Chernozhukov2020}  in the framework of causal trees using cross-fitted estimates of the mean function  $\widehat{m}^{(-i)}(X_i)$, propensity $\widehat{\pi}^{(-i)}(X_i)$ and the individual treatment effect $\widehat{\Delta}^{(-i)}(X_i)$ (here the superscript $(-i)$ indicates the $i$-th subject has been removed when estimating a given function). Their method applies to observational studies with unknown propensities. Specifically, the authors propose to fit a linear model for the centered outcome 
\begin{equation}\label{CF.HTE}
Y_i-\widehat{m}^{(-i)}(X_i)= \alpha \bar{\Delta}(T_i-\widehat{\pi}^{(-i)}(X_i)) + \beta (T_i-\widehat{\pi}^{(-i)}(X_i))(\widehat{\Delta}^{(-i)}(X_i)-\bar{\Delta}),
\end{equation}
where $\bar{\Delta}= n^{-1}\Sigma_{i=1}^n \widehat{\Delta}^{(-i)}(X_i)$ is the average treatment effect estimated by Causal Forest, and the coefficient $\beta$ for the interaction term captures the heterogeneity of treatment effect. If $\beta$ is significantly larger than zero, it can be taken as evidence of HTE as estimated by the Causal Forest, although the authors warn that the ``asymptotic results justifying such inference are not presently available.'' See also Crump et al \cite{Crump2008} who proposed a non-parametric test that applies to non-randomized trials.  

The R package \texttt{grf} for Causal Forest, provides an omnibus test for the presence of HTE based on the best linear fit of the target estimand (using the CF predictions on the held-out data via the \texttt{test\_calibration} function) based on the representation in \eqref{CF.HTE}. The p-value for this test on the case study data set was 0.949, which indicates lack of HTE.

 So far we only looked at approaches for detecting the presence of HTE explained by covariates (e.g., through CATE). Perhaps a more direct approach is to evaluate the  core characteristics of variability in ITEs, $\Delta_i=Y_i(1)-Y_i(0)$ such as its variance
 \[
 var(\Delta_i)=\sigma^2_{\Delta}=\sigma^2_{Y(1)}+\sigma^2_{Y(0)}-2\rho\sigma_{Y(0)}\sigma_{Y(1)},
 \]
 where $\rho=\text{cor}(Y(0),Y(1))$ is the correlation between potential outcomes $Y(0)$ and $Y(1)$. Using this representation, one could construct a confidence interval for $\sigma^2_{\Delta}$ to see if HTE can be ruled out based on the data at hand. Unfortunately the joint distribution of $Y(0), Y(1)$ is unobservable and their correlation can only be used as a sensitivity parameter. One direction is to consider a lower bound for $\sigma_{\Delta}$, $L=|\sigma_{Y(1)}-\sigma_{Y(0)}|$ by setting $\rho=1$, but this bound is often too low and not useful for any practical considerations. One can attempt to incorporate prognostic effects $X$ in hopes that it helps tighten the covariate-adjusted lower bound $L=|\sigma_{Y(1)|X}-\sigma_{Y(0)|X}|$.\cite{Gadbury2000} It can also be conjectured that conditional on covariates, PO's are independent\cite{Zhang2013}: $Y(0) \perp  Y(1) \mid X$, however it is more realistic to assume that independence can be achieved only after conditioning on all covariates and a random subject-specific effect $U$, $Y(0) \perp  Y(1) \mid X, U$.\cite{Zhang2013, Yin2018} For example, Zhang et al \cite{Zhang2013} proposes sensitivity analyses by manipulating the variance of random effect $U$ as a sensitivity parameter. 
  
  In addition to $\sigma_{\Delta}$, several researchers considered other more clinically interpretable measures of treatment effect heterogeneity, such as the treatment benefit rate (TBR) and the treatment harm rate (THR)\cite{Gadbury2000,Gadbury2001,Yin2018} that can be defined, for example, as $\text{TBR}=\text{Pr}(\Delta_i >0)$ and $\text{THR}=\text{Pr}(\Delta_i < 0)$. Lack of heterogeneity would manifest itself in TBR=1 and THR=0 (when the experimental treatment is uniformly superior to control). 
  
  It is natural to attempt to decompose the overall treatment effect heterogeneity into explainable (by the observed covariates) and unexplainable, and it is the latter that is tackled by some methods. For example, Yin, Liu, and Geng \cite{Yin2018} consider TBR($x$) as a function of several known effect-modifiers. In their illustrative example five known effect-modifiers form subgroups $A,B,C$ with potentially different average treatment effect across subgroups. Then they show that in addition to CATE($x$), treatment benefits rate TBR($x$) also varies substantially across subgroups (e.g., while about 65\% of patients from group $A$ would benefit from the active treatment, only about 40\% would benefit in groups $B$ and $C$) indicating there are additional unobserved factors that may explain the variation of treatment effect in these sub-populations. Of course, in general we do not know which variables are effect modifiers and it would be interesting to develop ML  methods that simultaneously estimate CATE($x$) and TBR($x$) based on a high dimensional set of candidate covariates $X$; consequently we could simultaneously test for the presence of explainable and unexplainable heterogeneity of treatment effect in the data. One may question how the awareness of presence of HTE (in a population or sub-population) that cannot be attributed to specific covariates may help physicians to allocate patients to the right treatment? Perhaps detecting presence of substantial heterogeneity after accounting for known covariates would help in our future attempts to identify missing covariates that could explain the remaining HTE. 
  
A different view on measuring the overall HTE is presented in Levy at al\cite{Levy2021} The authors focused only on explainable variation in the treatment effect, termed VTE, and defined as $\text{VTE}=\text{var}\{E(Y(1)-Y(0)|X)\}=\text{var}\{\Delta(X)\}$, and proposed a non-parametric approach for simultaneous inference on $\text{ATE}=E(Y(1)-Y(0))$ and VTE via a cross-validated targeted maximum likelihood estimator (CV-TMLE). They argued that, while a clinician's perception of highly varying outcomes may reflect a large $\sigma^2_{\Delta}$, VTE will allow the clinician to determine if the varying outcomes were due to the measured covariates, suggesting opportunities for precision medicine. Building on VTE, Hines et al\cite{Hines2023} proposed a measure of HTE variable importance, TE-VIM, quantifying the amount by which the VTE changes when a subset of covariates $X_s, s \subseteq \{ 1,..,p\}$, is excluded from modeling CATE: $\text{TE-VIM}=\text{var}\{\Delta(X)\}-\text{var}\{E(Y(1)-Y(0)|X_{-s})\}$, where $X_{-s}$ are the variables not included in the set $s$. They developed inference for TE-VIM using ML methods employing doubly robust estimators of CATE, which were subsequently extended to TMLE.\cite{Li2023}
  
\subsection{Inferences on ITE} 
Inferences on ITE estimates from ML methods received considerable attention over the last few years. Many approaches leverage recent advances in post-selection inference. For example, Ballarini et al \cite{Ballarini2018} conducted inferences on ITE, or PITE\textemdash predicted ITE in their terminology, i.e., $\Delta(x)$, estimated from penalized regression using recently developed post-selection inferences for LASSO.\cite{Lee2016, Taylor2018} Note that post-selection confidence intervals in Lee et al \cite{Lee2016} assume a fixed $L_1$ penalty. Choosing penalty in a data-driven fashion, e.g., by a popular $K$-fold cross-validation methods, would create yet another layer of selection. Another popular route for CATE inferences is based on using the infinitesimal jackknife method of Wager et al \cite{Wager2014} for constructing pointwise confidence intervals for Random Forests. These ideas were adopted in Random Forests of Interaction Trees (RFIT) by Su et al \cite{Su2018} and Causal Random Forests by Wager and Athey.\cite{WagerAthey2018} Recently Guo, Zhou, and Ma \cite{Guo.W.2021} proposed flexible semi-parametric modeling method offering simultaneous confidence bands for $\Delta(x)$. 

Lee, Okui, and Whang \cite{Lee2017} constructed simultaneous uniform confidence bands using a non-parametric kernel estimator of CATE as a function of a small number of predictors $X_{0}$ pre-selected by the user from a larger set of covariates, $X$. The authors argue that, while the full set $X$ may be needed to account for all confounders, a much smaller subset may be sufficient to capture the heterogeneity of treatment effect. Therefore, they proposed a two-stage procedure. At the first stage, doubly robust scores of CATE \eqref{eq.dr.scores} are constructed using parametric modeling of the outcome and treatment assignment utilizing all available covariates. At the second stage, the scores are smoothed out using a kernel estimator based only on the subset $X_0$. The authors showed that the uncertainty in the first stage does not need to be accounted for when computing the standard errors for the confidence band for $\Delta(X_0)$, provided the DR scores converge to their population counterparts at a fast parametric rate. These ideas have been further developed recently by methods allowing for high-dimensional nuisance functions at the first stage with low dimensional CATE at the second stage.\cite{Fan2022, Kato2024}

Bayesian regression models provide another route for performing inferences on ITEs. Many researchers used BART (Bayesian regression trees) as a building block for estimating individual causal effects. A notable application is that of Hill \cite{Hill2011} and non-parametric Bayesian accelerated failure time models by Henderson et al \cite{Henderson2020} (implemented in the R package \texttt{AFTrees}). Hahn et al \cite{HahnBART} noted regularization bias resulting from a straightforward application of BART to estimating the causal effects from non-randomized trials with strong confounding and proposed explicitly incorporating the propensity information in the BART priors for the outcome model. This allows separate degrees of shrinkage for prognostic and predictive effects in the outcome model (implemented in the R package \texttt{bcf}), not unlike similar proposals in the frequentest literature for penalized regression.\cite{Imai2013} 


 \subsection{Inferences on ITR} 
Inferences on the individualized treatment assignment rule $D(x)$ are typically concerned with its role in providing an answer to the following important question: what is the expected outcome if all patients in a population of interest would follow that rule? Specifically, we are interested in estimating $V(D(X))=E(Y(D(X))$, where the expectation is computed with respect to $Y$ and $X$. In the ITR literature, it is sometimes difficult to understand whether the target of inference is the value of the (unknown) optimal regimen/rule, $V({D}_{opt}(X))$, or the value of the \textit{estimated} regimen/rule  $V(\widehat{D}(X))$  involving a data-dependent parameter. It is important to note that both estimands are of clinical interest. The former provides a benchmark for the mean outcome if all patients are treated with the (possibly unattainable in real practice) optimal regimen. The latter tells us about the outcomes if all future patients are treated with the regimen estimated from the available data.

Inferences on the value of the optimal ITR are challenging because it is an irregular problem even in a relatively simple analytical setting of a single decision point when the optimal regimen is estimated by a two-stage procedure similar to Qian and Murphy \cite{qian2011} without variable selection, i.e., estimating the expected potential outcomes if treated and if untreated by linear regression models fitted to each treatment arm, $\widehat{E}(Y|X=x,T=t)$, and ``backsolving'' an optimal rule $\widehat{D}(x)$ as the rule maximizing outcomes for a given $x$. The $\max$ operator renders the estimation problem irregular (non-smooth). When using direct methods of ITR search by optimizing of the Value function, the resulting value estimator is also irregular (see Tsiatis et al.\cite{tsiatis2020book}) This occurs when there exists, with a positive probability, a subset of patients for whom the two treatment options are equally attractive (i.e., their outcomes are equal under either treatment choice) resulting in non-uniqueness of ITR; sometimes such data generation mechanism is referred to as ``exceptional law''.\cite{Luedtke2016} Several approaches were proposed to deal with the non-smooth nature of ITR. One remedy is the \textit{m-out-of-n} bootstrap, with $m<n$, see Chakraborty et al \cite{Chakraborty2013} who proposed a data adaptive choice of $m$. Note that using a smooth surrogate loss (\ref{eq.eloss.mod.smooth}) obviates the task of optimizing the non-smooth non-convex objective function (\ref{eq.value}), often producing an ITR very close to that maximizing (\ref{eq.value}). However, when estimating the value of the estimated ITR, we still need to plug the estimated regimen $\widehat{D}(X)$ in the Value definition \eqref{eq.value} to compute the expected outcomes under the optimal treatment, rather than using a smooth ``surrogate'' value, hence the estimand may still be a non-smooth function of the data. As a way to overcome this problem, Laber et al \cite{Laber2014a} proposed to actually define the Value of ITR using a smoothed version by replacing the indicator function in the Value definition with a smooth function of covariates. Another approach relies on truncating the data such that the remaining subjects do not exhibit exceptional behavior (thus changing the estimand).\cite{Laber2014c}
Luedtke and van der Laan \cite{Luedtke2016} proposed a framework for dealing with the non-uniqueness of the optimal ITR by imposing certain relatively weak conditions resulting in a regular and asymptotically linear estimator of the optimal value.


The non-smoothness of the ITR value is further exacerbated by using machine learning methods to estimate the outcome functions, which is likely to result in an over-optimism bias with respect to the estimated value $\widehat{V}(\widehat{D}(X))$ versus the true value of the estimated regimen, $V(\widehat{D}(X))$. Here, an important advance allowing for naturally incorporating ML methods relies on applying the Targeted Minimal Loss-based Estimator (TMLE) framework and treating the value of ITR (i.e., the mean outcome under the optimal regimen) as the targeted parameter.\cite{vanderLaan2013, vanderLaan2015} A consistent variance estimate of the TMLE estimator for the Value of optimal ITR can be obtained (under certain assumptions) by using the theory of doubly-robust efficient influence functions evaluated from the test data while all the nuisance parameters are estimated on the training data. This is achieved via cross-fitting, that divides the data set into $L$ non-overlapping subsets and treats in turn each set as the test data while the remaining $L-1$ sets as training sets (this approach is similar to the procedures described in Section~\ref{sec.post.inf.HTE}). See also Chu, Lu, and Yang\cite{Chu2022} who developed inferences for the value under a method for learning targeted ITRs using summary statistics.  

It is important to accurately estimate the value of the estimated regimen because it naturally affects the decision making process. The stakeholders want to know whether the optimal regimen estimated by ML algorithms is substantially better than a relevant benchmark regimen. Therefore, the value of an estimated regimen should be compared with that of the selected benchmark, e.g., a non-individualized treatment regimen such as a rule that assigns all patients to a control treatment\cite{AtheyWager2021} or to the treatment having the best average treatment effect. An example of a non-individualized rule applicable to RCTs is a random assignment of a fixed proportion of patients to the experimental treatment. Another approach (for observational data) is to use as the benchmark a specific ITR, such as that based on the current clinical practice. 

Imai and Li \cite{ImaiLi2021} proposed to compare the estimated ITR that assigns a proportion $p$ of patients to the experimental treatment with a non-individualized regimen that randomly assigns the same proportion of patients to the experimental treatment, and referred to the resulting contrast as PAPE (Population Average Prescriptive Effect). The PAPE is defined as:
\begin{equation}\label{eq.PAPE}
 \text{PAPE}=E[Y\left(D(X)\right)]-E[pY(1)+(1-p)Y(0)],
\end{equation}
where $p=\text{Pr}(D(X)=1)$.
They developed a procedure for constructing confidence intervals for PAPE using a cross-validation approach, similar to that used for estimating heterogeneous GATE effects in Imai and Li \cite{ImaiLi2022} described earlier in this section (and also implemented in their R package \texttt{evalITR}). They further extended PAPE to the setting when an optimal ITR has to obey a `` budget constraint'' $p$ that allows comparing different ITRs on the same footing. If an ITR is based on thresholding a predictive score, such as CATE, (i.e., $\widehat{\Delta}(x)>\delta$), then we can calibrate $\delta$ as a function of a pre-specified budget $p$ to ensure the resulting ITR assigns \textit{at most} the  proportion $p$ to the experimental treatment. Denote  the ITR for a given budget constraint $p$ by $D_p(X)=I(\widehat{\Delta}(X) > \delta(p))$, where $\delta(p)=\text{inf}\{\delta \in [\delta^*, +\infty): \text{Pr}(D_p(X)=1) \le p\}$; here $\delta^*$ quantifies the minimal required clinical benefit\textemdash it is the threshold that would have been used in absence of budget constraints (often $\delta^*=0$). In words, to allocate patients to the experimental treatment we use the lowest threshold for CATE within the budget on proportion treated, while not treating patients with no treatment benefits (or those who would be harmed by treatment). More generally, $\widehat{\Delta}(X)$ can be replaced by other scoring rules $s(X)$ determining the treatment prioritization, e.g., based on expert evaluations. Then PAPE($p$) is defined as a function of $p$:
\begin{equation}\label{eq.PAPEp}
 \text{PAPE}(p)=E[Y\left(D_p(X)\right)]-E[pY(1)+(1-p)Y(0)].
\end{equation}

The authors further proposed an area under the prescriptive effect curve (AUPEC), i.e., the area under $\text{PAPE}(p)$ plotted over the range of budget constraint $p$, as a metric of ITR performance. This  is a generalization of the QINI coefficient, which is used for uplift modeling in market research\cite{Radcliffe2007} and is related to an earlier measure of subgroup selection rule, $AD(\delta)$,\cite{Zhao2013} defined as the average treatment effect in the subpopulation selected by thresholding on the estimated CATE, $AD(\delta)= E(Y(1)-Y(0)|\widehat{\Delta}(x) > \delta)$. Recently Yadlowsky et al\cite{Yadlowsky2021a} proposed a related measure called rank-weighted average treatment effect (RATE), a general family of metrics for comparing treatment prioritization rules.

We evaluated the PAPE and AUPEC using \texttt{estimate\_itr} and \texttt{evaluate\_itr} functions of the \texttt{evalITR} package. The former function allows a wide range of ML algorithms available via the popular \texttt{caret} package to estimate the scoring rules based on CATE. Here we use Causal Forest, with the scoring rule for each patient estimated by 10-fold cross-validation. Because \texttt{evalITR} assumes that larger positive CATEs indicate superiority of the experimental treatment  over control, we used the reduction in the PANSS total score from baseline ($panss42.change \times -1$) as the outcome variable. The estimated PAPE based on the ITR with no budget constraint and the cutoff $\delta=0$, as in \eqref{eq.PAPE} and with $\widehat{p}=\widehat{\text{Pr}}(\widehat{D}(X)=1)$, resulted in a negative estimated PAPE=$-0.29$ (se=$2.2$), and p-value=$0.89$ suggesting that this ITR is no better than a rule randomly assigning the same proportion of subjects to the experimental treatment. The cross-validated estimate of the proportion treated under the estimated ITR was $\widehat{p}=0.43$. As another benchmark, we estimated PAPE$(p)$ in \eqref{eq.PAPEp} under a rather arbitrary budget constraint $p=0.49$ reflecting the actual proportion of patients randomized to the experimental treatment, which also resulted in a negative estimate (PAPE$(0.49)=-1.1$, se= $1.5$, p-value =$0.46$). The estimate of PAPE$(0.49)$ compares the average outcome under treatment regimen assigning up to 49\% of patients with largest estimated CATE to the experimental treatment with the value of a regimen assigning randomly 49\% of patients to the experimental treatment (as in our RCT). In real applications the selection of $p$ may be dictated by cost constraints, or we may be interested in comparing a regimen based on thresholding of CATE to a more interpretable alternative regimen (e.g., a tree-structured  regimen produced by CAPITAL in Figure~\ref{fig.CS3_CAPITAL}). In the latter case, for a fair comparison we may choose the cutoff for CATE to match the proportion of treated under the tree-structured regimen. 

\begin{figure}
\centering
\includegraphics[scale=0.5]{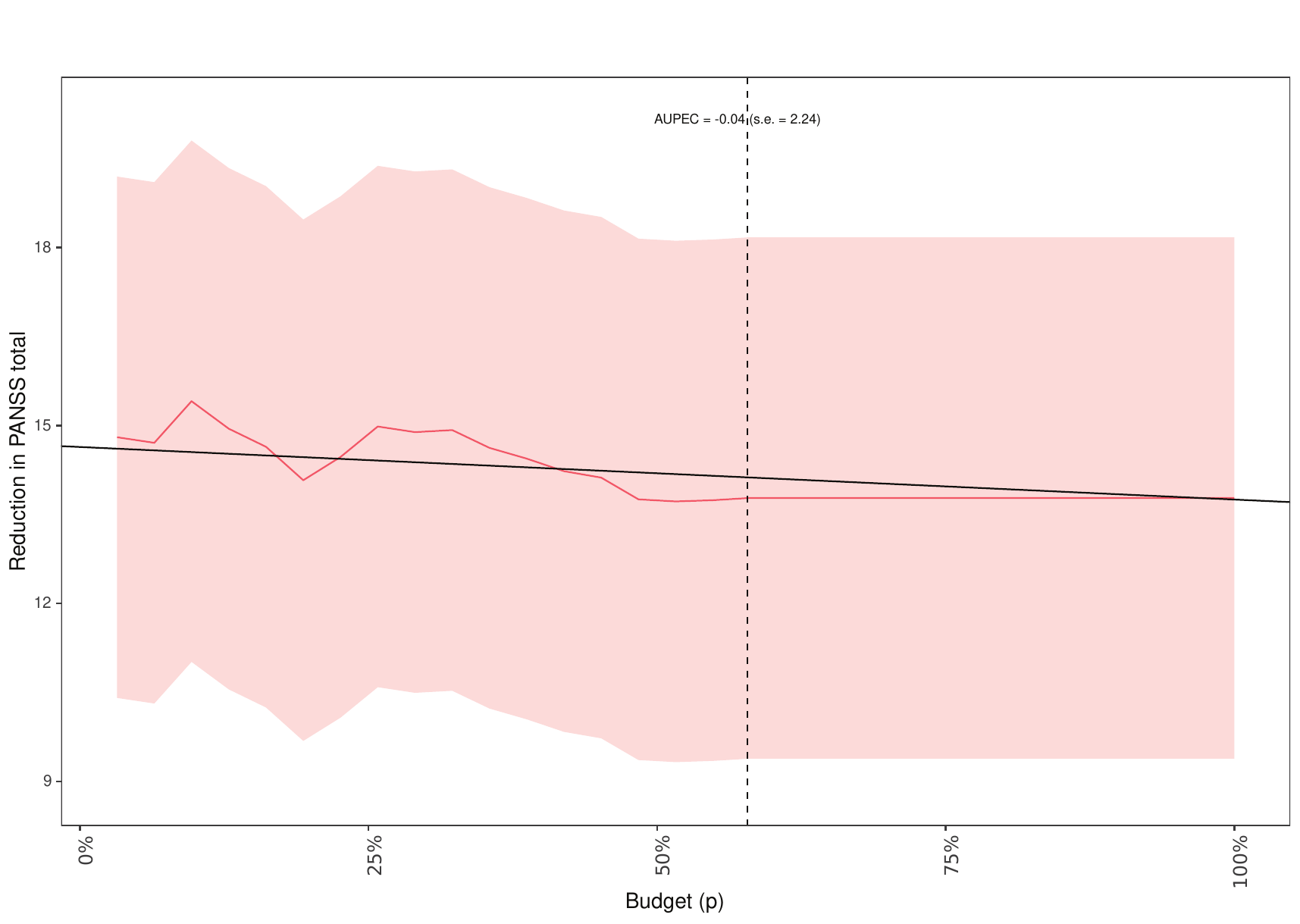} 
\caption{The plot of PAPE$(p)$ (Population Average Prescriptive Effect) as a function of budget constraint $p$ (maximum proportion treated) for the clinical trial example obtained by Causal Forest with 10-fold cross-validation. The vertical axis reflects the reduction in the PANSS total score from baseline at Day 42 (positive values indicating clinical improvement). The red curve represents the value of the ITR under a given budget $p$, and the black line represents the value of a random regimen assigning the proportion of $p$ patients to the experimental treatment. The dashed vertical line corresponds to  the proportion treated under the optimal rule $I(\widehat{\Delta}(X) >0)$ with no budget constraint.}
\label{fig.case.aupec}
\end{figure}

The plot of PAPE$(p)$ as a function of the budget $p$ is displayed in Figure~\ref{fig.case.aupec}, where the red solid curve is the value of the ITR under budget constraint (assigning patients with $\widehat{\Delta}(X) > 0$ to treatment, up to $100p\%$ of all patients) estimated by Causal Forest with 10-fold cross-validation; the black line represents the value of the rule that randomly assigns exactly $100p\%$ of the patients to the experimental treatment. The dashed vertical line is drawn at $p=0.58$, the proportion treated under the optimal rule $I(\widehat{\Delta}(X) >0)$ with no budget constraint (taken as the maximum proportion over cross-validation folds). The area between the red curve and the black line represents AUPEC estimated as $-0.037$ with standard error of $2.24$. The graph indicates lack of information about treatment effect heterogeneity that can be utilized via baseline covariates in constructing a treatment assignment rule that would ``beat'' a random rule (the pointwise confidence band covers the black line across all values of $p$).

\subsection{Inferences on identified subgroups} 
Post subgroup identification inferences are designed to control two sources of uncertainty related to the following two questions: (1) how close is the estimated subgroup $\widehat{S}(X)$ to the ``true'' subgroup $S_{true}(X)$, (2) how close is the estimated treatment effect in the estimated subgroup $\widehat{E}(\Delta(X)|\widehat{S}(X))$ to the true treatment effect in this subgroup, $E(\Delta(X)|\widehat{S}(X))$. These questions are related but different. For example, one can develop a very accurate procedure for identifying subgroups that are very similar to the true subgroups. However, the treatment effect estimated within that subgroup may be very biased (overly optimistic). 

The  first type of inferences is rarely considered in the literature. A notable example would be the works by Schnell et al \cite{Schnell2018} and Schnell \cite{Schnell2021}) on controlling the probability of selecting the correct subgroups $\hat{S}(X)$ vs $S_{true}(X)$. Here the true subgroup is understood as a subgroup with a beneficial effect. Schnell et al \cite{Schnell2018} provide Bayesian credible sets, i.e.,  $\text{Pr}(S_{lower} \subseteq S_{true} \subseteq S_{upper}) > 1-\alpha$, where $S_{lower}$ and $S_{upper}$ are the \textit{bounding} subgroups that are constructed in a way that the former (termed the \textit{exclusive} credible subset) is contained by the true subgroup and the latter (the \textit{inclusive} credible subset) contains the true subgroup, with a desired posterior probability $1-\alpha$. The true  subgroup is defined as a subset of the covariate points for which CATE is positive, i.e., $S_{true}(X)=\{x: \Delta(x)> \delta\}$, where $\delta \ge 0$ is the hypothesized value, often taken as $\delta=0$. The pair of credible subsets $S_{lower}$ and $S_{upper}$ are constructed from the simultaneous credible  bands on $\Delta(x)$: $\Delta_{lower}(x)$ and $\Delta_{upper}(x)$, as $S_{lower}(X)=\{x: \Delta_{lower}(x)> \delta\}$, and $S_{upper}(X)=\{x: \Delta_{upper}(x) \ge \delta\}$. The credible bands are based on posterior samples from $\Delta(x)$, and also have an asymptotic frequentist interpretation, therefore the credible sets methodology combines Bayesian and frequentist elements. This methodology  is extended in Schnell  \cite{Schnell2021} where a frequentist counterpart for Bayesian credible sets is proposed. The frequentist sets are constructed via a Monte-Carlo procedure. Ngo  et al \cite{Ngo2020} presents applications of this method to identifying patient subgroups with a beneficial treatment effect in a case study of prostate carcinoma patients.

The methodology of credible subsets is implemented in the R package \texttt{credsubs} that takes as its input the output of a broad range of Bayesian tools for estimating CATE, whether parametrically or non-parametrically. In the former  case, \texttt{credsubs} can assume posterior samples from the parameter space (such as a linear regression), and in the latter it takes posterior samples directly from CATE over the observed covariate space. The subgroups can be evaluated for any covariate/design space (an $X$-matrix representing a ``future trial'') provided by the user, or over the observed sample space. We applied the method to our case study with the objective to evaluate the hypothesis of no promising subgroups (i.e., that the exclusive credible subset is null). To this end, we estimated CATE using the \texttt{bcf} package and evaluated the membership of each patient in the exclusive set with the function \texttt{credsubs} for the true subgroup defined as $S_{true}=\{\Delta(X)>0 \}$ and assuming the credible level of 0.95. As we expected, the exclusive set was empty indicating no evidence of treatment effect heterogeneity in the data.

The approach of credible sets can be applied to CATE estimated by high-dimensional ML methods, therefore interpreting sets in terms of specific biomarker profiles/signatures may be difficult. On the other hand, providing an error control over the set of specific subjects suggests that the null hypothesis of no positive treatment effect may be tested at individual patient's level, $H_{0i}: \Delta_i \le 0$, rather than on the subgroup's level. This set-up  has been formally entertained recently\cite{Duan2023} in an approach that constructs a subgroup of patients having  positive treatment effects interactively (building on the ideas of interactive framework for false discovery rate\cite{Lei2020}) while controlling the false discovery at individual patient's level (i.e. the FDR is defined as the proportion of patients for whom the null of no treatment effect is falsely rejected among all patients in the identified subgroup).

Obtaining point and interval estimates for treatment effects in the selected subgroup(s) is often referred to as estimating an ``honest'' treatment effect, addressing the fact that simple summaries of treatment effects within the selected subgroups based on the same data that were used to identify the subgroups (i.e., ``resubstitution'' estimates) are likely to be biased upward (a.k.a. ``overoptimism'' or \textit{optimism bias}, the notion introduced by Efron\cite{Efron1983} in the context of the error rate of a prediction rule). A number of Bayesian approaches to performing inferences for heterogeneous treatment effects after model selection naturally shrink subgroup treatment effects towards the effect in the overall population via Bayesian hierarchical modeling\cite{Jones2011,Henderson2016, Henderson2020, WangLouis2018} or via Bayesian model averaging (presenting candidate subgroups as members of a ``model space'').\cite{Berger2014,Bornkamp2017} In the frequentist community, several early proposals for correcting for optimism bias in selected subgroups were resampling-based approaches such as methods based on the bootstrap and cross-validation. Foster et al \cite{Foster2011} and Faye et al \cite{Faye2011} proposed a bootstrap correction for selection bias that involved (i) generating multiple bootstrap samples $b=1,\dots B$, by sampling $n$ subjects with replacement from the original data set; (ii) applying the same subgroup selection procedure to each sample and selecting the subgroup $S^{(b)}$; (iii) estimating optimism bias from the difference between the two estimates of treatment effect in $S^{(b)}$: an estimate from the out-of-sample data (i.e., subjects not included in the $b$th sample) serving as a test set, and an estimate from the within-sample-data  (i.e., subjects in the $b$th sample) serving as a training set. Some procedures use bootstrap samples to ``harvest'' multiple subgroups to construct biomarker signatures by aggregating over a collection of subgroups while re-using the out-of-sample data for computing bias-corrected subgroup effects.\cite{Lipkovich2017a,TSDTCRAN2022} 

Permutation approaches are often used for constructing multiplicity adjusted p-values in a data-driven subgroup analysis,\cite{Foster2016,Chang2021} as was illustrated with the Adaptive SIDEScreen method\cite{Lipkovich2014,Lipkovich2017a} in Section~\ref{direct.SID} 

Several researchers proposed to compute bias-corrected estimates of treatment effect and associated $P$-values in selected subgroups using cross-validation.\cite{Freidlin2010, Chen2015, Huang2017} In this case, the full data set is randomly divided into $L$ sets (folds) of equal size, then each set $l=1,\cdots, L$ is considered in turn a test set, while the subgroup search algorithm is applied to the remaining $L-1$ subsets serving as a training set. The derived subgroup selection rule is used to classify patients in the test fold into ``subgroup-negative'' and ``subgroup-positive'', the latter forming a set $\widehat{S}_{test}^{(l)}$. After $L$ iterations, each patient appears once in a test set and is classified as ``subgroup-positive'' or ``subgroup-negative''. The final subgroup is defined as the union of subgroups from all test sets, $\widehat{S}_{test}=\cap_{l=1}^L \widehat{S}^{(l)}_{test}$ serving as the basis for computing an adjusted treatment effect and associated $P$-value. 
All resampling-based approaches considered so far remove an optimism bias induced by a given ML algorithm for subgroup search, therefore, strictly speaking, they aim not at predicting the treatment effect in a \textit{specific subgroup} (if applied to a future data set) but at predicting the treatment effect in a subgroup identified by a \textit{specific ML method} on a given data set and applied to future data. 

A different stream of literature on bootstrap-corrected inferences for selected subgroups is in the context of \textit{pre-defined} candidate subgroups. Although potentiality large, the set of candidate subgroups $S_1,..,S_K$ is different from subgroups generated by a ML algorithm in that they are indexed by a well-defined parameter or parameters. For example, it can be a list of subgroups based on $K$ binary biomarkers, or an infinite set of subgroups defined by thresholding a continuous biomaker, $S(c)=\{X \le c\}$. In either case, candidate subgroups can be indexed unambiguously on any other data set with covariates $X^*$, so that, for example, candidate subgroups $S(c)=\{X \le c\}$ will be mapped to the subgroups $S^*(c)=\{X^* \le c\}$. To fix ideas, let the true and estimated treatment effects in $K$ candidate subgroups be $\theta_1, \cdots, \theta_K$ and $\widehat{\theta}_1, \cdots, \widehat{\theta}_K$, respectively. The selected subgroup is $s=\argmax_{k \in \{1,..,K\}} \widehat{\theta}_k$ with the estimated effect given by $\widehat{\theta}_{\max}=\widehat{\theta}_{s}$. 

In this context, it is tempting to estimate optimism bias in $\widehat{\theta}_{\max}$ over the true $\theta_{\max}=\max_{k \in \{1,..,K\}}{\theta_k}$ using a straightforward bootstrap approach, as $\max_{k \in \{1,\cdots,K\}}\widehat{\theta}^{(b)}_k-\widehat{\theta}_{\max}$, averaged over $b=1,\cdots,B$ samples (here $\widehat{\theta}^{(b)}_k$ is the effect in the $k$th subgroup evaluated using the $b$th sample). Note that, unlike in many approaches, \cite{Foster2011, Faye2011} here out-of-sample subjects are not treated as a source of test data, only in-sample data records are used. These ideas were pursued in Rosenkranz \cite{Rosenkranz2016} implemented in the R package \texttt{subtee}.

However, while the aforementioned approach does provide a bias-reduced estimates of $\theta_{\max}$, there is a caveat. To see what the problem is, consider a toy example with only two subgroups, when the true effects are equal, $\theta_1=\theta_2$ (``homogeneity null hypothesis''). Clearly, in this null case the optimism bias will be most severe. Assume a data set generated under the null with an unusually large difference between the estimated effects, i.e., $\widehat{\theta}_2$ is appreciably greater than $\widehat{\theta}_1$. Trying to correct for bias with bootstrap samples extracted from such a data set may be problematic because the bootstrap distribution would mimic this non-null scenario with $\widehat{\theta}^{(b)}_2$ tending to be larger than $\widehat{\theta}^{(b)}_1$ in most bootstrap samples. In other words, if the ``harm'' is already committed to the observed data, how can it be fixed with bootstrap samples that reproduce the very problem that needs to be fixed? A key insight came from Guo and He \cite{GuoHe2021} who proposed to fix this issue using modified bootstrap estimates $\widehat{\theta}^{*(b)}_{k}=\widehat{\theta}^{(b)}_{k}+d_k$ computed by adding an adjustment term $d_k=(1-n^{r-0.5})(\widehat{\theta}_{\max}-\widehat{\theta}_k)$, where $r$ is a tuning parameter. In our toy example, the adjustment for $\widehat{\theta}^{(b)}_2$ will be 0, and the adjustment for $\widehat{\theta}^{(b)}_1$ will be positive, effectively pooling the bootstrap estimates of treatment effect towards the homogeneity null. Therefore the modified bootstrap estimate $\widehat{\theta}^{*(b)}_{\max}= max_{k \in \{1,..,K\}} \widehat{\theta}^{*(b)}_{k}$ for each bootstrap sample can be used to mimic the bias in $\widehat{\theta}_{\max}$ around the true $\theta_{\max}$. They proposed a bias-adjusted estimator for $\theta_{\max}$: 
\begin{equation}
    \widehat{\theta}^*_{\max}= \widehat{\theta}_{\max}-\Sigma_{b=1}^B (\widehat{\theta}^{*(b)}_{\max} - \widehat{\theta}_{\max} )/B.
\end{equation}
 
Guo and He (2020) proved that, under mild assumptions, the procedure is valid for any $r \in (0,0.5)$ and showed how to select $r$ by cross-validation. Also, based on the bootstrap distribution of $\widehat{\theta}^{*(b)}_{\max}$, they constructed a lower confidence limit for $\theta_{\max}$ (see also Fuentes, Casella, and Wells \cite{Fuentes2018}). An important consideration, when performing inferences on maximal subgroup effects, is the ability to distinguish between the following two estimands: the true maximal subgroup effect, $\theta_{\max}$, and the true effect in the selected subgroup, $\theta_{s}$. As shown in Guo and He (2020), their procedure covers both cases assuming that the selection is consistent, in the sense that $\theta_{s}$ converges to $\theta_{\max}$ in probability.

These ideas were extended to observational studies\cite{Guo2021} in the presence of confounding and high-dimensional covariates. Here the challenge is that the number of prognostic effects (potential confounders) and candidate subgroup effects can be large, possibly exceeding the sample size, thus requiring using sparse estimators of model parameters such as based on $L_1$ penalty.  
Therefore, in addition to correcting for over-optimism in the effect estimates in the maximal subgroup, one has to also correct for regularization bias when selecting prognostic and predictive effects in high-dimensional data. Guo et al \cite{Guo2021} addressed the regularization bias by using the de-sparsified (or de-biased) lasso\cite{Zhang2014,vandeGeer2014} when the number of subgroup effects is large and increasing with the sample size, while building on the repeated data splitting\cite{Wang2020} when the number of subgroup effects is fixed. The methods have been implemented in the R package \texttt{debiased.subgroup}.  

\begin{figure}[htb!] 
\centering
\includegraphics[scale=0.9, trim=100 120 50 140,clip]{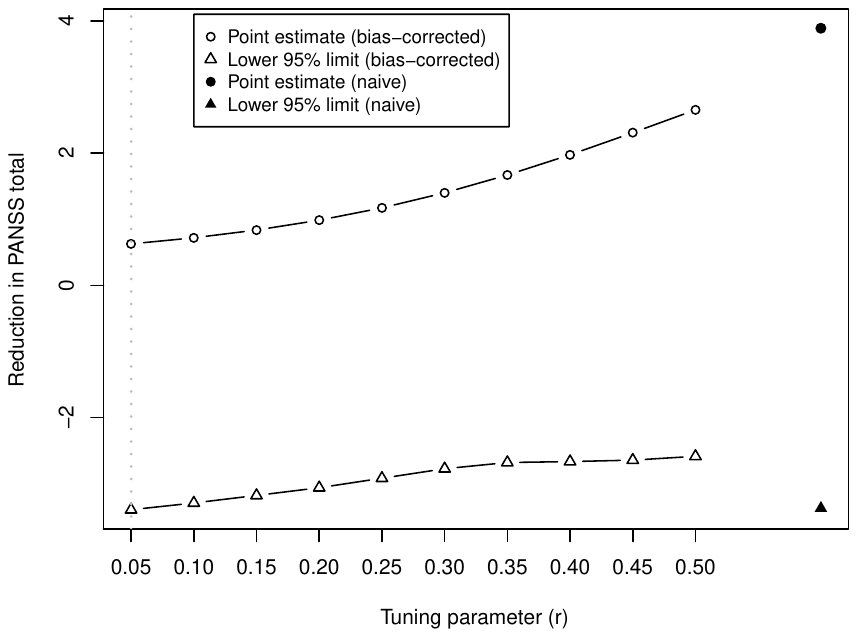} 
\caption{Application of the bias correction method of Guo and He\cite{GuoHe2021} to selecting the best subgroup from 20 candidate subgroups in the clinical  trial example. The filled circle and triangle indicate the na\"ive point estimates of treatment difference and the lower 95\% confidence limit, respectively for the best identified subgroup (gender='female'). The open circles and triangles indicate the bias-corrected estimates for tuning parameter ($r$) varying from 0.05 to 0.5. The value of 0.5 corresponds to the bootstrap estimate without modification. The cross-validation suggested selecting $r=0.05$.}
\label{fig.case.debias}
\end{figure}

We applied the method of Guo and He\cite{GuoHe2021} to our case study with 20 candidate subgroups representing the $gender$=Male, $race$=White, $race$=Black, $race$=Other, patients with $cgis=4$ and subsets of patients with continuous covariates $age$, $diagyears$, $pansspos$, $panssneg$, $panssgen$, $pansstotal$ less or equal than their median values; and their complements. A na\"ive approach based on the selection of the patient subgroup with the largest unadjusted treatment difference in $panss42.change$ produced $\{gender=Female\}$ with the treatment difference (defined here as the negative of the change in the PANSS total score from baseline to Day 42, to make larger positive values preferable) of $3.37$ that was not significant.  To illustrate the degree of bias adjustment in our case study, Figure~\ref{fig.case.debias} displays the na\"ive point estimate and the lower 95\% confidence limit for the best subgroup (open circles and triangles, respectively) along with the bias-adjusted estimates and associated lower 95\% confidence limits (filled circles and triangles, respectively) obtained by the modified bootstrap procedure with the tuning parameter $r$ varying from 0.05 to 0.5. The value of 0.5 corresponds to a usual bootstrap without modification, as the multiplier $(1-n^{r-0.5})$ for adjustment terms becomes zero. The cross-validation procedure suggested the value of $r=0.05$ (indicated by the vertical gray dashed line), which results in a considerable shrinkage of the adjusted treatment effect to 0.

We conclude this section by noting that the connection between HTE evaluation and multiple hypothesis testing is most obvious in the context of data-driven subgroup selection/identification. The multiplicity is often induced by multiple stages of subgroup selection. First, multiplicity is inherent in the inferences on a pre-defined family of $K$ possibly overlapping subgroups, where we may be concerned either with the simultaneous inference (e.g., providing the familywise Type I error control or the false discovery rate control on all subgroups),\cite{Wei2022, Wei2021, Luo2023} with selecting a set of subgroups satisfying certain constraints on the size of treatment effect (on the group average or, preferably, for each subject),\cite{Reeve2023, Muller2023} or with the inference for the  best subpopulation.\cite{GuoHe2021, Guo2021} Additional multiplicity challenges arise in the setting when these $K$ candidates are not pre-defined but selected in a data-driven fashion \textit{using the same data} from a larger pool of potential candidates defined by the search space (e.g., by incorporating subgroups as covariates within a penalized regression\cite{Guo2021,Zhao2023} or when subgroups are the terminal nodes of a tree-based search\cite{Lipkovich2011, Loh2015, Hsu2015, Seibold2016, Karmakar2018, Loh2019a}). The demarcation between these two settings (subgroups pre-defined or discovered on an independent data set versus subgroups discovered in a data-driven manner on the same data set) is often blurred in the subgroup identification literature, yet it is critical to account for all sources of multiplicity when conducting inferences on subgroup effects.

\section{Simulation experiments} \label{simdata.description}
It is common to utilize simulation approaches to assess the performance of selected methods for subgroup identification and HTE evaluation, e.g., a number of simulation studies for popular methods in various settings have been reported recently.\cite{Wendling2017,Alemayehu2018,Loh2019, Huber2019,Liu2019,Talisa2021,Sun2022} The results are mixed in the sense that there is no commonly established set of performance criteria and no method has been found to be uniformly superior to the others based on all criteria. To give an example, a study by Loh et al \cite{Loh2019} comparing 13 methods on 7 criteria using simulated data with a binary outcome concluded that most methods fared poorly on at least one criterion.

\subsection{Simulation set-up}

To further illustrate key methods for assessment of treatment effect heterogeneity and subgroup discovery presented earlier in this paper, we conducted a small simulation study with four scenarios reflecting different challenges encountered in HTE assessments (the scenarios are referred to as $S_1$, $S_2$, $S_3$, and $S_4$). Some aspects of our data generation models may appear too extreme and unrealistic. This is done to demonstrate that, if certain key assumptions are violated, it would no longer be feasible to estimate CATE even with very sophisticated methods.

\begin{figure}[htb!] 
\centering
\includegraphics[scale=0.5]{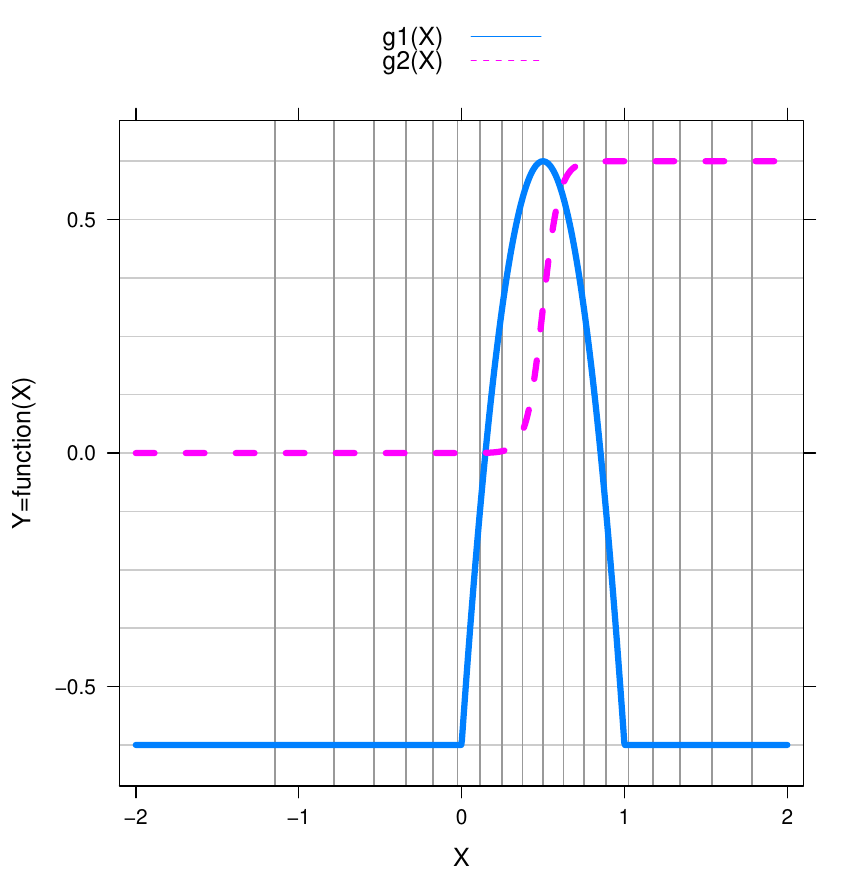} 
\includegraphics[scale=0.47]{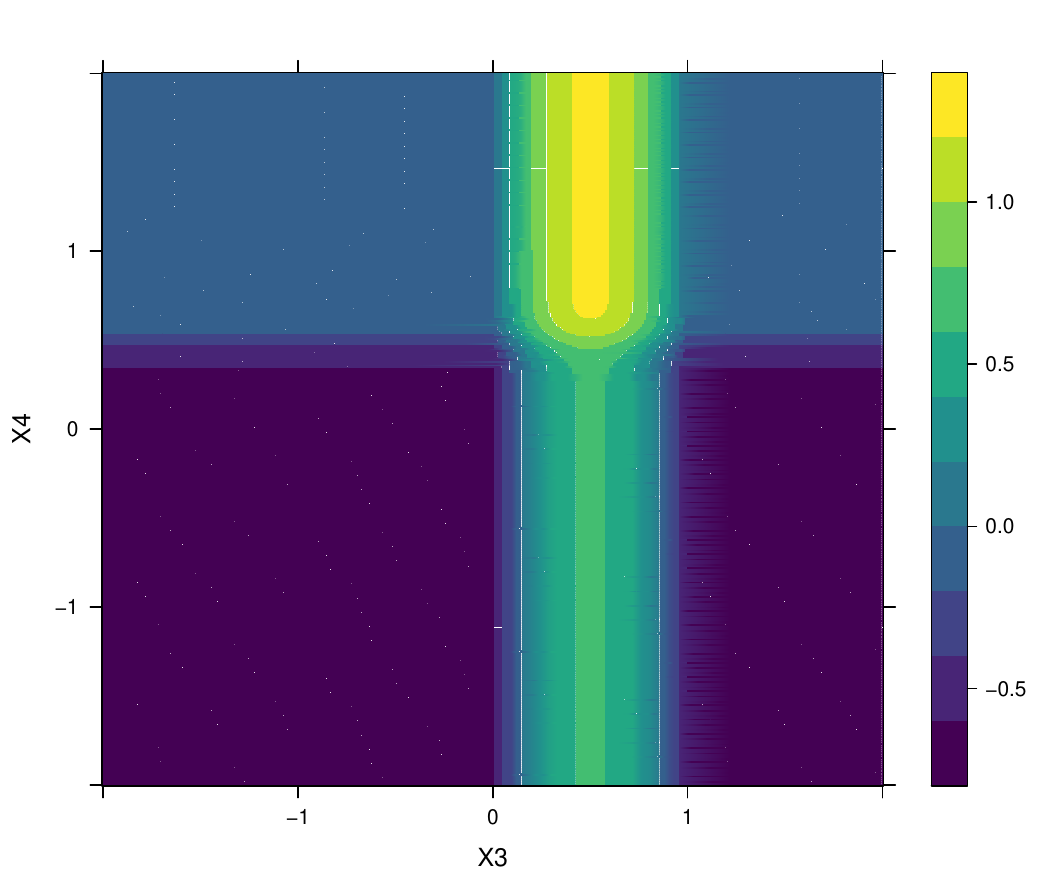} 	
\caption{The predictive part of the simulation model. Left: the two components $g_1(x)$ and $g_2(x)$ plotted marginally as a function of $x$, with vertical lines corresponding to $p$-percentiles of the underlying biomarkers $X_3$ and $x_4$ (with $p=0.05, 0.10, ...,0.95$). Right: True treatment effect $\Delta(X)$ as the joint function of $X_3$ and $X_4$, displaying areas with enhanced effect (lighter color).}
\label{fig.sim.te}
\end{figure}

Scenarios $S_1$ and $S_2$ represent data from an RCT, and $S_3$ and $S_4$ are observational data with the treatment choice generated using propensity models. The outcome ($Y$) is continuous, with larger values indicating treatment benefits. All data sets contain $N=1000$ observations. In RCT data sets, there are three times more subjects randomized to the treatment than to the control. In the observational data sets, there are three times more control subjects than treated.  In $S_1$ the outcome model is based on 4 covariates with non-linear predictive effects as will be detailed in the following.
\begin{equation}
 Y=100-(X_1+5X_2)+T \times (g_1(X_3)+g_2(X_4))+\epsilon,
\end{equation}
where $\epsilon \sim N(0,1)$, $T\in \{0,1\}$ is the treatment indicator with $T=1$ representing the treatment of interest and $T=0$ representing the standard of care; covariates $X_1, X_3, X_4$ are continuous generated from $N(0.5, 1)$ and $X_2$ is categorical with 3 levels generated with equal probabilities $p=1/3$. Therefore, $X_1, X_2$ form the prognostic component and $X_3, X_4$ the predictive part, with CATE given as $\Delta(x)=g_1(x_3)+g_2(x_4)$. 
Non-linearity is induced in CATE via a non-monotone $g_1(\cdot)$    
\begin{equation}\label{eq.g1}
  g_1(x) = 
     \begin{cases}
      a-b\cdot 0.25 & \text{if} \;\; x < 0\\
      a-b(x-0.5)^2 & \text{if} \;\; 0 \le x \le 1 \\
      a-b\cdot 0.25 & \text{if} \;\; x > 1\\
    \end{cases},       
\end{equation}
and a monotone $g_2(\cdot)$
\begin{equation}\label{eq.g1}
  g_2(x) = 
     \begin{cases}
      0 & \text{if} \;\; x < 0\\
      \frac{c}{1+\text{exp}(-d(x-0.5))} & \text{if} \;\; 0 \le x \le 1 \\
      c & \text{if} \;\; x > 1\\
    \end{cases}.       
\end{equation}
The constants $a=0.625,b=5, c=0.625, d=20$ are calibrated so as to make the overall treatment effect slightly  positive, $E[\Delta(X)]=0.0119$, the true signature $S_{true}=\{\Delta(X)>0\}$ has the proportion of subjects  $E[I(X \in S_{true})]=0.330$ and the true mean treatment effect in $S_{true}$ is $E[\Delta(X)|X \in S_{true}]=0.665$ (all estimated via a test data set of size $n=10^6$). Figure~\ref{fig.sim.te} shows the non-linear functions $g_1(x)$ and $g_2(x)$ contributing to the the CATE on the left, and the distribution of  $\Delta(X)$ as a function of $X_3, X_4$  on the right.  

Therefore, the challenges of modeling individual treatment effect in $S_1$ are: (1) non-linear and non-monotonic treatment effects resulting in non-smooth distribution of CATE; (2) presence of a subgroup of patients with  $\Delta(X)=0$, i.e., for whom the two treatment options are equally attractive (for $X_3 \ge 1$ and $X_4 \ge 1$); (3) the non-linear components are only contributing to outcome model in the experimental treatment arm ($T=1$) whereas the outcome  model in the control arm ($T=0$) is a simple linear function; (4) unequal sample sizes in treatment groups. The last item seems rather an advantage since the larger sample size is in the arm where the true outcome model is more complex. We  reverse this when simulating observational sets in $S3$, $S4$. 

Data under $S_2$ are simulated similarly to $S_1$, except the prognostic part is more complex: $E(Y)=100-(X_1 +5X_2) + 2(X_5+X_6+X_7+X_8+X_9)+T(g_1(X_3)+g_2(X_4))$, where additional 5 covariates 
are simulated from $N(0.5,1)$. 
Let $k_1(X)=-(X_1+5X_2)+ 2(X_5+X_6+X_7+X_8+X_9)$ and $k_2(X)=g_1(X_3)+g_2(X_4)$ so that $E(Y)=100+k_1(X)+k_2(X) \times T$. 

In addition to scenarios $S_1$ and $S_2$  representing RCTs, we created scenarios $S_3$ and $S_4$ mimicking observational studies by introducing propensity functions. While in RCTs we allocated more patients to the experimental treatment arm, in scenarios for  observational studies we assumed control groups are larger.
Scenario $S_3$ is similar to $S_2$, but treatment assignment mechanism now is driven by prognostic components of the outcome model:
\begin{equation}\label{eq.ps}
 \text{logit} \left( \text{Pr}(T=1|X) \right)=\alpha_1+\beta_1 \times k_1(X).
\end{equation} 
Coefficients $\alpha_1=-2.4$ and $\beta_1=-0.2$ were calibrated to ensure a 1:3 ratio for treated to control subjects and sufficient overlap in the distribution of propensity across the treatment and control arms. 

Since the logit of propensity is $\alpha_1 +\beta_1 \times k_1(X)$, setting $\beta_1 < 0$ results in patients with lower prognostic effect would be assigned treatment $T=1$ with higher probability. Therefore data $S_3$ mimics a physician who assigns patients to the active treatment if their prognosis under control (e.g., standard of care) is poor. 

\begin{figure*}[!ht] 
\centering
\includegraphics[scale=0.6]{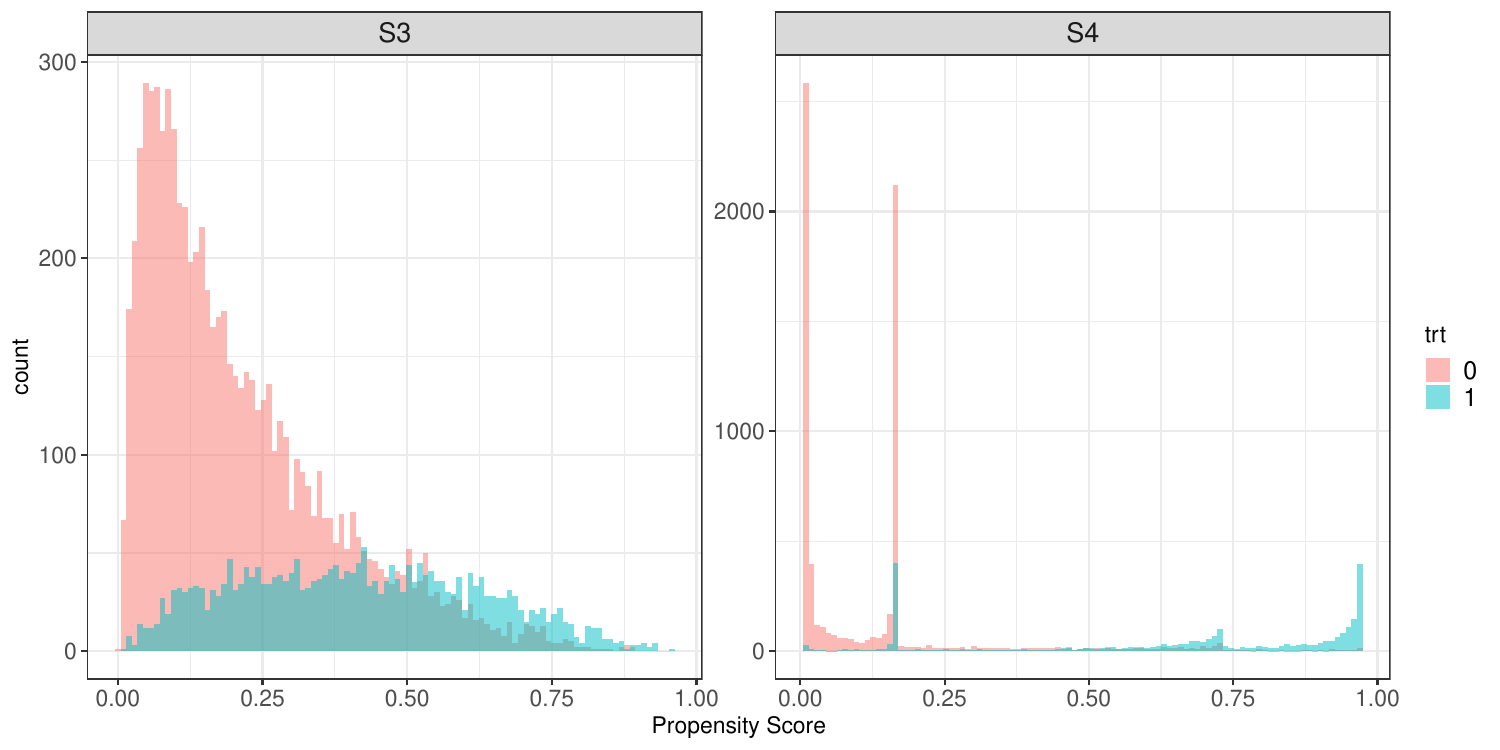} 
\caption{Overlap plot for cases $S_3$ and $S_4$, showing the distribution of the true underlying propensities by treatment arms for a single data set with $n=10,000$.}
  \label{fig.Overlap}
\end{figure*}

In the data set $S_4$, the treatment model is driven by the predictive component of the outcome model by setting the propensity
\begin{equation}\label{eq.ps}
\text{logit} \left( \text{Pr}(T=1|X) \right)=\alpha_2+\beta_2 \times k_2(X),
\end{equation}
 with $\alpha_2=-1.64$ and $\beta_2=4.2$ (again rendered 1-to-3 ratio regarding treated vs untreated). For scenario $S_4$, setting $\beta_2 > 0$ mimics a physician who somehow learned the predictive effect and uses a treatment assignment rule that makes a patients with $\Delta(X)=k_2(X)>0$ more likely to be assigned to $T=1$ while ignoring the prognostic score. Figure ~\ref{fig.Overlap} displays the distribution of propensity for a single large simulated set by treatment arm for scenarios $S_3$ on the left and $S_4$ on the right. While in $S_3$ the overlap in propensity scores across treatment arms is reasonable, for $S_4$ it is very poor suggesting that estimating true CATE may be very challenging.
 

The challenges of $S_3$ and $S_4$ are that (1) treatment is not random (2) there are less patients in the treatment arm where the outcome model model is more complex.   

To make analysis more challenging  we added in each data set additional 10 noise covariates independently drawn from the standard normal distribution.

\subsection{Performance metrics}\label{performance.metrix.sim}
The following metrics were used to summarize the performance of various methods for estimating CATE on simulated data. Our selection of metrics reflects our main focus on identifying the subgroup of patients with positive CATE therefore we did not look at the bias and MSE of CATE as such.
\begin{itemize}
 \item A general agreement between the true and estimated CATE was evaluated by their Pearson correlation; $cor(\widehat{\Delta}(X),\Delta(X))$
 \item The agreement between the true subgroup $S_{true}(X)=\{ x: \Delta(X)>0 \}$ and estimated subgroup $\widehat{S}(X)=\{x: \widehat{\Delta}(X)>0\}$ was computed using Jaccard similarity coefficient; $agree(\widehat{S},S_{true})=\frac{n \left( \widehat{S} \cap S_{true} \right)}{n \left( \widehat{S} \cup S_{true} \right)}$
 \item True average treatment effect on estimated subgroup; $ATE(\widehat{S})=E_X \{ \Delta(X)| \widehat{\Delta}(X)>0\}$. This expectation was approximated using independently simulated large test data ($N=10,000$). Specifically, the subgroup rule identified from the training data was evaluated on test data, $\widehat{S}(X_{test})= \{x:\widehat{\Delta}_{train}(X_{test})>0 \}$, and the true treatment effect averaged on selected test subjects.  
 \item Estimated average treatment effect on estimated subgroup, based on test data; $\widehat{ATE}(\widehat{S})=E_X \{ \widehat{\Delta}(X)| \widehat{\Delta}(X)>0\}$ 
 \item  $Bias\{ATE(\widehat{S})\}=\widehat{ATE}(\widehat{S})-ATE(\widehat{S})$ 
 \item Inherent variability in the estimator $\widehat{ATE}(\widehat{S})$ was evaluated as its simulation standard deviation $SD\{ \widehat{ATE}(\widehat{S})\}$ 
 \item Subgroup’s utility index computed as the true average treatment effect in identified subgroup per subject in the overall population, $\eta=ATE(\widehat{S}) \times \frac{n(\widehat{S})}{n}$. Note that for our simulation scenarios this measure is equivalent to the difference between the value of the estimated treatment assignment rule $\widehat{D}(X)=I(\widehat{\Delta}(X)>0)$ and that of a fixed regimen that assigns everyone to the control: $\eta=V(\widehat{D}(X))-V(0)$. 
\end{itemize}

\subsection{Applying selected methods to simulated data}\label{apply.methods.sim}
In this section we report the performance measures listed in Section~\ref{performance.metrix.sim} for selected methods of estimating CATE applied to $K=100$ data sets generated from the four simulation models described in section \ref{simdata.description}.  The selected methods are: T-, S-, X-, R-learning, Causal Forest, Bayesian Causal Forest, and Modified Loss Methods (Weighting and A-learning) with xgboost as the main learner (penalized regression with elastic net was used for fitting propensity and augmentation functions). Please refer to Section~\ref{seq.apply.casestudy} and supplementary materials for details of R software used.

For scenarios $S_1$ and $S_2$ (RCTs), the known propensities (based on randomization rates) were uses in MLM, CF, and BCF methods (like in the schizophrenia case study), and for $S_3$ and $S_4$ propensity scores where estimated for the MLM and CF methods,  using default settings available in the underlying software packages. For Bayesian Causal Forest\cite{HahnBART} we used $200$ and $50$ trees for the prognostic and predictive BART priors, respectively, as recommended by the authors.\cite{HahnBART}

\begin{figure*}[!ht] 
\centering
\includegraphics[scale=0.65]{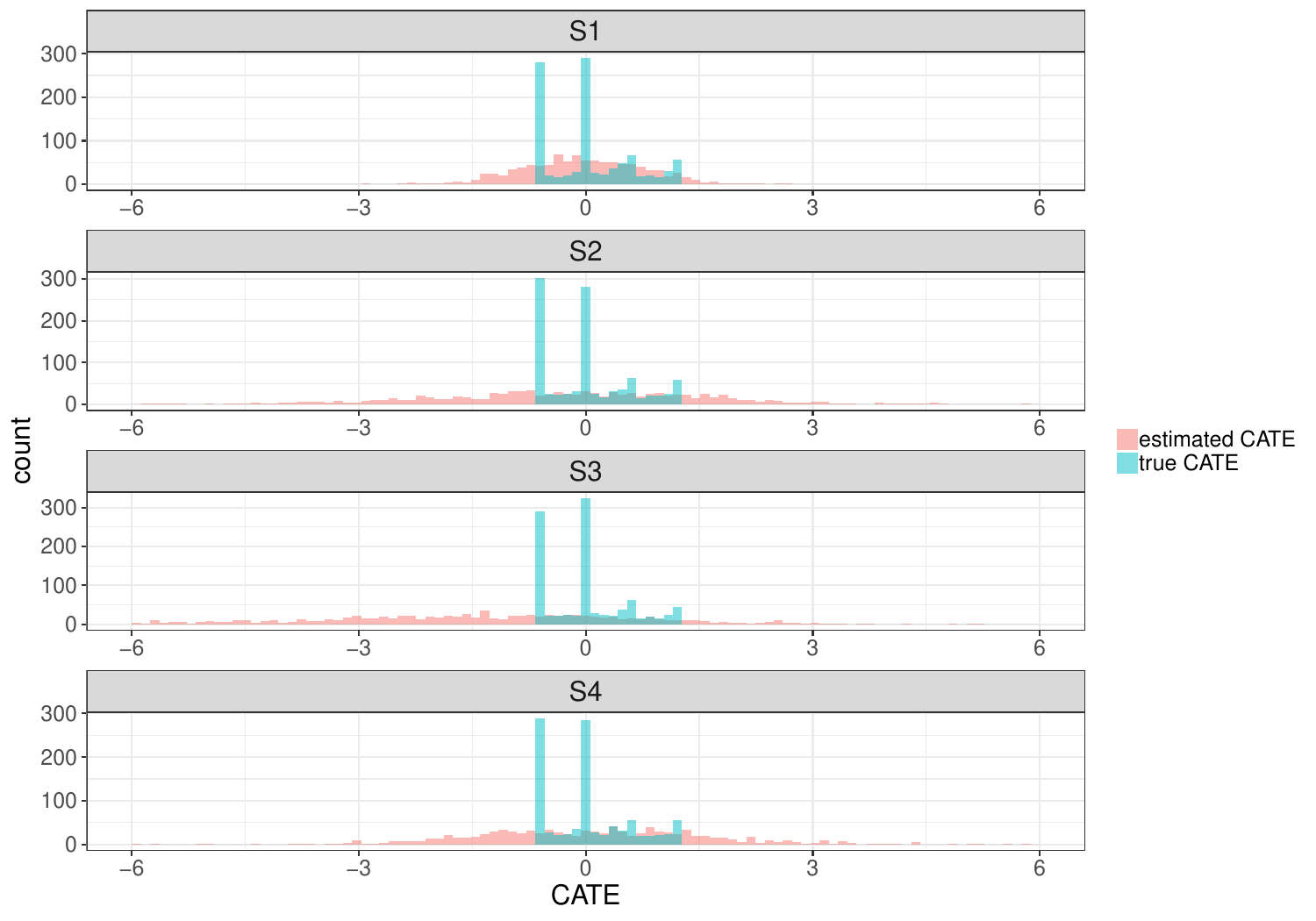} 
  \caption{Distribution of the true and estimated CATE in a single simulation iteration (sample size $n=1000$) for scenarios $S_1,S_2,S_3,S_4$; based on a T-learner using xgboost as the base learner. The x-axis was truncated at $x=-6$ (leaving out 33 observations).}\label{fig.HistCATES1}
  \end{figure*}

Figure~\ref{fig.HistCATES1} illustrates estimating CATEs displaying the distribution of estimated $\widehat{\Delta}(X_i)$ vs true CATEs $\Delta(X_i)$ for a single data set (with 1000 subjects) for T-learning across all scenarios. We can see that the T-learner (fitted with xgboost) is not able to capture the non-smoothness of the true CATE (e.g., the spikes). While producing a reasonably accurate representation of the distribution of positive CATE for $S_1$, it has excessive mass of positive CATEs for other scenarios suggesting overoptimism in identified subgroups. 

\begin{figure*}[!ht] 
\centering
\includegraphics[scale=0.8]{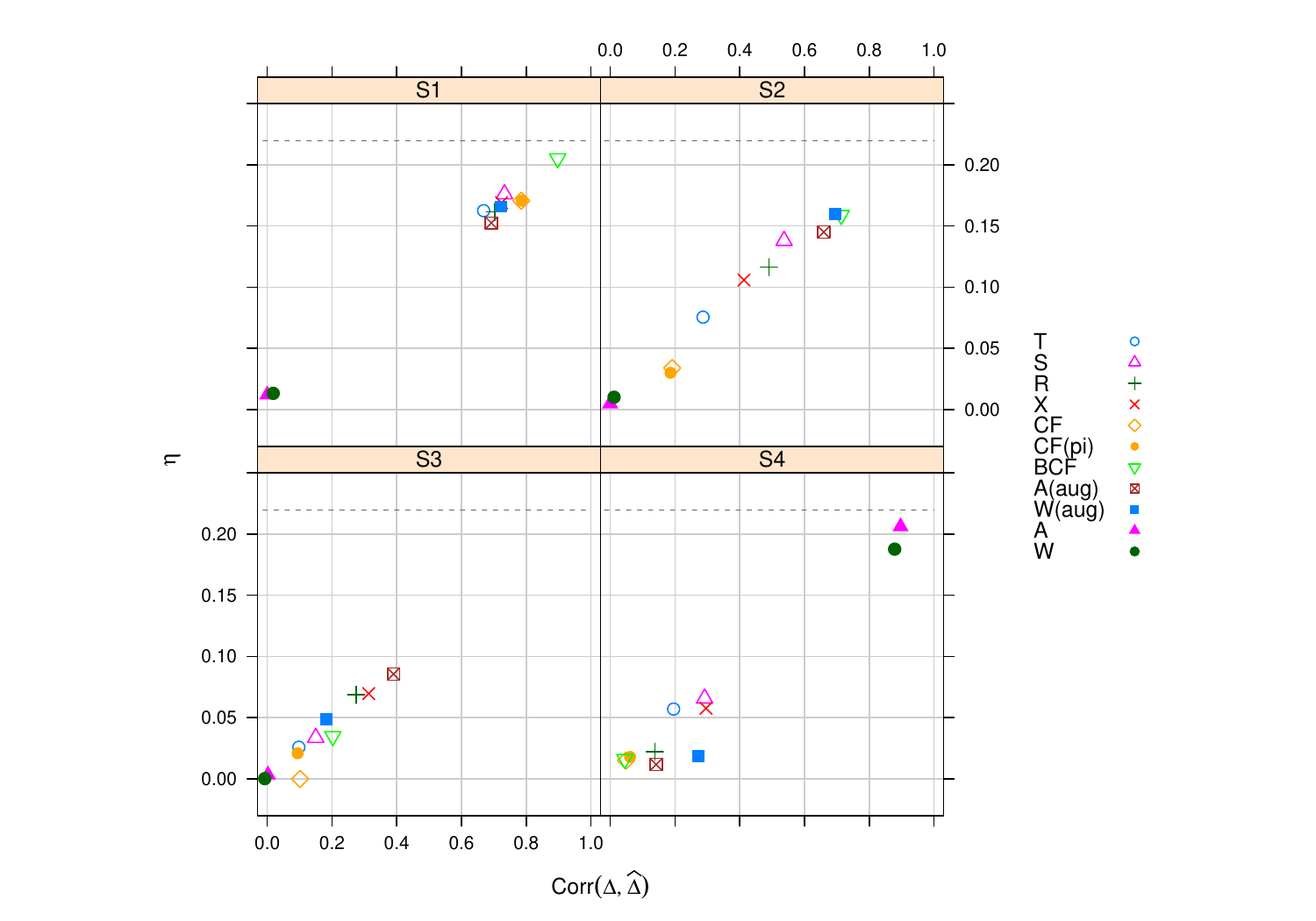} 
\caption{Benchmarking over $100$ iterations for each scenario; plotting the subgroup utility index ($\eta$ metric) against the Pearson correlation between estimated and true CATE. Methods producing high values on both metrics (which are highly related) indicate good ability to recover underlying CATE as well as the subgroup of patients truly benefitting from the active treatment. The horizontal dotted line indicates the theoretically largest attainable value of the metric $\eta=0.22$.}
  \label{fig.CORRETA100}
\end{figure*}

\begin{figure*}[!ht] 
\centering
\includegraphics[scale=0.8]{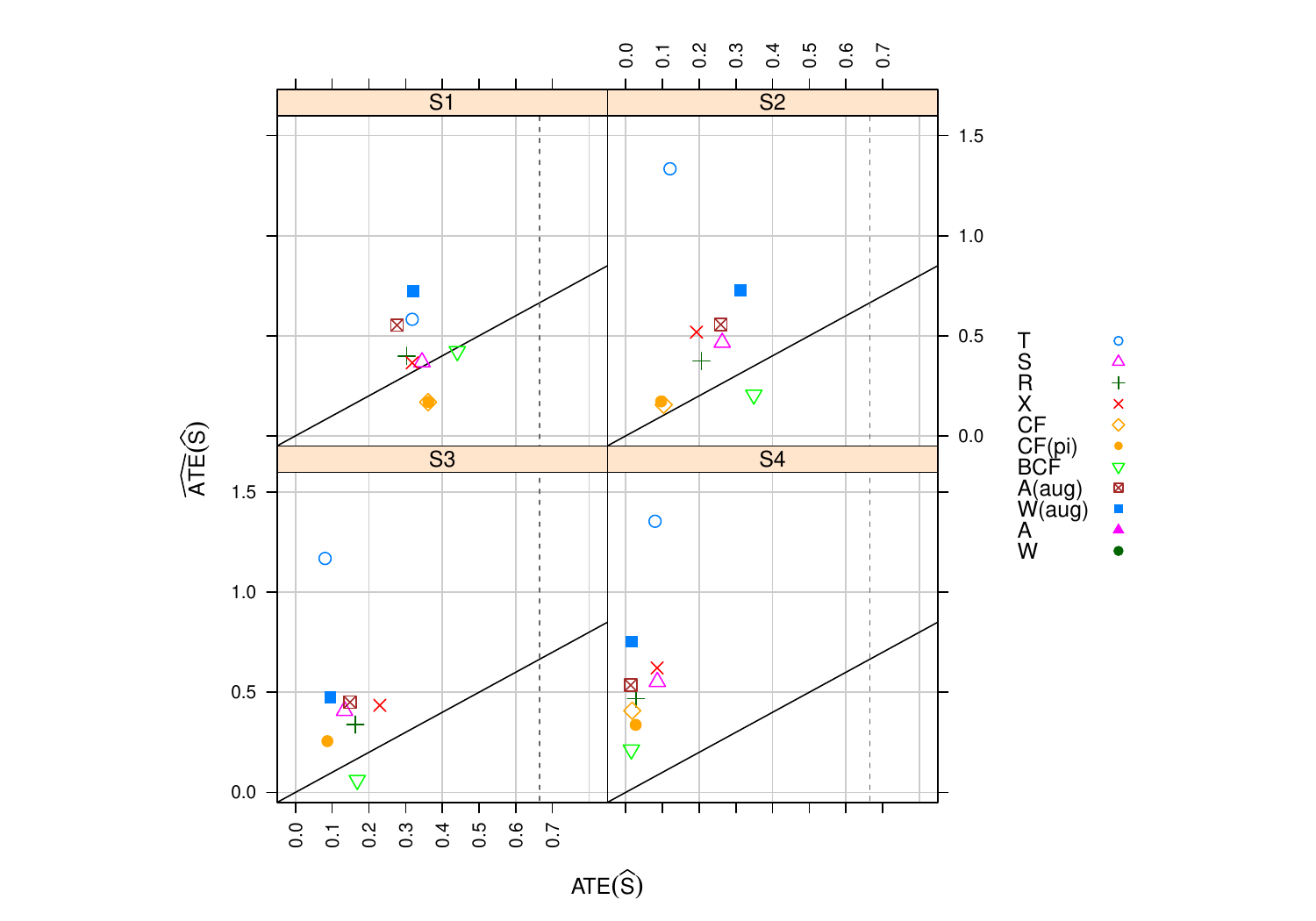} 
  \caption{Average treatment effect (ATE) in identified subgroups by different methods across 4 scenarios ($100$ simulations); Y-axis displays the estimated ATE in the estimated subgroup $\widehat{S}(X) =\{\widehat{\Delta}(X)>0\}$ vs the true treatment effect in $\widehat{S}$ (X-axis). Vertical dotted line marks expected average effect in the true subgroup  $S(X) =\{\Delta(X)>0\}$. Notably CF gave quite spurious results in scenario $S_3$ with every estimate below zero by a margin (hence it is off chart). The plot is truncated and does not display grossly outlying results for the none-augmented A-learning and W-learning across all scenarios.}  
  \label{fig.CATE_Shat100}
  \end{figure*}

\begin{figure*}[!ht] 
\centering
\includegraphics[scale=0.8]{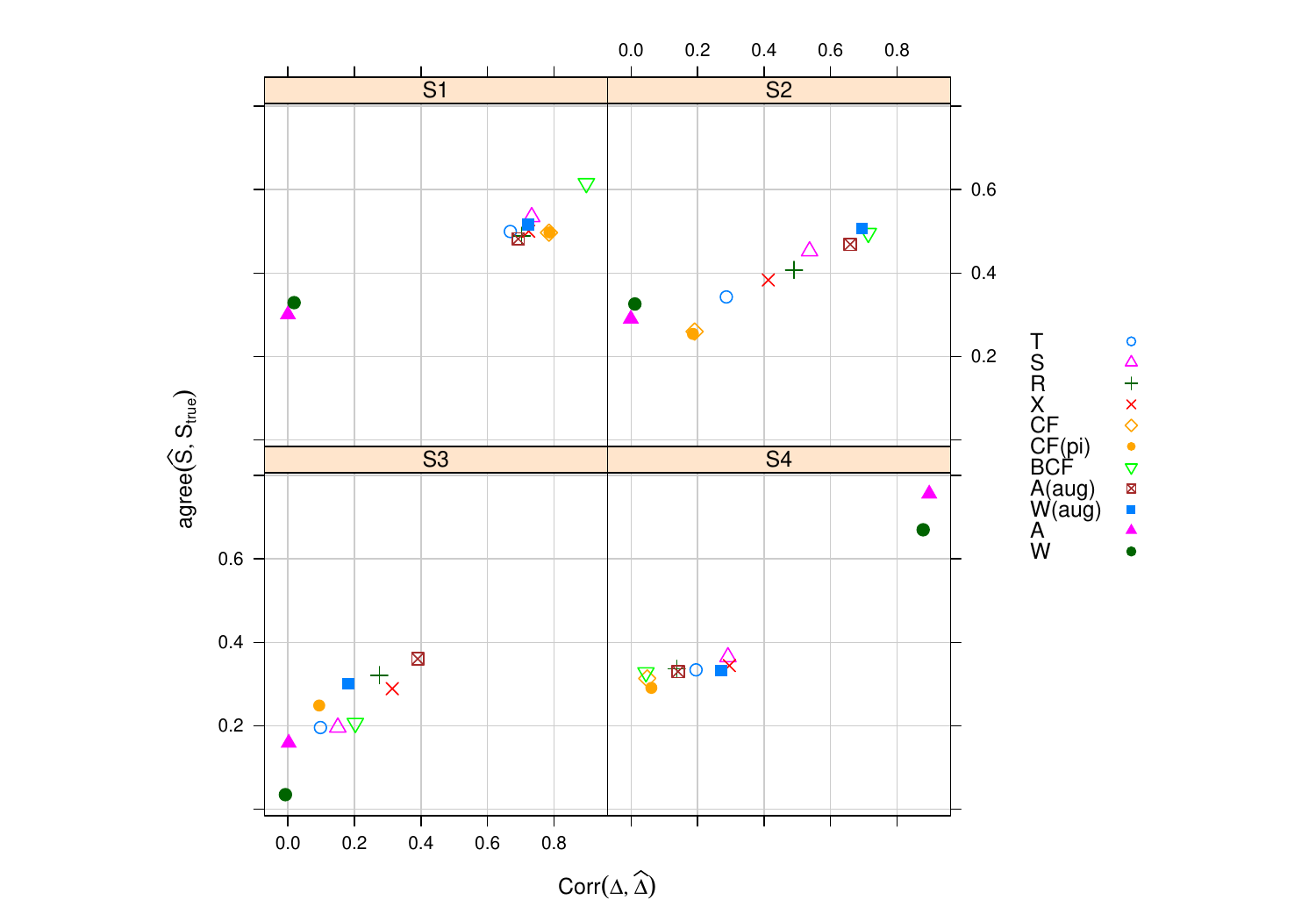} 
  \caption{The agreement between the true $S(X) =\{\Delta(X)>0\}$ and identified subgroup measured by the Jaccard coefficient vs. Pearson correlation between the true $\Delta(X)$ and estimated CATE.}  
  \label{fig.CATE_Jaccard100}
  \end{figure*}

Table~\ref{tab.valuesS1S2S3S4} summarizes results that are visualized in Figures \ref{fig.CORRETA100}, \ref{fig.CATE_Shat100}, and \ref{fig.CATE_Jaccard100}. 

Figure \ref{fig.CORRETA100} plots the subgroup utility index $\eta$ against correlation between the estimated and true CATE $corr(\Delta(X), \widehat{\Delta}(X))$. Vertical dotted lines indicate the maximal attainable $\eta=0.22$ (for the optimal regimen based on $\Delta(X)>0$). For $S_1$ all methods performed reasonably well (BCF being the best), except non-augmented versions of weighting and A-learning. Scenario $S_2$ illustrates challenges of identifying CATE caused by adding 5 prognostic variables. As expected, the performance of T-learning was reduced considerably. Surprisingly, CF was most affected by adding prognostic variables, although unlike T-learning it is furnished with heavy machinery for dealing with nuisance parameters. The performance of all methods was further worsened for non-RCT scenario $S_3$. A relatively better performance is observed for A-learning with augmentation, R-learning, and X-learning. For $S_4$, the performance was even worse except, surprisingly, for non-augmented MLM. This can be explained by the special feature of this scenario, i.e.,  the  treatment assignment is  driven by the true CATE and, as a result, the observed treatment indicator variable is highly correlated with the true CATE. In this situation, the MLM methods essentially model a response variable that inherently integrates the true CATE (recall that it is formed by multiplying the outcome by the treatment indicator). This special setting then gives an advantage to the A- and W-learning methods, which is otherwise not present in general.

Figure~\ref{fig.CATE_Shat100} plots the estimated ATE for the estimated best subgroup vs the true ATE for the estimated subgroup. The 45 degree line helps to see whether the estimated ATE is biased upward or downward. In general the plots tell the same story as Figure~\ref{fig.CORRETA100}, indicating (on the horizontal axis) worsened performance of estimated subgroup vs the true subgroup in terms of the  true ATE captured within the subgroup. The vertical dotted line indicates the ATE for the true subgroup, $E(\Delta(X)|\Delta(X)>0)=0.665$. 

We note that non-augmented versions of MLM (A-learning and Weighting) renders estimates of CATE with very high variability, therefore these are excluded from the graph (the results are provided  in Table~\ref{tab.valuesS1S2S3S4}). This suggests that augmentation is quite critical for the MLM approach, despite our initial thinking that direct modeling of ITEs (without modeling of prognostic effects) may still produce reasonable estimates. Interestingly, for the base scenario $S_1$ when all methods produced reasonable estimate of the subgroup, T-learning and Modified Loss methods overestimated the ATE in identified subgroup, whereas CF underestimated the true treatment effect. Interestingly for scenario $S_3$ CF method produced all CATEs < 0 therefore failing to identify any subgroup. This is largely die to its poor ability to model propensity score which was surprising as other methods (such as xgboost, results not shown) produced much better estimates. Fitting CF with improved propensities estimated by xgboost or true propensities (known to simulator) resulted in improved estimates of subgroup by CF (these are given in the row labeled $CF(\pi)$ in Table~\ref{tab.valuesS1S2S3S4}). 

From Figure~~\ref{fig.CATE_Shat100} we could see that the ATE in the identified subgroups is in general substantially lower than that in the true subgroup, while the estimated ATE is in general overestimating the true ATE captured in the identified subgroups. Figure~\ref{fig.CATE_Jaccard100} allows us to see how closely the identified subgroups represent the true subgroup using Jaccard's similarity index as a simple measure of proximity (displayed on the vertical axis). It shows that for the RCT scenarios $S_1,S_2$, subgroups identified by BCF are fairly close to the truth despite substantial deterioration in terms of the ATE. Also, for the RCT scenarios, the non-augmented versions of A- and W-learning produced reasonably accurate subgroups, even though their estimated CATE were poorly correlated with the true CATE (displayed on the horizontal axis). Apparently, despite their high variability, they captured the sign of the CATE quite well. 

\section{Summary and discussion}\label{summary.discussion}


In this tutorial we provided an overview of methods for HTE assessment focusing on most recent innovations that appeared since an earlier review.\cite{Lipkovich2017} Following our previous typology, we considered methods for estimating CATE (both directly and indirectly, methods for estimating ITR, and direct methods for subgroup identification.  

We note a shift from ad-hoc ``subgroup chasing'' methods towards principled methods of personalized/precision medicine utilizing ideas from causal inference, machine learning and multiple testing emerged in last 10 years producing a vast number of diverse approaches. 

Methods for HTE naturally incorporate multiple steps of model fitting, each step often implemented via machine learning. For naïve multistage methods (based on fitting the response surface $m(x,t)$), the regularization bias can be large, as each step is optimized for prediction rather than for the final estimation target. While methods based on ideas of modified outcomes/covariate/loss that estimate CATE directly, thus obviating the need to fitting prognostic effects, may appear attractive, their performance may be very poor without augmenting the loss function with prognostic effects. There is growing understanding that correctly modeling individualized treatment effects requires pre-fitting multiple nuisance functions and there is a great deal of developments regarding how to properly regularize these to optimize the estimation of the target HTE parameters. Substantial efficiency can be gained by using doubly-robust methods, such as utilizing augmented inverse propensity weighted scores, even in the context of RCT where propensities are known.\cite{AtheyWager2021,Kennedy2020} When the data are high-dimensional, as is often the case in HTE applications, ideas from doubly/debiased machine learning\cite{chernozhukov2018} seems promising  ensuring faster convergence rates to the target parameters by cleverly combining the outcome and propensity functions. However, there is also increasing understanding that in the context of high dimensional data accurate assessment of all aspects of CATE distribution may be unattainable and researchers offer focusing on useful summaries of CATE instead.\cite{Chernozhukov2020, ImaiLi2022}

There is also understanding of usefulness of holistic approach to modeling HTE utilizing ideas from model averaging, e.g., via Super-learners that may help overcame limitations of specific approaches. As we illustrated with our small simulation study, even cleverly designed methods (such as Causal forests) may fail for specific scenarios because one of their sub-tasks (e.g propensity modeling) may perform poorly and other methods that would deal with this sub-task much batter can be leveraged.  

Inference for HTE is another challenging area. One the one hand, there is a need to produce reliable ``gate-keeping'' strategies that would allow a researcher/sponsor/regulator to rule out HTE by failing to reject  the overall hypothesis of the presence of heterogeneity in treatment effects. Unfortunately, the power of such procedures seems low\cite{ImaiLi2022, Sun2022} making such strategies unattainable. Another important task is inference on ITR contrasting proposed ITR with the appropriate benchmark (such as ``current practice'' in the real world setting). The operating characteristics of existing approaches is not well documented and understood and need improvement. In many cases na\"ive inference for target causal parameters ignoring estimation of nuisance parameters at previous stages leads to incorrect inference, however, there are clever and principled ways of estimating causal parameters while ``ignoring'' (or being immune to) uncertainty in fitting nuisance functions, e.g., using cross-fitting estimates of efficient influence functions/curves for a target estimand such as in direct estimation of ITR by the \textit{value search}\cite{vanderLaan2015} and in some other approaches.\cite{Lee2017, Huang2020,  Guo2021} In many situations efficient root-$n$ consistent estimators of the causal parameters (including those targeting HTE) can be constructed (under feasible assumptions) using the ``targeted learning'' or ``debiased'' machine learning paradigm.\cite{Hines2021, Boileau2022, Li2023} 

There is interest in developing ITRs respecting constraints on costs, adverse events, sample size.\cite{Wang2018, AtheyWager2021, CaiCAPITAL2022} Also there is a need in interpretable personalized solutions (ITR’s) within a predefined policy class, e.g tree-structured or rectangular areas.\cite{laber2015, CaiCAPITAL2022, Doubleday2022} Multi-stage machine learning solutions with separate modeling of nuisance functions (i.e., for prognostic effects and treatment assignment mechanism) and HTE, such as in R-learning and many other proposals,\cite{vanderLaan2013,Lee2017,AtheyWager2021} allows extra flexibility in choosing different sets of covariates. For example, accounting for confounding may require a broader set of variables that can be used in fitting nuisance functions, the assessment of HTE may rely on a smaller set of predefined potential effect modifiers and use an interpetable representation such as decision trees. Also it is common that certain variables such as economic status or insurance type cannot be used as part of treatment policy while they may be accounted for as potential confounders.
  
A number of special topics were not covered by this tutorial: methods for data-driven selection of patient subpopulations for adaptive signature\cite{Freidlin2010, Zhang2016, Diao2018} and adaptive enrichment\cite{Zhang2017, Simon2017, Xu2018, Joshi2020} designs of clinical trials, subgroups identification with longitudinal data\cite{Loh2016,Cho2017} and for multiple outcomes,\cite{Zhao2024} change point methods for subgroup identification,\cite{Li2021} methods for ITR for continuous treatments, \cite{Zhang2020A} ITRs for time-to-event outcomes,\cite{bai2017, diaz2018, Liu2023,Cui2023}  transfer learning allowing researchers to generalize ITRs from experimental to real-world data,\cite{Wu2022} combining estimates of HTE across heterogeneous data sources (where the analyst is not granted access to the individual patient data) by methods of federated learning,\cite{Brantner2023} selection of important predictors of treatment effect while preserving the Type I error or false discovery rates (e.g., via knockoff variables\cite{Sechidis2021,Zimmermann2024}).
   
Another important topic not covered, largely because of the scarcity of the literature, is sensitivity analyses to potential violations of assumptions that may result in biased assessment of HTE (e.g., due to unmeasured confounders in non-randomized data, or due to omitted time-modifiers in both RCT and observational data).\cite{Rose2023}


\section*{Acknowledgements}
We thank Drs. Hengrui Cai, Xinzhou Guo, Xuanyao He, Xin Huang, Jared Huling, Michael Li, Patrick Schnell, Shu Yang; Mr. Anthony Zagar and the EFSPI special interest working group on subgroups for numerous discussions; Mr. Victor Dmitrienko for his work on the software tools. This tutorial is dedicated to the memory of Prof. Ralph B. D'Agostino Sr. who provided  encouragement and motivation for us.   






\bibliography{main_arxiv.bbl}

\newpage

\begin{table}[h!]
\centering
\begin{tabular}{rlllrl}
  \hline
 & Biomarker & Type & Values (Range)  & Mean & Proportion (\%) \\ 
  \hline
1 & age & Continuous & (22-63) & 43.75 & \\ 
  2 & gender & Categorical & Female (0)/Male (1)&  & 26.9/73.1 \\ 
  3 & race & Categorical &  Black/White/Other&  &  44.2/52.8/3.0\\ 
  4 & diagyears & Continuous & (2-48) & 18.99 &  \\ 
  5 & cgis & Ordinal  & (4-6) & 4.79 &  \\ 
  6 & pansspos & Continuous & (15-37) & 25.67 &  \\ 
  7 & panssneg & Continuous & (11-42) & 21.36 &  \\ 
  8 & panssgen & Continuous & (29-66) & 43.46 &  \\ 
   \hline
\end{tabular}
\caption{Baseline covariates in the schizophrenia case study\label{tab.bslCS3}}
\end{table}

\newpage

\begin{table}[h!]
\centering
\begin{tabular}{rrrrrrrrr}
  \hline
 & 1 & 2 & 3 & 4 & 5 & 6 & 7 & 8 \\ 
  \hline
Subgroup size & 14 & 47 & 4 & 63 & 77 & 18 & 14 & 64 \\ 
  Na\"ive estimate & -10.63 & 3.40 & 24.33 & -6.94 & 8.71 & -17.05 & -7.67 & 7.16 \\ 
  RF estimate & -4.50 & 2.19 & 5.30 & -3.94 & 2.82 & -2.94 & -1.53 & 3.07 \\ 
   \hline
\end{tabular}
\caption{Summary of terminal nodes of the CAPITAL tree from Figure~\ref{fig.CS3_CAPITAL} (left to right). The ``subgroup size'' (in the first  row) displays the number of patients in each leaf (terminal node); ``na\"ive estimate'' is the mean treatment difference based on \textit{panss42.change} in each terminal node; ``RF estimate" is the averaged ITEs from the T-learning based on the random forest predictions \eqref{eq.RF.pred}. Larger negative values indicate treatment benefit.\label{tab.CAPITAL}} 
\end{table}

\newpage

\begin{table}[h!]
\centering
\begin{tabular}{llccccccc}
  \hline
  \\
  $Scenario$ & $Method$ & $corr(\Delta,\widehat{\Delta})$ & $agree(S,\widehat{S})$ & $\widehat{ATE}(\widehat{S})$ & $ATE(\widehat{S})$ & $SE\{\widehat{ATE}(\widehat{S})\}$ & $bias\{ATE(\widehat{S})\}$ & $\eta$ \\  \hline
  S1 & T & 0.67 & 0.50 & 0.58 & 0.32 & 0.062 & 0.26 & 0.162 \\ 
  S1 & S & 0.73 & 0.53 & 0.37 & 0.34 & 0.063 & 0.02 & 0.176 \\ 
  S1 & R & 0.70 & 0.49 & 0.40 & 0.30 & 0.081 & 0.10 & 0.161 \\ 
  S1 & X & 0.72 & 0.50 & 0.36 & 0.32 & 0.072 & 0.05 & 0.169 \\ 
  S1 & CF & 0.78 & 0.50 & 0.17 & 0.36 & 0.067 & -0.19 & 0.170 \\ 
  S1 & CF($\pi$) & 0.79 & 0.50 & 0.17 & 0.36 & 0.068 & -0.19 & 0.171 \\ 
  S1 & BCF & 0.90 & 0.62 & 0.42 & 0.44 & 0.135 & -0.02 & 0.205 \\ 
  S1 & A & -0.00 & 0.30 & 3.32 & 0.01 & 4.863 & 3.31 & 0.012 \\ 
  S1 & A(aug) & 0.69 & 0.48 & 0.55 & 0.28 & 0.086 & 0.28 & 0.152 \\ 
  S1 & W & 0.02 & 0.33 & 2.85 & 0.01 & 1.531 & 2.84 & 0.013 \\ 
  S1 & W(aug) & 0.72 & 0.52 & 0.72 & 0.32 & 0.059 & 0.40 & 0.166 \\ \hline
  S2 & T & 0.29 & 0.34 & 1.33 & 0.12 & 0.081 & 1.21 & 0.076 \\ 
  S2 & S & 0.54 & 0.45 & 0.47 & 0.26 & 0.055 & 0.20 & 0.138 \\ 
  S2 & R & 0.49 & 0.41 & 0.37 & 0.21 & 0.081 & 0.17 & 0.116 \\ 
  S2 & X & 0.41 & 0.38 & 0.52 & 0.19 & 0.085 & 0.33 & 0.106 \\ 
  S2 & CF & 0.19 & 0.26 & 0.15 & 0.10 & 0.094 & 0.05 & 0.034 \\ 
  S2 & CF($\pi$) & 0.19 & 0.25 & 0.17 & 0.10 & 0.110 & 0.08 & 0.030 \\ 
  S2 & BCF & 0.71 & 0.50 & 0.20 & 0.35 & 0.122 & -0.14 & 0.159 \\ 
  S2 & A & -0.00 & 0.29 & 4.90 & 0.01 & 6.258 & 4.89 & 0.005 \\ 
  S2 & A(aug) & 0.66 & 0.47 & 0.56 & 0.26 & 0.086 & 0.30 & 0.145 \\ 
  S2 & W & 0.01 & 0.33 & 3.41 & 0.01 & 2.402 & 3.39 & 0.010 \\ 
  S2 & W(aug) & 0.69 & 0.51 & 0.73 & 0.31 & 0.059 & 0.42 & 0.160 \\ \hline
  S3 & T & 0.10 & 0.20 & 1.17 & 0.08 & 0.110 & 1.09 & 0.026 \\ 
  S3 & S & 0.15 & 0.20 & 0.41 & 0.13 & 0.079 & 0.27 & 0.034 \\ 
  S3 & R & 0.27 & 0.32 & 0.34 & 0.16 & 0.088 & 0.18 & 0.069 \\ 
  S3 & X & 0.31 & 0.29 & 0.43 & 0.23 & 0.111 & 0.20 & 0.070 \\ 
  S3 & CF & 0.10 &  &  &  &  &  & 0.000 \\ 
  S3 & CF($\pi$) & 0.09 & 0.25 & 0.26 & 0.09 & 0.190 & 0.17 & 0.021 \\ 
  S3 & BCF & 0.20 & 0.21 & 0.06 & 0.17 & 0.066 & -0.11 & 0.035 \\ 
  S3 & A & 0.00 & 0.16 & 4.57 & 0.02 & 3.020 & 4.55 & 0.003 \\ 
  S3 & A(aug) & 0.39 & 0.36 & 0.45 & 0.15 & 0.091 & 0.30 & 0.086 \\ 
  S3 & W & -0.01 & 0.03 & 10.37 & 0.00 & 3.329 & 10.37 & 0.000 \\ 
  S3 & W(aug) & 0.18 & 0.30 & 0.47 & 0.09 & 0.073 & 0.38 & 0.049 \\ \hline
  S4 & T & 0.20 & 0.33 & 1.35 & 0.08 & 0.082 & 1.27 & 0.057 \\ 
  S4 & S & 0.29 & 0.36 & 0.55 & 0.09 & 0.076 & 0.47 & 0.066 \\ 
  S4 & R & 0.14 & 0.34 & 0.47 & 0.03 & 0.068 & 0.44 & 0.022 \\ 
  S4 & X & 0.30 & 0.34 & 0.62 & 0.09 & 0.075 & 0.54 & 0.058 \\ 
  S4 & CF & 0.05 & 0.31 & 0.41 & 0.02 & 0.202 & 0.39 & 0.015 \\ 
  S4 & CF($\pi$) & 0.06 & 0.29 & 0.34 & 0.03 & 0.183 & 0.31 & 0.018 \\ 
  S4 & BCF & 0.04 & 0.33 & 0.21 & 0.02 & 0.110 & 0.20 & 0.016 \\ 
  S4 & A & 0.90 & 0.76 & 74.19 & 0.74 & 5.690 & 73.46 & 0.206 \\ 
  S4 & A(aug) & 0.14 & 0.33 & 0.54 & 0.01 & 0.051 & 0.52 & 0.012 \\ 
  S4 & W & 0.88 & 0.67 & 103.33 & 0.80 & 8.274 & 102.53 & 0.188 \\ 
  S4 & W(aug) & 0.27 & 0.33 & 0.75 & 0.02 & 0.061 & 0.74 & 0.019 \\ 
   \hline
\end{tabular}
\caption{Performance of simulation scenarios, $100$ runs per scenario\label{tab.valuesS1S2S3S4}}
\end{table}
\newpage

\appendix
\section{. Summary of HTE evaluation methods}\label{sec.summary.tab}
Here we provide high-level summaries of most important methods for HTE evaluation described in  these tutorial that would help the reader to navigate in this diverse and crowded space.

Table~\ref{tab.appendix.keyfeatures} characterises the ML methods that identify heterogeneous treatment effects using their key features (in the columns of the table) listed below (some of them were introduced in Section~\ref{key.features} and are, with some additional categories.
The last column contains information on available software for subgroup identification. For convenience the  methods are grouped within the four types described in Section~\ref{tech.details}.

\begin{enumerate}
\item \textit{Modeling type}: Freq (Frequentist), Bayes (Bayesian), P (parametric), SP (semiparametric), NP (nonparametric). Note that many methods can choose between using parametric or generic ML methods as building blocks therefore any type of models (P/SP/NP) can be used 
\item \textit{Application type}: RCT (randomized controlled/clinical  trials) or OS (observational studies), if the method can accommodate both types of data, we will list it as RCT, OS.
\item \textit{Dimensionality of the covariate space}: Low ($p \le 10$), Medium ($ 10 < p \le 100$), H (high; $p > 100$). Note that some methods can handle high dimensional prognostic effects but assume the low dimensional set of predictive (effect modifying) variables
\item \textit{Results} produced by the method: B (list of biomarkers identified as predictive or a list of candidate biomarkers with associated variable importance scores or rankings), CATE (individual benefit scores, i.e., estimated CATE, as a function of covariates), ITR (optimal treatment regimens as a function of covariates), S (subgroup signatures as a function of covariates). Note that any method producing CATE can be also used to construct ITR and S by thresholding on CATE, however we only label as ITR or S methods that directly target such
\item \textit{Inferential support} provided by the method (Yes[type]/No). This can relate to various components of HTE; therefore if provided we further classify the HTE inference by type [in brackets] using the same letter codes as for the associated types of outputs: S (bias-corrected treatment effect estimates and confidence intervals in subgroups), ITR (Value of the estimated treatment regimen and CI), CATE (bias-correct estimator and confidence intervals for conditional treatment effect). Additionally, we use letter G for testing the \textit{global} hypothesis of presence of HTE.
\item \textit{Availability of software} implementation: R [name] (R package available on the CRAN or github website), B (R code or reference to implementation by the developers available on the Biopharmaceutical Network website at \texttt{http://biopharmnet.com/subgroup-analysis/}. 
\end{enumerate}

As we see from Table~\ref{tab.appendix.keyfeatures} most  methods for HTE discovery do not provide any inferential support as part of their implementation. Fortunately, there are generic methods (some are described in Section~\ref{sec.post.inf}) that can be used as ``wrappers'' or ``stand-alone'' tools assuming input from various generic ML methods or invoke ML methods from within using available ML libraries (such as \texttt{caret}). Table~\ref{tab.appendix.inference} describes some of such methods.


\newpage

\begin{table}[t!]
\begin{center}
\begin{tabular}{lcccccc} \hline
Method & Modeling & Application  & Dimen- & Results & Inferential  & Software \\
   &  type (1)  &  type (2)  & sionality (3) & type (4)   & support (5) & (6) \\\hline
   \multicolumn{6}{c}{Global outcome modeling} \\ \hline
Virtual Twins\cite{Foster2011} & Freq/NP & RCT, OS & High & CATE, S & No & R [\textbf{aVirtualTwins}] \\
S-,T- X-learner\cite{Kuenzel2019} & Freq/SP,NP & RCT,OS & High & CATE, S & No & R [\textbf{rlearner}]$^1$ \\
\\
\hline
  \multicolumn{6}{c}{Global treatment effect modeling} \\ \hline
Interaction Trees\cite{Su2009}  & Freq/NP & RCT & Medium & S & No & B\\
GUIDE\cite{Loh2015}  & Freq/NP & RCT & Medium & B,S & Yes & B\\
Model-based trees and forests\cite{Seibold2016}  & Freq/NP & RCT & High & B,S & Yes & R [\textbf{model4you}] \\
Causal forests\cite{WagerAthey2018,AtheyGRF2019}  & Freq/NP & RCT,OS & High & B,CATE & Yes[G, CATE] & R [\textbf{grf}] \\
Bayesian causal forests\cite{HahnBARTCRAN,Hill2011} & Bayes/NP & RCT,OS & High & CATE & Yes[CATE] & R [\textbf{bcf, bartCausal}] \\
Bayesian linear models\cite{Jones2011, WangLouis2018} & Bayes/P & RCT & Low & CATE & Yes[CATE, S] & R [\textbf{DSBayes,beanz}] \\
Modified loss methods\cite{Huling2021} & Freq/P,SP,NP & RCT, OS & High & CATE & No  & R [\textbf{personalized}] \\
R-learner\cite{Nie2020} & Freq/P,SP,NP & RCT,OS & High & CATE & No  & R [\textbf{rlearner}]$^1$ \\
\hline
    \multicolumn{6}{c}{Direct modeling of ITR} \\ \hline
AIPW estimator\cite{Zhang2012} & Freq/SP & RCT,OS & Medium & ITR & No &  R [\textbf{DynTxRegime}] \\  
OWL\cite{zhao2012}, RWL\cite{Zhou2017}, AOL\cite{Liu2018} & Freq/P,SP,NP & RCT,OS & High & ITR & No & R [\textbf{DTRlearn2}]\\ 
Tree-based ITR\cite{laber2015,Tao2018,AtheyWager2021} & Freq/SP,NP & RCT, OS & Medium & ITR & No & R [\textbf{T-RL$^2$,policytree}]\\ 
\hline
    \multicolumn{6}{c}{Direct subgroup identification} \\ \hline
SIDES and SIDEScreen\cite{Lipkovich2011,Lipkovich2014} & Freq/NP & RCT & Medium & B,S & Yes[G,S] & B,R [\textbf{SIDES, rsides}]\\
TSDT\cite{TSDTCRAN2022}  & Freq/NP & RCT & Medium & B,S & Yes[S] & R [\textbf{TSDT}]\\
PRIM\cite{Chen2015} & Freq/NP & RCT & Medium & S & No & R [\textbf{SubgrpID}]\\
Sequential-Batting\cite{Huang2017} & Freq/NP & RCT & Medium & S & Yes[S] & R [\textbf{SubgrpID}]\\
CAPITAL\cite{CaiCAPITAL2022} & Freq/NP & RCT & Medium & S & No & R [\textbf{policytree}]\\
Bayesian Model Averaging\cite{Bornkamp2017}  & Bayes/NP & RCT & Low & S & Yes[S] & R [\textbf{subtee}]\\ 
\hline
\end{tabular}
{\footnotesize \begin{flushleft} 
1: available at GitHub \textbf{xnie/rlearner}\\
2: available at GitHub \textbf{Team-Wang-Lab/T-RL}
 \end{flushleft}}
\caption{Key features of selected HTE evaluation methods}\label{tab.appendix.keyfeatures}
\end{center}
\end{table}

\newpage

\begin{table}[h!]
\centering
\begin{tabular}{lccccc}
  \hline
 Method & Modeling  & Application  & Target of inference  & Software \\ 
  &  (Freq/Bayes) & (RCT/OS) & (G/B/CATE/S) & implementation \\ 
  \hline
  Inference on the overall HTE\cite{Chernozhukov2020,ImaiLi2022} & Freq & RCT & G & R [\textbf{evalITR,GenericML}] \\
  Inference on ITR\cite{ImaiLi2021} & Freq & RCT & ITR & R [\textbf{evalITR}] \\
  Credible subgroups\cite{GuoHe2021} & Bayes & RCT,OS & ITR & R [\textbf{credsubs}] \\
  Inference on selected subgroups in RCT\cite{GuoHe2021} & Freq & RCT & S & B \\
  Inference on selected subgroups in OS\cite{Guo2021} & Freq & OS & S & R [\textbf{debiased.subgroup}]$^{1}$ \\
  FDR controlled biomarker identification\cite{Sechidis2021,Zimmermann2024} & Freq & RCT,OS & B &  R [\textbf{knockofftools}]$^{2}$\\
   \hline
\end{tabular}
{\footnotesize \begin{flushleft} 
1: available at GitHub \textbf{WaverlyWei/debiased.subgroup}\\
2: available at GitHub \textbf{Novartis/knockofftools} \end{flushleft}}
\caption{``Stand-alone'' inferential tools for HTE\label{tab.appendix.inference}}
\end{table}

\end{document}